\documentclass[a4paper,11pt]{article}

\usepackage{amsmath}
\usepackage{amssymb}
\usepackage{bm}
\usepackage{bbm}
\usepackage{subcaption}
\usepackage{color}
\usepackage{longtable,booktabs}
\usepackage[colorlinks, linkcolor=red]{hyperref}
\usepackage{cite}
\usepackage{verbatim}
\usepackage{float}
\usepackage{graphicx}
\usepackage{tikz}
\usetikzlibrary{positioning,arrows}
\usepackage[left=2cm,right=2cm,top=2cm,bottom=2cm]{geometry}

\usepackage{setspace}

\numberwithin{equation}{section}

\newcommand{\D}{d}

\newcommand{\N}{\mathbb{N}}

\renewcommand{\S}{\mathcal{S}}

\newcommand{\fne}{\tilde{f}}

\renewcommand{\xi}{\boldsymbol{f}}

\renewcommand{\zeta}{\boldsymbol{f}}

\newcommand{\1}{\mathbbm{1}}

\newcommand{\K}{\mathcal{K}}

\newtheorem{theorem}{Theorem}[section]
\newtheorem{lemma}{Lemma}

\newtheorem{corollary}{Corollary}[section]
\newtheorem{definition}{Definition}[section]
\newtheorem{conjecture}{Conjecture}[section]
\newtheorem{example}{Example}[section]

\newcommand{\M}{\mathcal{M}}

\newcommand{\awd}{asymptotically well designed }
\newcommand{\aw}{asymptotically well  }

\newcommand{\Wuu}[1]{{#1}}
\newcommand{\Wuuu}[1]{{#1}}
\newcommand{\Wuuuu}[1]{{#1}}

\title{Sefishness need not be bad\footnote{Zijun Wu (\texttt{zijunwu1984a@163.com}) and Rolf H. M{\"o}hring (\texttt{rolf.moehring@me.com}) are with Department of Computer Science at Hefei University, Hefei, Anhui, China;
Yanyan Chen (\texttt{cdyan@bjut.edu.cn}) is with College of Metropolitan Transportation at Beijing University of Technology, Beijing, China; Dachuan Xu
(\texttt{xudc@bjut.edu.cn}) 
is with College of Applied Science at Beijing University of Technology, Beijing, China.}}
\author{Zijun Wu, Rolf H. M{\"o}hring, Yanyan Chen and Dachuan Xu}

\begin{document}
	
	\maketitle

	\begin{center}
		\textbf{Abstract}\\[5ex]
		\begin{minipage}{0.95\textwidth}
			We investigate the price of anarchy (PoA) in non-atomic congestion games when the total demand $T$ gets very large.
			First results in this direction have recently been obtained by \cite{Colini2016On, Colini2017WINE, Colini2017arxiv} for routing games and show that the PoA converges to 1 when the growth of the total demand $T$ satisfies certain regularity conditions. We extend their results by developing a \Wuuu{new} framework for the limit analysis of \Wuuuu{the PoA that offers strong techniques such as the limit of games and applies to arbitrary growth patterns of $T$.} \Wuuu{We} show that the PoA converges to 1 in the limit game regardless of the type of growth of $T$ for a large class of cost functions that contains all polynomials and all
			regularly varying functions. 
			For routing games with BPR \Wuu{cost} functions, we show in addition that socially optimal strategy profiles converge to \Wuu{equilibria} in the limit game, and that  PoA$=1+o(T^{-\beta})$, where  $\beta>0$ is the degree of the \Wuu{BPR} functions. However, the precise convergence rate depends crucially on the the growth of $T$, which shows that a conjecture proposed by \cite{O2016Mechanisms} need not hold.\\[1ex]
			
			\begin{flushleft}
			\end{flushleft}
		\end{minipage}
	\end{center}
	




\maketitle

\newpage

\tableofcontents

\newpage

\section{Introduction}

Traffic congestion has become a serious problem in many cities of China.  In 2017, more than 71\% of the major Chinese cities  have been suffering from congestion during rush hours, see
\cite{amap2017}.
Congestion does not only considerably enlarge travel times, but also causes serious economic
losses. In 2017, \Wuu{this loss in Beijing amounted} to 3.1\% of its  annual GDP, see \cite{baidu2017}.

This raises the natural question if and how much traffic conditions would improve if the users would follow the socially optimal routing pattern instead of letting them selfishly choose their quickest route.

This topic has been studied intensively during the last two decades, both in routing networks and the more general (atomic and non-atomic) congestion games. Selfish routing leads to a Wardrop equilibrium \cite{Wardrop1952ROAD} in routing \Wuu{games.} 
The ratio between its cost and the socially optimal cost is known as the {\em price of anarchy} (PoA), see \cite{Papadimitriou1999}, and measures the inefficiency of selfish routing. 

In principle, this inefficiency can get arbitrarily large in routing \Wuuu{games}, as was already shown by \cite{Pigou1920} in his famous example, see Figure~\ref{fig:Pigou}(a). Traffic from $o$ to $t$ may choose between the lower arc with constant travel time $1$ and the upper arc with travel time $x^{\beta}$, where $x$ is the amount of traffic on that arc and 
$\beta\ge 0$ is a constant.

\begin{figure}[!htb]
	\centering
	\begin{subfigure}{0.32\textwidth}
			\centering
		\Wuu{\begin{tikzpicture}[
			>=latex
			]
			\node[scale=0.4,circle,fill=black,label=left:$o$](o){};
			\node[right =1.1of o](m){};
			\node[scale=0.4,circle,fill=black,label=right:$t$,right =1.1of m](t){};
			\node[above of=m](xbeta){$x^\beta$};
			\node[below of=m](One){$1$};
			\draw[-,thick] (o) to [out=90,in=180] (xbeta);
			\draw[->,thick] (xbeta) to [out=0,in=90] (t);
			\draw[-,thick] (o) to [out=-90,in=180] (One);
			\draw[->,thick] (One) to [out=0,in=-90] (t);
			\end{tikzpicture}}
		\subcaption{\Wuu{Pigou's game}}
	\end{subfigure} 
\begin{subfigure}{0.65\textwidth}
		\centering
	\begin{tikzpicture}[
	domain=-1:7,>=latex,scale=0.7
	]
		\draw[->,thick] (-0.25,0) to (6.7,0) node[right,scale=0.7]{$T$};
		\draw[->,thick] (0,-0.25) to (0,3) node[left,scale=0.7]{PoA};
		\draw[smooth,thick,color=blue,domain=1:6.5] plot (\x,{\x/(\x-0.8*pow(5,-0.25))});
		\draw[-,dashed,very thin,color=black] (0,1) node[left,scale=0.7]{$1$} to (6.5,1);
		\draw[-,dashed,very thin,color=black] (0,2) node[left,scale=0.7]{$2$} to (6.5,2);
		\draw[-,dashed,very thin,color=black] (1,0) node[below,scale=0.7]{$1$} to (1,2.8);
		\draw[-,dashed,very thin,color=black] (2,0) node[below,scale=0.7]{$2$} to (2,2.8);
		\draw[-,dashed,very thin,color=black] (3,0) node[below,scale=0.7]{$3$} to (3,2.8);
		\draw[-,dashed,very thin,color=black] (4,0) node[below,scale=0.7]{$4$} to (4,2.8);
		\draw[-,dashed,very thin,color=black] (5,0) node[below,scale=0.7]{$5$} to (5,2.8);
		\draw[-,dashed,very thin,color=black] (6,0) node[below,scale=0.7]{$6$} to (6,2.8);
		\draw[smooth,thick,color=blue,domain=0:{pow(5,-0.25)}]
		plot (\x,1);
		\draw[smooth,thick,color=blue,domain={pow(5,-0.25)}:1]
		plot (\x,{pow(\x,5)/(\x-pow(5,-0.25)+pow(5,-1.25))});
	\end{tikzpicture}
	\subcaption{\Wuu{The plot of the PoA for $\beta=4$}}
\end{subfigure}
	\caption{Pigou's example}
	\label{fig:Pigou}
\end{figure}

The PoA equals $\frac{T}{T-(\beta+1)^{-1/\beta}\cdot \big(1-(\beta+1)^{-1}\big)},$ when the total travel demand $T\ge 1$. Obviously,
fixing $T = 1$ and considering all possible $\beta$, the  PoA 
 tends to $\infty$ as $\beta \to \infty$. But if we consider $\beta$ as fixed (which is natural in routing networks) and $T$ as a variable, then we obtain the plot of the PoA in
 Figure~\ref{fig:Pigou}(b) as a function of the demand $T$. 
It shows that the PoA is only large in a small neighborhood of $T = 1$, and tends to 1 with growing demand $T$. 

This ``convergence''  of the PoA to 1 in traffic networks with large demand has been observed before in experiments by \cite{Youn2008Price}, \cite{O2016Mechanisms}, and \cite{Monnot2017}. When the travel time
 functions  are BPR functions of degree $\beta>0$, \cite{O2016Mechanisms} even conjectured that the PoA obeys a power law of the form $1+O(1/T^{-2\beta})$ and thus converges to 1 very fast.

A theoretical analysis of the convergence of the PoA has been missing until the recent seminal work of \cite{Colini2016On,Colini2017WINE,Colini2017arxiv}. They established for the first time conditions under which the PoA converges to 1 for growing demand. 

One condition guarantees a kind of ``regular growth'' of the arc travel times, another a certain ``tightness"  of paths and  origin-destination pairs, and a third asks that there are tight origin-destination pairs \Wuuu{that} ``route a non-negligible amount of normalized demand'' in the limit for the given sequence of growing demands.

Loosely speaking, ``regular growth'' means that the travel time functions $\tau_a(x)$ grow with increasing traffic $x$ at constant rates in comparison with a certain global {\em benchmark function} $g(x)$, i.e., the ``normalized'' travel times $\frac{\tau_a(x)}{g(x)}$ have a (possibly infinite) limit $\alpha_a := \lim_{x\to\infty}\frac{\tau_a(x)}{g(x)}$. These growth rates $\alpha_a$ are used to classify arcs $a$ into \emph{fast} ($\alpha_a = 0$), \emph{slow} ($\alpha_a = \infty$), or \emph{tight} ($0 < \alpha_a < \infty$). 

Paths are likewise fast, slow or tight, based on their slowest edge; and an origin-destination pair is tight if its fastest path is tight. The ``tightness'' condition then requires that every origin-destination pair has a path that is not slow and that at least one \Wuu{origin-destination} pair is tight. 

Finally, the third condition asks that all tight \Wuu{origin-destination} pairs together carry a non-negligible normalized demand for the given sequence $(d^{(n)})_{n\in \N}$ with growing total demand. 

When these three conditions hold for the traffic \Wuuu{game} and the sequence $(d^{(n)})_{n\in \N}$ of growing total demand, then the PoA converges to 1. We will come back to their results in sections~\ref{sec:relatedWork}, \ref{sec:coliniFirstView} and \ref{sec:coliniSecondView} for more details. These results undoubtedly form a milestone in the analysis of the PoA, but the requirements of ``regular growth'' and  ``non-negligible'' traffic on tight origin-destination pairs seem strong and somewhat limiting.

\subsection{Our contributions}

We aim to deepen this limit analysis of the PoA. To that end, we develop a new framework that can cope with non-regular growth patterns of the demand sequence. 
An important role in our approach is played by the notion of \emph{limit games}. These are games that arise as the limit of growth sequences of the demands. \cite{Colini2016On, Colini2017WINE, Colini2017arxiv} consider demand sequences in which---viewed \Wuuu{according to our approach}---every subsequence has the same unique limit game. 

In general, there may be multiple limit games of the same demand sequence. We introduce an ``asymptotic decomposition" technique to capture these different limits of a game. This technique is  crucial to show that the convergence of the PoA to 1 does not depend on a ``regular'' growth of the demands, but on the existence of these limit games. 
When they exist, the PoA is 1 in the limit and we call the game \emph{asymptotically well designed} to reflect the surprising property that selfish routing in these games already leads to the social optimum in the limit for every sequence with growing total demand. 

We develop this theory for general non-atomic congestion games and
show that non-atomic congestion games with arbitrary regularly varying cost functions are asymptotically well designed---and that without any restrictions on the growth of the  demand sequence. The class of regularly varying functions is very extensive, \Wuu{which includes, e.g.,}
polynomials, logarithms and logarithmic polynomials,
and is closed under finite sums and products, see \cite{Bingham1987Regular}.

Some of our results can be strengthened for routing games with BPR \Wuu{cost (travel time)} functions. 
Socially optimal strategy profiles in such games \Wuuu{approximate equilibria} 
as the total travel demand $T$ increases, and 
the cost of \Wuu{equilibria} and the social optimum can be
efficiently approximated with the use of the limit game for large $T$.

Also, the PoA follows the power law $1+o(T^{-\beta})$, where  $\beta$ is the degree of the \Wuu{BPR} functions, which is usually $4$ in practice. 
Our detailed analysis shows that the decay rate can vary with the demand sequence. For each $\beta\ge 1,$ there is an instance with multiple 
origin-destination pairs such that, for each $\theta\in [\beta+1,2\cdot\beta),$ 
there is a sequence $(d^{(n)})_{n\in \N}$ with growing total demand $T(d^{(n)})$ for which PoA$(d^{(n)})=1+\Theta(T(d^{(n)})^{-\theta})$. 
 So the above-mentioned conjecture by \cite{O2016Mechanisms} that PoA$(d^{(n)})=1+O(T(d^{(n)})^{-2\beta})$
need not hold.

Finally, to empirically verify our theoretical findings, we have analyzed real traffic data within the 2nd ring road of Beijing in an experimental study. Our empirical results definitely validate our findings. They show that the current traffic in that area of Beijing is already far beyond the point at which the PoA is 1. So no route guidance policy can reduce the total travel time without significantly reducing the current huge total travel demand.

\subsection{Related work}\label{sec:relatedWork}

\cite{Papadimitriou1999} proposed to quantify the inefficiency of equilibria in arbitrary congestion games from a
{\em worst-case} perspective. \Wuuu{This} resulted in the concept of the {\em price of anarchy} (PoA)  
that is usually defined as the ratio of cost of the worst case Nash equilibrium over the cost of the social optimum \cite{Papadimitriou2001Algorithms}.

A wave of research has been started with the pioneering paper of \cite{Roughgarden2000How} on the PoA of routing networks with affine linear cost functions. Examples are  \Wuu{\cite{Roughgarden2001Designing,Roughgarden2000How,Roughgarden2004Bounding,Roughgarden2005Selfish,Roughgarden2015Intrinsic,CorreaSchSti04,Correa2005On,Perakis2007}.} They investigated the worst-case upper bound of the PoA 
for different types of cost functions
$\tau_a(\cdot),$ and analyzed the influence of the
network topology on this bound. 
In particular, they
showed that this bound 
is $\tfrac{4}{3}$ when
all $\tau_a(\cdot)$ are affine linear (\cite{Roughgarden2000How}), and
$\Theta(\rho/\ln \rho)$ when all $\tau_a(\cdot)$ are polynomials with maximum degree $\rho>0$ (\cite{Roughgarden2004Bounding} and \cite{Roughgarden2015Intrinsic}).
Moreover, they proved that this bound
is independent of the network topology, \Wuu{see,
e.g., }
\cite{Roughgarden2002The}. 
They also developed a $(\lambda,\mu)$-smooth
method by which one can obtain 
a {\em tight} and {\em robust} worst-case upper bound 
of the PoA for a large class of
cost functions, \Wuu{see, e.g.,} \cite{Roughgarden2002The}, \cite{Roughgarden2004Bounding} and \cite{Roughgarden2015Intrinsic}.
This method was then reproved by \cite{Correa2005On} from
a geometric perspective. 
See
\cite{Roughgarden2007Introduction} for a comprehensive overview of the early development of that research.  
\Wuuu{\cite{Perakis2007} generalized the worst-case analysis to routing games with
non-separable, asymmetric and nonlinear cost functions.}

Recent papers have also empirically studied the PoA in traffic networks with real data.   
\cite{Youn2008Price}
observed that the empirical PoA depends crucially on the total travel demand. Starting from~1, it grows with some oscillations, and
ultimately becomes 1 again as the total demand increases. A similar observation was made by \cite{O2016Mechanisms}.
They also conjectured that the PoA converges to $1$ in the power law $1+O\big(T^{-2\beta}\big)$ when the total travel
demand $T$ becomes large. \cite{Monnot2017} showed that routing choices of commuting students in Singapore are near-optimal and that the empirical PoA is much smaller than known worst-case upper bounds. Similar observations have been reported by \cite{Jahn2004System}.

The closest to our paper are the results by \cite{Colini2016On,Colini2017WINE,Colini2017arxiv}. They were the first to theoretically analyze the convergence of the PoA in network games with growing total demand.

In a first step, \cite{Colini2016On} considered networks with only one 
origin-destination pair $(o,t)$ and identified two special cases in which the PoA converges to 1. They generalized that substantially  in \cite{Colini2017WINE,Colini2017arxiv} to arbitrary network games with multiple origin-destination pairs 
$(s_k,t_k)$, ${k\in \K}$. In these networks, they considered demand sequences $(d^{(n)}=(d_k^{(n)})_{k\in \K})_{n\in \N}$ with total demand $T(d^{(n)}) = \sum_{k  \in \K}d_k^{(n)}$ satisfying $\lim_{n \to \infty}T(d^{(n)})=\infty.$ 

Their main result then states that the PoA converges to 1 for such a sequence  $\big(d^{(n)}\big)_{n\in \N}$  if the above-mentioned conditions of (i) ``regular growth''of the arc travel time functions $\tau_a(x)$, (ii)  ``existence of a ``non slow'' path for every user in the limit'', and (iii) ``non-negligible'' normalized demand on ``tight'' origin-destination pairs are satisfied. We will now make (i)--(iii) more precise.

Condition (i) (``regular growth'') means that the above-mentioned global benchmark function $g(x)$ for the arc travel time functions 
$\tau_a(\cdot)$ is \emph{regularly varying} and that \Wuu{$\lim_{x \to \infty} \frac{\tau_a(x)}{g(x)}$} is a non-negative constant $\alpha_a$ or $\infty$ for each arc $a.$

The existence of such a benchmark function is a strong requirement. A function $g(\cdot)$ is regularly varying iff \Wuuu{$\lim_{t \to \infty} \frac{g(t\cdot x)}{g(t)} = q(x) \in (0,\infty)$} for all $x>0$. Karamata's Characterization Theorem (see \cite{Bingham1987Regular}) then implies that $g(x) = x^\rho\cdot Q(x)$ where $Q(x)$ is slowly varying, i.e.,  \Wuu{$\lim_{t \to \infty} \frac{Q(t\cdot x)}{Q(t)} = 1$} for all $x>0$. The constant $\rho$ is called the \emph{regular variation index} of $g(\cdot)$.

Condition (ii) (existence of of a ``non slow'' path for every origin-destination pair) has already been discussed above and seems natural.

Condition (iii) (routing a non-negligible amount of normalized demand on tight \Wuu{origin-destination} pairs) is again strong and requires  \Wuu{a particular demand growth pattern depending crucially on} the travel time functions and \Wuuu{the} benchmark function. Recall that an origin-destination pair $(o_k,t_k)$ is ``tight'' if its fastest path is tight, i.e., the largest $\alpha_a$ of the arcs 
$a$ contained in the $(o_k, t_k)$-path is finite and positive. Condition (iii)  requires that 
$
	\varliminf_{n\to \infty} \sum_{(o_k, t_k)  \text{ is tight}}\frac{d_k^{(n)}}{T(d^{(n)})}>0.
$

Other results in \cite{Colini2016On,Colini2017WINE,Colini2017arxiv} are an example of a routing game with
non-regularly varying travel time functions for which the PoA {\em diverges}
when total demand $T \to \infty,$ and  a special convergence result 
of 
PoA$(d^{(n)})=1+O\big(\frac{1}{T(d^{(n)})}\big)$ when the travel time functions $\tau_a(\cdot)$ are polynomials and the demand sequence
$(d^{(n)})_{n\in \N}$ fulfills the above condition (iii).

Besides, they also considered the convergence of the PoA to 1 under conditions similar with (i)--(iii)
when $\lim_{n \to \infty}T(d^{(n)})=0$ (the ``light'' traffic case).

\subsection{Outline of the paper}

The paper is organized as follows. We develop our approach for general non-atomic congestion games. These games and the class of asymptotically well designed games are introduced in
Section~\ref{sec:MODEL}.
Section~\ref{sec:PoA_Limit} presents our techniques and main results. Finally, Section \ref{sec:RoutingPoA} contains  our analysis of routing games with \Wuu{BPR cost} functions and the empirical study about the traffic in Beijing. We conclude with a short summary in Section \ref{sec:conclusion}.
All proofs have been moved to an Appendix.

\section{Model and preliminaries} \label{sec:MODEL}

\subsection{Non-atomic congestion games, routing games, and the price of anarchy}
\label{sec:Model_Formalization}

A \emph{non-atomic congestion game} $\Gamma$ consists of a finite non-empty set $\K$ of groups $k\in \K$ of \emph{players} or \emph{users} who compete for a finite set $A$ of \emph{resources}, \Wuu{see, e.g.,} \cite{Dafermos1969} and \cite{Rosenthal1973A}. Each group $k\in \K$ has a \emph{demand} $d_k > 0,$  
and must satisfy that by choosing one or more strategies $s$ from a finite set $\S_k$ of {\em candidate strategies} that are {\em only}  available to  group $k$. Each strategy $s\in \S:=\bigcup_{k\in \K}\S_k$ is a {\em non-empty} subset of $A,$ and users of group $k$ distribute their
demand to the strategies $s\in \S_k$ in \Wuuu{arbitrary} quantities
$f_s\ge 0,$ so $d_k=\sum_{s \in S_k} f_s.$ Users choose their strategies independently, and their joint decisions 
result in a \emph{strategy profile} or simply \emph{profile} $f=(f_s)_{s\in \S}.$ The
 joint consumption or use of resource $a \in A$ is obtained as $f_a := \sum_{k \in \K}\sum_{s \in \S_k : s \ni a}f_s$.
 \Wuuuu{Technically, the term ``non-atomic'' means that  the demand $d_k$ is arbitrarily splittable, and is usually interpreted that 
 	each individual user is too infinitesimal to influence others.}

 Competition happens through \Wuuu{the cost of} jointly used resources. 
 Each resource $a\in A$ 
 has a {\em non-negative, non-decreasing and
 	continuous} (unit) cost function $\tau_a(\cdot)$ 
 depending only on the amount $f_a$ consumed by its users. 
 The \emph{cost of a strategy $s\in \S$} is  $\tau_s(f):=
 	\sum_{a\in A: a\in s}
 	\tau_a(f_a)$ and the 
 \emph{(social) cost of a strategy profile} $f$ is  $C(f):=\sum_{k \in \K}\sum_{s \in \S_k}f_s\cdot \tau_s(f) = \sum_{a\in A}f_a\cdot \tau_a(f_a)$.

We illustrate these notions on routing games that \Wuuuu{form standard examples} of non-atomic congestion \Wuuu{games} in which each \Wuu{individual} user \Wuu{controls
	an infinitesimal fraction of traffic and has no influence
    on the routing choice of others}, see \cite{Roughgarden2007Introduction-routing} for an introduction. 

Routing games are used to model static traffic in networks. \Wuu{We} are given a directed graph $G=(V,A)$ (the traffic network), and {\em origin-destination (O/D) pairs} $(o_k,t_k)$ with (traffic) demands $d_k, k \in \K$. 
This defines a non-atomic congestion game $\Gamma$ as follows. The groups of $\Gamma$ are the O/D pairs, the demand of group $k$ is $d_k$,  the resources are the arcs in $A$, the cost function $\tau_a(\cdot)$
	is the travel time function of
	arc $a\in A,$ the strategies of group $k\in\K$ are the $(o_k,t_k)$-paths leading
from the origin $o_k$ to the destination $t_k$,  a strategy profile 
$f=(f_s)_{s\in \S}$ is a multi-commodity flow satisfying the demands $d_k,$ and $f_s$ is the flow value along path $s$. Finally, 
the joint consumption $f_a$ of resource $a \in A$ is the flow value of arc $a$.

Let us \Wuu{now continue discussing} general non-atomic congestion games.
A profile with minimum cost
is called a {\em social optimum} of $\Gamma$, abbreviated as \emph{SO profile}. Such a profile $f^*$ is considered as an {\em ideal} state of
the game, as $C(f^*)\le C(f)$ for
each profile $f,$ and so resources are consumed in 
the  globally most efficient way.  Due to our assumptions on the cost functions, every non-atomic congestion game has an SO profile.

Obviously, users will choose their strategies $s\in\S$ selfishly and want to minimize their own cost $\tau_s(f)$, 
\Wuuu{which}
leads to a \emph{Wardrop equilibrium} (\cite{Wardrop1952ROAD}) $\tilde{f}=(\tilde{f}_s)_{s\in \S}$
in which every group $k$ uses only strategies $s \in \S_k$ (i.e., $\tilde{f}_s > 0$) satisfying $\tau_s(\tilde{f}) \le \tau_{s'}(\tilde{f})$ for all $s' \in \S_k$. 
	\Wuu{Wardrop equilibria} have the same resource cost, \Wuu{i.e.,} $\tau_a(\tilde{f}_a)=
\tau_a(\tilde{f}'_a)$ for each $a\in A$ for any
two
Wardrop equilibria $\tilde{f}$ and $\tilde{f}'$ of
$\Gamma$, see, e.g., \cite{Smith1979The}
and \cite{Roughgarden2000How}. \Wuu{Hence}, all Wardrop equilibria $\tilde{f}$ have the same cost
$C(\tilde{f}).$ \Wuuu{Under our assumptions,} Wardrop equilibria are pure Nash equilibria of $\Gamma$ from a game-theoretic perspective, \Wuu{see, e.g.}, \cite{Roughgarden2000How}. We will therefore denote these equilibria also as \emph{NE profiles} in the sequel.

Consider now a given demand vector $d=(d_k)_{k\in \K}$ of $\Gamma$. The price of anarchy for $d$ is defined as the ratio of the worst cost of a NE profile over the optimum cost. Since all NE profiles have the same cost in our case, the PoA is given as
$
\text{PoA}(d):=
\frac{C(\tilde{f})}{C(f^*)},
$
where $\tilde{f}$ is an NE profile of $\Gamma$ for $d$ and $ 
f^*$ is an SO profile of  $\Gamma$  for $d,$
\Wuu{see, e.g.,} \cite{Papadimitriou2001Algorithms}.
The notation indicates that the PoA
depends on the vector $d.$

In the sequel we will use the term \emph{game} to mean a non-atomic congestion game.
	Moreover, $\tilde{f}$ and
	$f^*$ will always denote an NE profile and an SO profile, respectively. 
To avoid degenerate cases \Wuu{in proofs}, we \Wuu{assume} that every resource $a$ is part of some strategy, and that every strategy $s$ is non-empty, i.e.,
	\begin{equation}
	\label{eq:StrategiesScarce}
	\{s\in \S: 
	a\in s \}  \ne \emptyset \quad \forall a \in A \quad 
	\text{and} \quad s \ne \emptyset \quad \forall s\in \S.
	\end{equation}
\Wuu{Moreover,} we assume that the {\em total demand}
$T(d):=\sum_{k  \in \K}d_k>0$ and that each cost function $\tau_a(x) \neq 0$
for some $x>0.$ Otherwise, the demand
may	have no influence on the total cost of users.
Finally, we call a demand sequence $\big(d^{(n)}=(d_k^{(n)})_{k\in\K}\big)_{n\in \N}$ {\em unbounded} if its  total demand $T(d^{(n)})=\sum_{k\in \K}d^{(n)}_k \to \infty$ as $n\to \infty$.

\subsection{Asymptotically well designed games}

There are games $\Gamma$ in which PoA$(d) = 1$ for all possible demand vectors $d$. This happens, e.g., when the cost functions of $\Gamma$ have the form $\tau_a(x)=\alpha_a\cdot  x^{\beta}$ with $\beta \ge 0$, as 
$\tau_a(x)=\frac{1}{\beta+1}\cdot \frac{\partial\ x\cdot \tau_a(x)}{\partial\ x}$
for each $a\in A,$ 
see, e.g., \cite{Roughgarden2000How}.
We call such games $\Gamma$ \emph{well designed}. 
This motivates a similar definition for the PoA in the limit.
\begin{definition}
		\label{def:AWDG}
		We call a game $\Gamma$ \emph{asymptotically well designed} if the PoA$(d^{(n)})$ converges to 1 for \emph{all} unbounded demand sequences $(d^{(n)})_{n\in \N}$.
\end{definition}
Pigou's game (see Figure~\ref{fig:Pigou} above) is asymptotically well designed but not \Wuu{well designed}. Definition~\ref{def:AWDG} has the advantage that the convergence of the PoA does no longer depend on the sequence of demands (as in \cite{Colini2017WINE, Colini2017arxiv}), but only on the game.
Our limit analysis of the PoA can then be seen as investigating the class of
\awd games. 

We will not work with  Definition~\ref{def:AWDG} directly, but with the equivalent characterization in Lemma~\ref{lma:ConvergenceOfSequences} below. 
	Despite its triviality, it will be extremely helpful in our discussion, as it allows to discuss only subsequences $(d^{(n_i)})_{i\in \N}$ with specified properties.
\begin{lemma}\label{lma:ConvergenceOfSequences}
	A game 
	$\Gamma$ is \awd  iff 
	each unbounded sequence 
	$(d^{(n)})_{n\in \N}$ has
	an infinite subsequence 
	$(n_i)_{i\in \N}$ s.t. 
	$\lim_{i\to \infty}$PoA$(d^{(n_i)})=1.$
\end{lemma}

We will start with the main result of \cite{Colini2017WINE, Colini2017arxiv}.
Afterwards, we will refine their conditions and identify
a large class
of \awd games that is quite extensive and contains, e.g., 
all games with arbitrary
regularly varying cost functions.

\subsection{A first view on the results by Colini-Baldeschi et al.}\label{sec:coliniFirstView}

Let us reconsider their most general result. We first state their central definitions.

\begin{definition}[Tight game, see \cite{Colini2017WINE}]
	\label{def:TightGame}
	A game $\Gamma$ is called \emph{tight} if there
	is a regularly varying 
	benchmark function $g(\cdot)$ such that:
	\begin{itemize}
		\item[(T1)]  $\lim_{x\to\infty}\tau_a(x)/g(x)= \alpha_a$ for each $a\in A,$ where 
		$\alpha_a\in [0,\infty]$ is a constant.
		\item[(T2)] For each group $k\in \K,$ there is a strategy $s\in \S_k$ such that
		$
		\alpha_s:=\max\{\alpha_a: a\in s\} \in [0,\infty).
		$
		Such a strategy is called {\em tight}, and so (T2) says that every group has a {\em tight} strategy.
		\item[(T3)] There is {\em at least} one group $k\in \K$
		that is {\em tight}, \Wuu{i.e.,}
		$
		\alpha_k:=\min_{s\in \S_k}\alpha_s=\min_{s\in \S_k}\max\{\alpha_a: a\in s\}\in (0,\infty).
		$
	\end{itemize}
\end{definition}

\begin{definition}[Gaugeable sequence]
	\label{def:Salience}
	Let $\Gamma$ be a tight game. An unbounded sequence $(d^{(n)})_{n\in \N}$ is called  
	{\em gaugeable} w.r.t.\ $\Gamma,$ if 
	$
	\varliminf_{n\to \infty} \sum_{k  \text{ is tight}}\frac{d_k^{(n)}}{T(d^{(n)})}>0.
	$
\end{definition}

Their main result then reads as follows.

\begin{theorem}[\cite{Colini2017WINE,Colini2017arxiv}]
	\label{lma:GaugeableTheorem}
	Let $\Gamma$ be a tight game and $(d^{(n)})_{n\in \N}$ be a gaugeable sequence w.r.t.\ $\Gamma$.
	Then $\lim_{n\to \infty}$PoA$(d^{(n)})=1$.
\end{theorem}

The condition of being gaugeable relates the growth of the demands $d^{(n)}$ to the tight groups and thus 
to the benchmark function $g(x)$,
which is a restriction of the sequence. 
There are in fact unbounded sequences $(d^{(n)})_{n\in \N}$
	without gaugeable subsequences, even if
	the game is tight. 
We illustrate this in Example~\ref{example:GaugeableCounterExample}. 
\begin{example}[Tight games with a non-gaugeable
	unbounded sequence]
	\label{example:GaugeableCounterExample}
	Consider the routing game $\Gamma$ shown in Figure~\ref{fig:NonTightGroups}.
	$\Gamma$ has two  O/D pairs with non-overlapping strategies. 
	O/D pair $(o_1, t_1)$ has
	two strategies with cost functions $2x+1$ and $3x+1,$ and
	O/D pair $(o_2, t_2)$
	has also two strategies with cost functions $4x^2+1$ and $5x^2+1.$
	\begin{figure}[!htb]
		\centering
		\begin{tikzpicture}[
		>=latex, scale=0.2,node distance=1cm
		]
			\node[scale=0.4,circle,fill=black,label=left:$o_1$](o1){};
			\node[above right =of o1](xbeta){$2 x+1$};
			\node[scale=0.4,circle,fill=black,label=right:$t_1$,below right =of xbeta](t1){};
			\node[below right =of o1](One){$3 x+1$};
			\draw[-,thick] (o1) to [out=90,in=180] (xbeta);
			\draw[->,thick] (xbeta) to [out=0,in=90] (t1);
			\draw[-,thick] (o1) to [out=-90,in=180] (One);
			\draw[->,thick] (One) to [out=0,in=-90] (t1);
			\node[scale=0.4,circle,fill=black,label=left:$o_2$,right =2of t1](o2){};
			\node[above right =of o2](xbeta2){$4 x^2+1$};
			\node[scale=0.4,circle,fill=black,label=right:$t_2$,below right =of xbeta2](t2){};
			\node[below right =of o2](One2){$5 x^2+1$};
			\draw[-,thick] (o2) to [out=90,in=180] (xbeta2);
			\draw[->,thick] (xbeta2) to [out=0,in=90] (t2);
			\draw[-,thick] (o2) to [out=-90,in=180] (One2);
			\draw[->,thick] (One2) to [out=0,in=-90] (t2);
		\end{tikzpicture}
		\caption{Tight games with non-tight groups}
		\label{fig:NonTightGroups}
	\end{figure}
	$\Gamma$ is obviously tight, since any regularly varying 
	function $g(x)\in \Theta(x^2)$ is
	a benchmark fulfilling (T1)--(T3).  Moreover, if a regularly varying function $g(x)$ fulfills (T1)--(T3), then
	$g(x)\in \Theta(x^2).$ Otherwise, O/D pair $(o_2, t_2)$ 
	has either no tight strategies, or $\Gamma$ has no tight O/D pair.  So $g(x)\in \Theta(x^2)$ and 
    $(o_2,t_2)$ is the {\em unique} tight O/D pair. Therefore, an unbounded sequence  $\big(d^{(n)}=(d_1^{(n)},d_2^{(n)})\big)_{n\in \N}$ is gaugeable w.r.t.\ $\Gamma$ 
	iff $\varliminf_{n\to\infty}\frac{d_2^{(n)}}{T(d^{(n)})}>0,$  where $d_k^{(n)}$ denotes the
demand of O/D pair $(o_k,t_k)$, $k=1,2$.
So, there are many unbounded sequences that are not gaugeable w.r.t.\ $\Gamma,$ 
e.g., all sequences with $\varliminf_{n\to\infty} \frac{d_2^{(n)}}{T(d^{(n)})}=0.$
\end{example}

In fact, there is no direct relationship between tight games and \awd games. There are games that are \aw designed, but not tight, and vice versa. This is shown in 
Examples~\ref{example:NonRegularlyVaryingBenchmark} and \ref{exam:TightNeedNotAWDG} below.

\begin{example}[Not tight, but asymptotically well designed]
	\label{example:NonRegularlyVaryingBenchmark}
	Consider the routing game $\Gamma$ shown in Figure~\ref{fig:GaugeableNonCompleteness}.
	Except for the shared arc in the middle, all cost functions are constant (1 or 2) and displayed next to the arcs.
	If we choose an exponential cost function for the shared arc, e.g., $e^{x}$, then $\Gamma$ is not tight,
	since there is no regularly varying benchmark function $g(\cdot)$
	fulfilling (T1)--(T3). But this
	\begin{figure}[!htb]
		\centering
		\begin{tikzpicture}[
		>=latex
		]
			\node[scale=0.4,circle,fill=black,label=left:$o_1$](o1){};
			\node[scale=0.4,circle,fill=black,below right =of o1](s){};
			\node[scale=0.4,circle,fill=black,label=left:$o_2$,below left =of s](o2){};
			\node[scale=0.4,circle,fill=black,right =1.5of s](h){};
			\node[scale=0.4,circle,fill=black,label=right:$t_1$,above right =of h](t1){};
			\node[scale=0.4,circle,fill=black,label=right:$t_2$,below right =of h](t2){};
			\draw[->,thick] (o1) to [out=0,in=90] node[right]{$\ 1$} (s);
			\draw[->,thick] (o1) to [out=-90,in=165] node[left]{$2\ $} (s);
			\draw[->,thick] (o2) to [out=0,in=-90] node[right]{$\ 2$} (s);
			\draw[->,thick] (o2) to [out=90,in=195] node[left]{$1\ $} (s);
			\draw[->,thick] (s) to (h);
			\draw[->,thick] (h) to [out=90,in=180] node[left]{$2\ $} (t1);
			\draw[->,thick] (h) to  [out=15,in=-90] node[right]{$\ 2$} (t1);
			\draw[->,thick] (h) to  [out=-15,in=90] node[right]{$\ 2$} (t2);
			\draw[->,thick] (h) to [out=-90,in=-180] node[left]{$2\ $} (t2);
		\end{tikzpicture}
		\caption{An \awd game without a regularly varying benchmark function}
		\label{fig:GaugeableNonCompleteness}
	\end{figure}
game is still asymptotically well designed, see  Example~\ref{example:NontightSSD}.
\end{example}

\begin{example}[Tight, but not  asymptotically well designed]\label{exam:TightNeedNotAWDG}
	Consider the routing game $\Gamma$ in Figure~\ref{fig:TightIsNotAWDG} with two O/D pairs
	and parallel arcs. $\Gamma$ is tight
	with regularly varying benchmark function $g(x)=x^5$ and tight group $(o_2,t_2).$
	\Wuu{By} Theorem~\ref{lma:GaugeableTheorem},
	the PoA$(d^{(n)})$ converges to $1$ for
    each gaugeable sequence $(d^{(n)})_{n\in \N}.$
	However, this convergence does not
	carry over to an arbitrary unbounded sequence, and so $\Gamma$ is not asymptotically well designed.
	\begin{figure}[!htb]
		\centering
		\begin{tikzpicture}[
		>=latex, scale=0.5,node distance=1cm
		]
			\node[scale=0.4,circle,fill=black,label=left:$o_1$](o1){};
			\node[right =1.8of o1](f2){$x^2$};
			\node[above =0.4of f2](f1){$x^2\!+\!\frac{x^2}{2}\sin(\ln x)$};
			\node[below =0.4of f2](f3){$x^2\!+\!\frac{x^2}{2}\cos(\ln x)$};
			\node[scale=0.4,circle,fill=black,label=right:$t_1$,right =1.8of f2](t1){};
			\draw[-,thick] (o1) to [out=90,in=180] (f1);
			\draw[->,thick] (f1) to [out=0,in=90] (t1);
			\draw[-,thick] (o1) to (f2);
			\draw[->,thick] (f2) to (t1);
			\draw[-,thick] (o1) to [out=-90,in=180] (f3);
			\draw[->,thick] (f3) to [out=0,in=-90] (t1);
			\node[scale=0.4,circle,fill=black,label=left:$o_2$,right =1.5of t1](o2){};
			\node[above right =of o2](f4){$x^5$};
			\node[below right =of o2](f5){$x^6$};
			\node[scale=0.4,circle,fill=black,label=right:$t_2$,below right =of f4](t2){};
			\draw[-,thick] (o2) to [out=90,in=180] (f4);
			\draw[->,thick] (f4) to [out=0,in=90] (t2);
			\draw[-,thick] (o2) to [out=-90,in=180] (f5);
			\draw[->,thick] (f5) to [out=0,in=-90] (t2);
		\end{tikzpicture}
		\caption{Tight games may not be asymptotically well designed}
		\label{fig:TightIsNotAWDG}
	\end{figure}
	To see this, we consider the non-gaugeable unbounded sequence
	$\big(d^{(n)}=(d_1^{(n)}=n, d_2^{(n)}=0)\big)_{n\in \N},$ where $d_k^{(n)}$ is again the demand of O/D pair 
	$(o_k,t_k),$ $k=1,2.$ 
	The PoA$(d^{(n)})$ equals the \Wuuu{PoA$(d^{(n)}_{1})$
	for the routing game consisting
	only of O/D pair $(o_1,t_1)$,}
	 since $d_2^{(n)}\equiv 0.$ \cite{Colini2017WINE,Colini2017arxiv}
	showed that \Wuuuu{PoA$(d^{(n)}_{1})$}
	oscillates as $n\to \infty$ because of the periodicity of the multiplicative factors
	$1+\frac{\sin(\log x)}{2}$ and
	$1+\frac{\cos(\log x)}{2}$ in
	the cost functions.
\end{example}

\Wuu{Tightness 
is thus neither {\em sufficient} nor {\em necessary} for a game to be asymptotically well designed.
The convergence} of PoA$(d^{(n)})$ in Theorem~\ref{lma:GaugeableTheorem}
need not hold for arbitrary 
unbounded sequences $(d^{(n)})_{n\in \N}$, even if the
game is tight, and even if it holds
for gaugeable sequences. 
Nonetheless, Theorem~\ref{lma:GaugeableTheorem}
reveals that a tight game is \awd if 
all unbounded sequences are gaugeable.
Games with polynomials of the same degree as
cost functions have this property. 
However, games with arbitrary polynomial cost functions generally do not, 
although they are tight.

\section{Our main results}\label{sec:PoA_Limit} 

\subsection{Scalable games}\label{sec:scalableGames}

We now introduce our first class of asymptotically well designed games. It is based on a refinement of the ideas of \cite{Colini2017WINE,Colini2017arxiv}. They introduced the normalization or scaling of the cost functions \Wuuu{$\tau_a(x)$ to} $\frac{\tau_a(x)}{g(x)}$ and use a regularly varying function $g(\cdot)$ for it. We will just use constants $g>0$ instead and introduce a scaling of the whole game by $g.$
\begin{definition}[Scaled Game]
	\label{def:ScaledGame}
	Consider a game $\Gamma=\big(A,\S,\K,\tau=(\tau_a)_{a\in A},d=(d_k)_{k\in \K}\big)$ and a constant $g > 0$. Then 
	$
	\Gamma^{[g]}=\Big(A,\S, \K, \tau^{[g]}=(\tau^{[g]}_a)_{a\in A}, \frac{d}{T(d)}=\big(\frac{d_k}{T(d)}\big)_{k\in \K}\Big)
	$
	is called the {\em scaled game} of $\Gamma$ w.r.t.\ the {\em scaling factor}
	$g>0,$ if $\tau^{[g]}_a(x)=\frac{\tau_a(T(d)\cdot x)}{g}$
	for each $a\in A$ \Wuu{and each $x\in [0,1].$}
\end{definition} 

The scaling does not change the PoA, since all groups and strategies are the same and the scaled cost of 
resource $a\in A$ in $\Gamma^{[g]}$ equals its original cost in $\Gamma$ divided by the
scaling factor $g$. This follows directly by observing that the demand $f_a$ assigned to resource $a$ by strategy profile $f$ in $\Gamma$ transforms under scaling into $\frac{f_a}{T(d)}$ in $\Gamma^{[g]}$ and so its scaled cost is $\tau^{[g]}_a(\frac{f_a}{T(d)}) = \frac{\tau_a(f_a)}{g}$. We summarize this in Lemma~\ref{lma:ScaledPoA=PoA}.

\begin{lemma}\label{lma:ScaledPoA=PoA}
	Consider a game $\Gamma$ and its scaled game $\Gamma^{[g]}$ for a 
	scaling factor $g>0$. 	Then \Wuu{$\text{PoA}(d)=\text{PoA}^{[g]}(\frac{d}{T(d)})$ for each $d=(d_k)_{k\in\K},$ where $\text{PoA}^{[g]}(\frac{d}{T(d)})$
	denotes the PoA of $\Gamma^{[g]}$ for $\frac{d}{T(d)}=(\frac{d_k}{T(d)})_{k\in\K}.$}
\end{lemma}

So it suffices to apply our limit analysis of the PoA to scaled games instead. In the scaled game $\Gamma^{[g]}$ of $\Gamma$, the total demand is 
 $T(\frac{d}{T(d)})
	=\sum_{k  \in \K}\frac{d_k}{T(d)}=1$, the demand vector $\frac{d}{T(d)}=\big(\frac{d_k}{T(d)}\big)_{k\in \K}$ is a distribution of the demand over the groups (called the {\em demand distribution} of $d$), the vector $\frac{f}{T(d)} := (\frac{f_s}{T(d)})_{s\in \S}$ is a distribution of the total demand $T(d)$ over the strategy profiles (called the {\em strategy distribution} of the profile $f$), and the vector $ \big(\frac{f_a}{T(d)}\big)_{a\in A}$ is a distribution of the total demand $T(d)$ over the resources $a \in A$ (called the \emph{consumption distribution} of the profile $f$ over the resources).

Scaled games
have the advantage that the limit analysis of PoA$(d^{(n)})$ for an arbitrary unbounded
sequence $\big(d^{(n)}\big)_{n\in \N}$ 
transforms to that of a joint sequence
$\Big(\big(\tau_a^{[g_n]}(\cdot)\big)_{a\in A}, \frac{d^{(n)}}{T(d^{(n)})}\Big)_{n\in \N}$ for a suitably chosen
sequence $\big(g_n\big)_{n\in \N}$ of scaling factors. By Lemma~\ref{lma:ConvergenceOfSequences},
one can assume additionally that the demand distribution
$\frac{d^{(n)}}{T(d^{(n)})}=\big(\frac{d_k^{(n)}}{T(d^{(n)})}\big)_{k\in \K}$
converges to a limit 
$d^{(\infty)}=(d_k^{(\infty)})_{k\in \K}$ as $n\to \infty$. 
Then, the limit analysis of PoA$(d^{(n)})$
transforms further to the analysis
of the function sequence $(\big(\tau_a^{[g_n]}(\cdot)\big)_{a\in A})_{n\in \N},$
which may in turn transform to the analysis of the PoA$(d^{(\infty)})$
of a ``limit" game $\Gamma^{(\infty)}.$ We illustrate this below on Pigou's game.

\begin{example}[Limit Game]
	\label{exam:LimitGame}
	Let $\big(d^{(n)}=(d_1^{(n)})\big)_{n\in \N}$ be an unbounded sequence in Pigou's game $\Gamma$ from Figure~\ref{fig:Pigou} and set the scaling factor $g_n := 1$ for each $n\in \N.$ Call the upper arc $u$ and the lower arc $\ell$. Then the total demand $T(d^{(n)})= d_1^{(n)}$ and the scaled game $\Gamma^{[g_n]}$ has cost functions $\tau^{[g_n]}_u(x)=\frac{\tau_u(d_1^{(n)}\cdot x)}{1} = (d_1^{(n)}\cdot x)^\beta$  and 
	$\tau^{[g_n]}_{\ell}(x) = 1$. We can interpret $\Big(\tau^{[g^{(n)}]}=\big(\tau^{[g_n]}_a(x)\big)_{a\in \{u,\ell\}}\Big)_{n\in \N}$  as a sequence of cost functions with  ``limit cost functions'' $\tau^{(\infty)}_u(x) = \infty$ (for $x>0$) and  $\tau^{(\infty)}_{\ell}(x) = 1$. Similarly, the demand
	distribution $(\frac{d^{(n)}}{T(d^{(n)})})_{n\in \N}$ has the ``limit'' \Wuu{$d^{(\infty)}=(d_1^{(\infty)})$ with $d^{(\infty)}_1 = \lim_{n \to \infty} \Wuu{\frac{d^{(n)}_1}{T(d^{(n)})}} = 1$.} So one can say that the sequence of scaled games $\Gamma^{[g_n]}$  has a ``limit game'' $\Gamma^{(\infty)}$ with total demand \Wuu{$T(d^{(\infty)})=d^{(\infty)}_1 = 1$} and cost functions $\tau^{(\infty)}_u(x) = \infty$ and $\tau^{(\infty)}_{\ell}(x) = 1$. Both, NE and SO profiles of $\Gamma^{(\infty)},$
	will use only the lower arc, and have both a cost of 1.  
 So, PoA$(d^{(\infty)})=1$. 
 \Wuu{Figure~\ref{fig:Pigou}(b)}
 	seems to indicate that PoA$(d^{(n)})$
 	converges to the PoA$(d^{(\infty)})$ of the limit game
 	$\Gamma^{(\infty)}.$
 	We will confirm this in Lemma~\ref{lma:PriceConvergenceOfStrategies}e).
\end{example}

This convergence  does of course not hold for every game and demand sequence. 
Finding for every unbounded sequence a
suitable subsequence in view
of Lemma~\ref{lma:ConvergenceOfSequences} is the main goal of this paper. In this analysis, we will consider demand and resource consumption as variables in $\Gamma$ and study scaled and limit cost functions as functions of the variable demand and resource consumption.

Recall that the consumption distribution of a strategy profile $f$ of a scaled game $\Gamma^{[g]}$ is $(\frac{f_a}{T(d)})_{a \in A}$.
We denote by $I_a(d) := \{ \frac{f_a}{T(d)} \mid f \text{ is a strategy  profile for $d$} \}$ the set of all
possible {\em consumption rates} of resource $a\in A$ in $\Gamma^{[g]}$ for fixed demand $d = (d_k)_{k \in \K}$.  
Then $I_a:=\bigcup_{d} I_a(d)$ is the {\em range}
of consumption rates of resource $a\in A$ for variable demand $d$. For simplicity, we may  call 
$I_a$ also the  {\em consumption domain} or simply \emph{domain} of resource $a.$

$I_a$ is obviously independent of $d$ and $T(d)$. Our assumption~\eqref{eq:StrategiesScarce} that every resource $a$ is needed by some strategy and the fact that demands $d_k$  can be assigned \Wuuu{in} arbitrary amounts to strategies $s \in \S_k$ yields that 
\begin{equation}\label{eq:DomainI_a}
I_a \ne \emptyset \text{ and is either a closed non-empty subinterval of } [0,1] \text{ or a singleton } \{u\}, 0 < u \le 1. 
\end{equation}


An unbounded sequence
$(d^{(n)})_{n\in \N}$ is called {\em regular}, if $\lim_{n\to \infty}d_k^{(n)}\in [0,\infty]$ exists for each $k\in \K,$ and if its
distribution sequence converges, i.e.,  
$d_k^{(\infty)}:=\lim_{n\to \infty}\frac{d_k^{(n)}}{T(d^{(n)})}
\in [0,1]$ exists for
each $k\in \K$. We call $d^{(\infty)}=(d_k^{(\infty)})$
the {\em limit distribution} of the sequence $(d^{(n)})_{n\in \N}.$
Trivially, each unbounded demand sequence has
a regular subsequence. By Lemma~\ref{lma:ConvergenceOfSequences}, a game $\Gamma$ is \awd iff the PoA$(d^{(n)})$ converges to 1 for all regular demand sequences $(d^{(n)})_{n\in \N}$. \Wuu{We summarize this in Lemma~\ref{lma:Convergence_iff_RegularSequence} below.
\begin{lemma}\label{lma:Convergence_iff_RegularSequence}
	The following statements are equivalent.
	\begin{itemize}
		\item[a)] A game $\Gamma$ is \awd. 
		\item[b)] $\lim_{n \to \infty}$PoA$(d^{(n)})=1$ for all regular sequences $(d^{(n)})_{n\in \N}$.
		\item[c)] For each regular sequence 
		$(d^{(n)})_{n\in \N},$ $\lim_{i \to \infty}$PoA$(d^{(n_i)})=1$ for an infinite subsequence $(n_i)_{i\in \N}.$  
	\end{itemize}
\end{lemma}}

\Wuuuu{Lemma~\ref{lma:Convergence_iff_RegularSequence} allows us to focus for an arbitrary regular sequence on one of its subsequences
with particular properties.}

Definition~\ref{def:LimitGameAndStrongScalable}
below introduces the limit of a game $\Gamma$ w.r.t.\
a regular sequence $(d^{(n)})_{n\in \N}$ and domains $I_a$,
and formalizes the idea sketched in  Example~\ref{exam:LimitGame}.
\begin{definition}[Limit game]
	\label{def:LimitGameAndStrongScalable}
	Consider a game $\Gamma=(A,\K, \S,
	 (\tau_a)_{a\in A}, d)$, a regular sequence $(d^{(n)})_{n\in \N}$ with limit distribution
	$d^{(\infty)}=(d_k^{(\infty)})_{k\in \K},$ and a scaling sequence $(g_n)_{n\in \N}$. Then
	$
	\Gamma^{(\infty)}=\Big(A,\K, \S, \Wuu{\tau^{(\infty)}=}(\tau^{(\infty)}_a)_{a\in A}, d^{(\infty)}\Big)
	$ is called the \emph{limit} game of $\Gamma$ 
	or the {\em limit} of the scaled games $\Gamma^{[g_n]}$ for $(d^{(n)})_{n\in \N}$ and $(g_n)_{n\in \N}$,
	if its cost functions $\tau^{(\infty)}_a(x)$ fulfill (L1)--(L4) below.
	\begin{itemize}
		\item[(L1)] For each
		$a\in A,$ \Wuu{$\tau^{(\infty)}_a(x) =\lim_{n\to \infty} \tau_a^{[g_n]}(x)= \lim_{n\to \infty} \frac{\tau_a\big(T(d^{(n)})\cdot x\big)}{g_n}\in [0,\infty]$}
			for each $x\in I_a\setminus\{0\},$ and 
			$\tau^{(\infty)}_a(0):=\lim_{x\to 0^+}\tau_a^{(\infty)}(x)$
			if $0\in I_a.$ 
		\item[(L2)] $\tau^{(\infty)}_a(x)$ is either \Wuu{the constant}
			$\infty$, or finite and continuous on $I_a$ for each $a\in A$.
		\item[(L3)] Each group $k\in \K$ has a strategy $s\in \S_k$ that is {\em tight} w.r.t.\ $(g_n)_{n\in \N},$ i.e., 
			$\tau^{(\infty)}_a(x)$ is finite and continuous on $I_a$ for each $a\in s$.
		\item[(L4)] NE profiles of $\Gamma^{(\infty)}$ have positive cost, and the PoA$(d^{(\infty)})$ of $\Gamma^{(\infty)}$
			is $1.$   
	\end{itemize} 
\end{definition}

The functions $\tau^{(\infty)}_a(\cdot)$ \Wuuu{are called}
the \emph{limit cost functions} of $\Gamma$
under the scaling and \Wuuu{the regular sequence.}
%
We shall write $\Gamma^{(\infty)} = \lim_{d^{(n)}\to \infty} \Gamma^{[g_n]},$ when 
	$\Gamma^{(\infty)}$ is the limit of 
	$\Gamma$ w.r.t. scaling sequence $(g_n)_{n\in \N}$
	and regular demand sequence $(d^{(n)})_{n\in \N}.$
	This notation indicates that \Wuu{the limit distribution 
	$d^{(\infty)}$ and} the limit cost functions \Wuuu{$\tau_a^{(\infty)}$ of
the limit game $\Gamma^{(\infty)}$} depend \Wuu{crucially} on
	the regular sequence $(d^{(n)})_{n\in \N}.$
It may of course happen that the limit cost functions are
independent
of $(d^{(n)})_{n\in \N}.$ Then we shall simply write 
$\Gamma^{(\infty)}=\lim_{n \to \infty}\Gamma^{[g_n]}$
and say that the limit of $\Gamma$ is
{\em essentially unique}, though the limit
distribution $d^{(\infty)}=(d_k^{(\infty)})_{k\in \K}$
still depends on \Wuu{the regular sequence} $(d^{(n)})_{n\in \N}$.
We will see in Lemma~\ref{lem:Tight<=Scalable} that the limit
is essentially unique when $\Gamma$ is tight and all
\Wuuu{unbounded sequences} are gaugeable.

Definition~\ref{def:LimitGameAndStrongScalable} above involves some technicalities that need more explanation.  

The first technicality involves the limit cost functions $\tau^{(\infty)}_a(\cdot)$. They  are defined 
only on the domains $I_a$  
because the
	$x\notin I_a$ play no role 
	in the scaled games $\Gamma^{[g_n]}$ and thus
	have no influence on the PoA$(d^{(\infty)})$
	of the limit game $\Gamma^{(\infty)}.$ Moreover,
$\varlimsup_{n \to \infty}\tau_a^{[g_n]}(0)=\varlimsup_{n \to \infty}\frac{\tau_a(0)}{g_n}\le 
\tau_a^{(\infty)}(0)=\lim_{y\to 0+}\tau^{(\infty)}_a(y)$ for each $a\in A$ with $0\in I_a$.
So, (L1)--(L2) imply that all $\tau^{(\infty)}_a(\cdot)\not\equiv\infty$ 
are continuous, non-decreasing and non-negative on 
$I_a$.

The second technicality involves non-tight strategies. 
By (L3), each group $k\in \K$ has a tight strategy. So,
neither NE nor SO profiles of the limit game $\Gamma^{(\infty)}$
will use a non-tight strategy $s\in \S_k$ because its cost is $\tau_s^{(\infty)}(f^{(\infty)}) =
\sum_{a\in A: a\in s} 
\tau_a^{(\infty)}(f_a^{(\infty)})=\infty$
for each feasible profile
$f^{(\infty)}=(f_s^{(\infty)})_{s\in \S}$
of $\Gamma^{(\infty)}$ with $f_s^{(\infty)}>0.$ These strategies are thus {\em negligible} when
we consider the PoA$(d^{(\infty)})$ of $\Gamma^{(\infty)}$. So,
	loosely speaking, $f^{(\infty)}_s\cdot \tau_s^{(\infty)}(f^{(\infty)}) = 0 \cdot \infty = 0$ in the limit game
	when $f^{(\infty)}_s=0$ for
	a non-tight 
	strategy $s\in \S$, and therefore we will w.l.o.g.\ make the convention that $0\cdot \infty=0.$ 
	
	\Wuu{The last technicality involves the condition in (L4) that PoA$(d^{(\infty)})$=1.}
	\Wuuu{We have included it here since it ensures that a game is asymptotically well designed. One can of course define the limit without that condition and obtain a well justified notion of sequences of games and limit games. But then one has to exclude the case that 
	PoA$(d^{(\infty)})\ne 1$ for further analysis \Wuuuu{elsewhere} since there are
   games and regular sequences $(d^{(n)})_{n\in \N}$ that satisfy (L1)--(L3)
and have PoA$(d^{(\infty)})\ne 1.$ For instance, the game consisting only of the O/D pair $(o_1,t_1)$ in 
Figure~\ref{fig:TightIsNotAWDG} is not \awd and has a limit 
$\Gamma^{(\infty)}$ with PoA$(d_1^{(\infty)})\ne 1$ for 
the regular sequence 
$(d_1^{(n)}=e^{2\cdot \pi\cdot n})_{n\in \N}$ and \Wuuuu{the} scaling 
sequence $(g_n=e^{4\cdot\pi\cdot n})_{n\in \N}.$}

\Wuuu{We will see} 
$\Gamma^{(\infty)}$ as a game, although 
some of its cost functions may be the constant $\infty$. NE and SO strategy profiles 
 will not use resources with that cost.  
\Wuuu{Condition (L4) then} implies
that $\Gamma^{(\infty)}$ has a unique positive
NE cost for the limit distribution $d^{(\infty)}$\Wuu{ and
its NE profiles are also socially optimal}. 
Example~\ref{exam:LimitGame} above has already demonstrated these effects.

We now define scalable games, and
show in Theorem~\ref{thm:StrongScalabilityTheorem}
that they form a class of \aw designed games.
\begin{definition}[Scalable sequence and scalable game]
	\label{def:StrongScalabilityNew}
A regular sequence  $(d^{(n)})_{n\in \N}$ is {\em scalable} w.r.t.\ a game $\Gamma$, if 
there is a scaling sequence $(g_n)_{n\in \N}$
such that $\lim_{d^{(n)}\to \infty}\Gamma^{[g_n]}=\Gamma^{(\infty)}$
for a limit game $\Gamma^{(\infty)}$.
A game $\Gamma$ is \emph{scalable}, if each regular
	sequence $(d^{(n)})_{n\in \N}$ has a 
	{\em scalable subsequence} $(d^{(n_i)})_{i\in \N}$.
\end{definition}

Pigou's game in Example~\ref{exam:LimitGame} is scalable, as 
it is tight and each regular sequence itself is scalable.

We now present our main result for scalable games.

\begin{theorem}\label{thm:StrongScalabilityTheorem}
	A scalable game $\Gamma$ is \aw designed.
\end{theorem}

Theorem~\ref{thm:StrongScalabilityTheorem} follows from \Wuu{Lemma~\ref{lma:Convergence_iff_RegularSequence}c)} and Lemma~\ref{lma:PriceConvergenceOfStrategies}e) below,
since $\lim_{n \to \infty}$PoA$(d^{(n)})=1$
	for each scalable sequence $(d^{(n)})_{n\in \N}$ by 
	Lemma~\ref{lma:PriceConvergenceOfStrategies}e),
	and since each unbounded sequence of a scalable game
has a scalable subsequence by Definition~\ref{def:StrongScalabilityNew}.

Lemma~\ref{lma:PriceConvergenceOfStrategies} presents a kind of scaling 
theory for games and generalizes 
Theorem~\ref{lma:GaugeableTheorem}, as 
it no longer requires a  regularly varying benchmark function. 
Lemma~\ref{lma:PriceConvergenceOfStrategies}a)
states that the scaled cost of tight
strategies converge to their limit cost in the limit game $\Gamma^{(\infty)}$, while
Lemma~\ref{lma:PriceConvergenceOfStrategies}b) and d) ensure
that NE or SO profiles do not use non-tight strategies in $\Gamma^{(\infty)}$. Lemma~\ref{lma:PriceConvergenceOfStrategies}c)
shows that the limit 
	$\tilde{f}^{(\infty)}$ of a sequence of strategy distributions $\frac{\tilde{f}^{(n)}}{T(d{(n)})}$ of NE profiles $\tilde{f}^{(n)}$
is an NE profile of $\Gamma^{(\infty)}$.
Thus, if $\Gamma^{(\infty)}$ exists and has a unique NE profile, then strategy distributions of NE profiles converge to 
that unique NE profile of $\Gamma^{(\infty)}$. 
Lemma~\ref{lma:PriceConvergenceOfStrategies}e)
follows immediately from Lemma~\ref{lma:PriceConvergenceOfStrategies}a)--d) and shows that $\lim_{n\to\infty}$PoA$(d^{(n)})=1$ for a scalable sequence $(d^{(n)})_{n\in \N}$.

\begin{lemma}[Scaling Properties]\label{lma:PriceConvergenceOfStrategies}
	Consider a game $\Gamma,$ a regular sequence $(d^{(n)})_{n\in \N}$ with 
	limit distribution $d^{(\infty)}=(d_k^{(\infty)})_{k\in \K}$,
	and a scaling sequence $(g_n)_{n\in \N}.$ Suppose that 
	$\lim_{d^{(n)}\to \infty}\Gamma^{[g_{n}]}=\Gamma^{(\infty)}$
for a limit game $\Gamma^{(\infty)}$. Let 
	$f^{(n)}$, $\tilde{f}^{(n)}$, and $f^{*(n)}$ be an arbitrary strategy 
	profile, an NE profile, and an SO profile of game $\Gamma$ for demand
	$d^{(n)}$, respectively.
	Then the following statements hold.
	\begin{itemize}
		\item[a)]  Let $(n_i)_{i\in \N}$
		be an infinite subsequence of $\N$ s.t. $f^{(\infty)} 
		:= \lim_{i\to\infty}\frac{f^{(n_i)}}{T(d^{(n_i)})}
		= (f_s^{(\infty)})_{s\in \S}$, and let $s\in \S$ be a tight strategy. 
		Then $\varlimsup_{i\to\infty}\frac{\tau_{s}(f^{(n_i)})}{g_{n_i}}\le \tau^{(\infty)}_{s}(f^{(\infty)})<\infty$ and
		$\lim_{i\to \infty}  \frac{ f_{s}^{(n_{i})}\cdot\tau_{s}(f^{(n_i)})}{T(d^{(n_i)})\cdot  g_{n_i}}
		=f_{s}^{(\infty)}\cdot \tau^{(\infty)}_{s}(f^{(\infty)})<\infty.$ Furthermore, if $f^{(\infty)}_s>0$, then $\lim_{i\to\infty}\frac{\tau_s(f^{(n_i)})}{g_{n_i}}=\tau^{(\infty)}_s(f^{(\infty)})<\infty.$
		
		\item[b)]  For each non-tight \Wuuu{strategy} $s\in \S$, 
		$\lim_{n\to\infty}
		\frac{\tilde{f}_s^{(n)}}{T(d^{(n)})}=0$ and 
		$\lim_{n\to \infty}  \frac{\tilde{f}_s^{(n)}\cdot \tau_s\big(\tilde{f}^{(n)}\big)}{T(d^{(n)})\cdot g_n}=0.$
		
		\item[c)]
		If $\tilde{f}^{(\infty)}=(\tilde{f}^{(\infty)}_s)_{s\in \S}$ is
		a limit distribution of 
		$(\tilde{f}^{(n)})_{n\in \N}$, i.e.,
		$\lim_{i\to\infty}\frac{\tilde{f}^{(n_i)}}{T(d^{(n_i)})}=\tilde{f}^{(\infty)}$
		for some infinite subsequence
		$(n_i)_{i\in \N},$
		then $\tilde{f}^{(\infty)}$ 
		is both an NE and an SO profile of the
		limit game $\Gamma^{(\infty)}.$
		
		\item[d)] For each non-tight \Wuuu{strategy} $s\in \S$, 
		$\lim_{n\to\infty}
		\frac{f_s^{*(n)}}{T(d^{(n)})}=0.$
		
		\item[e)] 
		Let $\tilde{f}^{(\infty)}$ be an NE profile of $\Gamma^{(\infty)}$
		w.r.t.\ $d^{(\infty)}.$ Then, $\lim_{n\to\infty}
		\frac{C(\tilde{f}^{(n)})}{T(d^{(n)})\cdot g_n}
		=:C_{\Gamma^{(\infty)}}(\tilde{f}^{(\infty)})\in (0,\infty)$ and $\lim_{n\to\infty}$PoA$(d^{(n)})=$PoA$(d^{(\infty)})=1.$ Here 
		$C_{\Gamma^{(\infty)}}(\cdot)$ denotes the cost function of strategy profiles in the limit $\Gamma^{(\infty)}$.
	\end{itemize}
\end{lemma}

The proof of Lemma~\ref{lma:PriceConvergenceOfStrategies} has been moved to Appendix~\ref{proof:PriceConvergenceOfStrategies}.

\subsection{A second view on the results by Colini-Baldeschi et al.}\label{sec:coliniSecondView}

Our notion of scalable sequences generalizes gaugeable sequences introduced by \cite{Colini2017WINE,Colini2017arxiv}. Every gaugeable sequence $(d^{(n)})_{n\in \N}$ in a tight game is scalable and the PoA$(d^{(n)})$ tends to 1.

\begin{lemma}\label{lem:Tight<=Scalable}
Let $\Gamma$ be a tight game with regularly 
	varying benchmark function $g(\cdot)$ and $(d^{(n)})_{n\in \N}$ be a gaugeable sequence. Then $(d^{(n)})_{n\in \N}$ is scalable with scaling factors $g_n:=g(T(d^{(n)}))$ and
		the limit $\Gamma^{(\infty)} = \lim_{n \to \infty} \Gamma^{[g_n]}$ is well designed, and essentially unique.
\end{lemma}

This can be seen as follows.  Condition
(T1) and the regular variation of $g(\cdot)$ imply that the limit cost functions
$\tau^{(\infty)}_a(\cdot)$ are obtained as $\tau^{(\infty)}_a(x)=\alpha_a\cdot x^{\rho}$, where $\rho>0$ is the regular variation index of
$g(\cdot)$ introduced in Section~\ref{sec:relatedWork} and $\alpha_a\in [0,\infty]$ is a constant.
So, (L1) holds, and the limit cost functions $\tau^{(\infty)}_a(\cdot)$
are monomials of the same degree $\rho$
determined uniquely by 
the benchmark function $g(\cdot)$.
Trivially, condition (T2) implies (L2)--(L3), i.e.,
each group has a tight strategy. Condition (T3),
condition (T1) and the gaugeability
of $(d^{(n)})_{n\in \N}$ then imply (L4).

The regular variation of the benchmark function $g(\cdot)$ and  gaugeability of the unbounded sequence $(d^{(n)})_{n \in \N}$ play a crucial role in the work of \cite{Colini2017WINE,Colini2017arxiv}. Regular variation of $g(\cdot)$ implies
that $\lim_{t \to \infty}\frac{\tau_a(t\cdot x)}{g(t)}
=\alpha_a\cdot x^{\rho}$ for each $a\in A$ and $x>0$ when (T1)--(T3) hold. This is the basic
premise for their work.
\Wuuu{Gaugeability}
guarantees that resources $a\in A$ with $\alpha_a=\infty$ are {\em negligible} in the limit 
when the game is tight and cost functions are scaled by the benchmark function. 
So \cite{Colini2017WINE,Colini2017arxiv} 
showed implicitly---without having the notion of limit game---\Wuuu{that tight} games scaled by a regularly varying benchmark function w.r.t a gaugeable sequence converge to a game with monomials of \Wuuu{the} same degree, which is known to have \Wuuu{a} PoA of 1.

The difference between tight games and scalable games becomes now clear. A scalable game permits different 
scaling sequences for different regular demand sequences, while a tight game requires the same
benchmark function for all unbounded demand sequences and thus has 
an essentially unique limit.   
Limits of scalable games w.r.t.
different regular demand sequences thus need not
be essentially unique, i.e., 
the limit cost functions may be different
w.r.t. different regular demand sequences.
So,
it is not surprising that there are scalable games that are not tight.
Moreover, a tight game need not be scalable, as we have
demonstrated in Example~\ref{exam:TightNeedNotAWDG} with a tight game that is not \aw designed.

{In fact,  tight games and scalable games may even differ, when the
	limit game is essentially unique.
We illustrate this in 
Example~\ref{example:NontightSSD}. 

\begin{example}[A non-tight scalable game with essentially unique limit]\label{example:NontightSSD}
	Consider the routing game $\Gamma$ in Figure~\ref{fig:GaugeableNonCompleteness} from
	Example~\ref{example:NonRegularlyVaryingBenchmark}.
	We assume now that the cost function of the shared arc is $e^{x}.$ Let $(d^{(n)})_{n \in \N}$ be an arbitrary regular
	sequence  and let 
	$g_n:=e^{T(d^{(n)})}$ for each $n\in \N$. The shared arc (resource) \Wuuu{has the constant} $1$ as limit cost function, as its domain is
	the singleton $\{1\}$. All
	 other limit cost functions are
	$0$ on their domains. So this game has an essentially
	unique limit and is scalable. However, it is not
	tight.
\end{example}

Example~\ref{example:SSDDivergentSequence} below illustrates that a scalable game may have multiple
	limits for an unbounded demand sequence $(d^{(n)})_{n\in \N}$, and that
	these limits are different. In particular, it illustrates also that
	a well designed game need not be tight.
\begin{example}[A scalable game  with
	multiple limit games]
	\label{example:SSDDivergentSequence}
	Consider the Pigou-like game $\Gamma$ in Figure \ref{fig:WeaklyScalability}(a).
	The only O/D pair is $(o,t)$, and the
	resources are the two parallel arcs with the same cost function
	$\tau(\cdot)$ defined as follows.
	\begin{figure}[!htb]
		\centering
		\begin{subfigure}{0.3\textwidth}
			\centering
			\begin{tikzpicture}[
			>=latex
			]
			\node[scale=0.4,circle,fill=black,label=left:$o$](o){};
			\node[above right =of o](f1){$\tau(x)$};
			\node[below right =of o](f2){$\tau(x)$};
			\node[scale=0.4,circle,fill=black,label=right:$t$,below right =of f1](t){};
			\draw[-,thick] (o) to [out=90,in=180] (f1);
			\draw[->,thick] (f1) to [out=0,in=90] (t);
			\draw[-,thick] (o) to [out=-90,in=180] (f2);
			\draw[->,thick] (f2) to [out=0,in=-90] (t);
			\end{tikzpicture}
			\subcaption{}
		\end{subfigure}
	\begin{subfigure}{0.6\textwidth}
		\centering
		\includegraphics[width=0.5\linewidth,height=0.153\textheight]{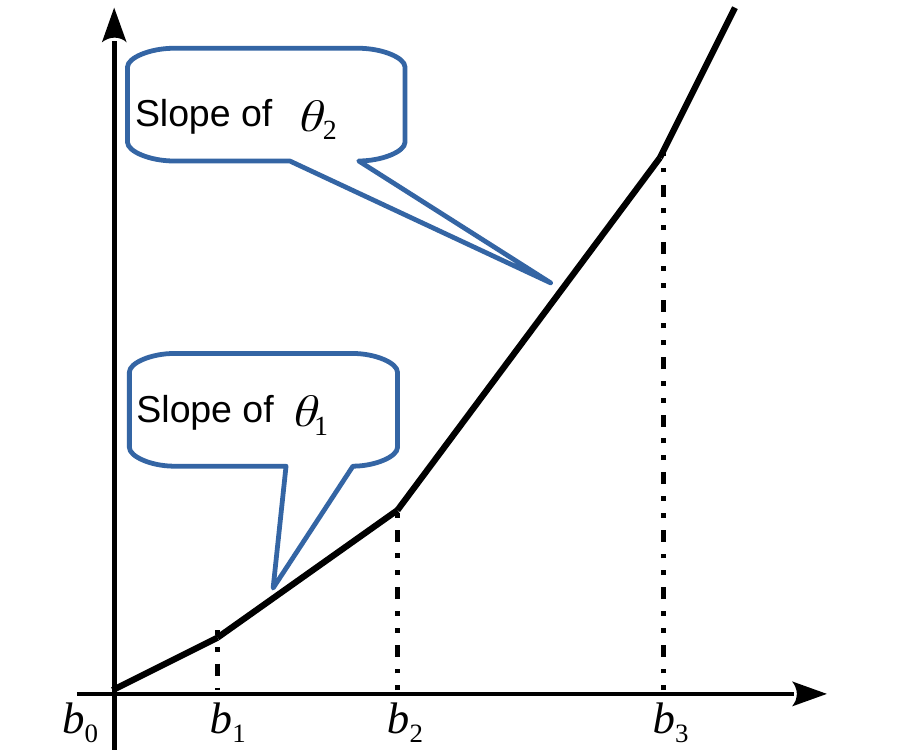}
		\subcaption{}
	\end{subfigure}
		\caption{A scalable game with multiple
		limits}
		\label{fig:WeaklyScalability}
	\end{figure}

	Let $\epsilon>1$
	be a constant, and let $(b_{n})_{n\in \N}$ and $(\theta_n)_{n\in \N}$ be two {\em strictly increasing} sequences of
non-negative reals s.t. $b_0=0,$
	$\lim_{n \to \infty}\frac{b_{n+1}-b_n}{b_{n+1}}=1,$
	$\lim_{n \to \infty}\frac{\theta_{n-1}\cdot b_n}{b_{n+1}}=0$
and $\lim_{n \to \infty}\frac{\theta_{n+1}}{\theta_n}=\epsilon>1.$
We define $\tau(\cdot)$
	piecewise as follows:
	$
		\tau(x)=
		b_0=0\text{ if }x=b_0,\ \text{and } 
		\tau(x)=\theta_n\cdot (x-b_n)+\tau(b_n)\text{ if } x\in  [b_{n},b_{n+1})
	$
		for all $n\in \N$,
	see Figure~\ref{fig:WeaklyScalability}(b).
	\Wuuu{$\Gamma$ is} well designed, since
	the two arcs have the same strictly convex cost function 
$\tau(\cdot).$ However, $\tau(\cdot)$ is not regularly varying,
since $\frac{\tau(t\cdot x)}{\tau(t)}$ diverges for each $x>0$ as $t\to\infty.$
So, $\Gamma$ is not tight, although (T1)--(T3) hold when
we put $g(\cdot)=\tau(\cdot).$
 	Appendix~\ref{sec:proofExampleSSDDivergentSequence} shows that this game is
 	scalable, although
 	{\em not every regular sequence of 
 		the game is
 	scalable.}
 But each non-scalable regular sequence
	$(\D^{(n)})_{n\in \N}$ is shown to 
	contain scalable subsequences with  essentially
	{\em different} limits. 
	
\end{example}

\subsection{Asymptotic decomposition of games}\label{sec:asymptoticDecomposition}

Showing that a game is scalable can be difficult, as it may be hard to construct a scaling sequence such that every group has a tight  strategy, in particular when the game has many
groups and largely differing cost functions. To overcome this, we will develop a ``decomposition'' technique that permits to consider subclasses $\K' \subseteq \K$ of groups $k\in \K$ of a game independently in an ``asymptotic'' analysis of the PoA. We call it the ``asymptotic decomposition'' of a game.
In Sections~\ref{sec:scalableByDecompostion}--\ref{sec:ClassRelations}, we will combine this decomposition with scalability of its parts and show that these properties together lead to a rich class of \aw designed games that contains all scalable games. 


\Wuuu{We consider first a simple non-scalable \awd game  
in Example~\ref{example:NoScalableSubsequences}} below, and
   \Wuuu{illustrate  how to obtain and use the decomposition
   	on this simple game. } 

\begin{example}[A non-scalable game and its ``asymptotic decomposition'']
	\label{example:NoScalableSubsequences}
	Let $\Gamma$ be the routing game in Figure~\ref{fig:NoScalableSubsequences}(a) with cost functions displayed next to the arcs.
	$\Gamma$ has two groups given by the O/D pairs
	$(o_k,t_k)$,  $k = 1,2$. \Wuuuu{The regular} sequence $(d^{(n)})_{n\in \N}=\big(d_1^{(n)}= \sqrt[3]{n},\;d_2^{(n)}=n)\big)_{n\in \N}$
	with limit distribution
	$d^{(\infty)}=(d_1^{(\infty)},
	d_2^{(\infty)})=(0,1)$ \Wuuu{has no scalable subsequences
		$(n_i)_{i\in\N}$,
	since any scaling sequence 
	$(g_i)_{i\in \N}$
	fulfilling (L1)--(L3) 
	does not fulfill (L4) because the NE profile of the limit game
	has zero cost. Hence, $\Gamma$
    is not scalable by Definition~\ref{def:StrongScalabilityNew}. However, using \Wuuuu{the} asymptotic decomposition, 
    we will see  by Theorem~\ref{thm:AsymptoticDecomposition} below that the PoA$(d^{(n)})$ still converges to $1$
    for this regular sequence.} 
	\begin{figure}[!htb]
		\centering
		\begin{subfigure}{0.3\textwidth}
			\centering
			\begin{tikzpicture}[
			>=latex
			]
				\node[scale=0.4,circle,fill=black,label=left:{$o_1$}](o1){};
				\node[scale=0.4,circle,fill=black,label=left:$H$,below right =of o1](s){};
				\node[scale=0.4,circle,fill=black,label=left:{$o_2$},below left =of s](o2){};
				\node[scale=0.4,circle,fill=black,label=right:$F$, right =of s](h){};
				\node[scale=0.4,circle,fill=black,label=right:{$t_1$},above right =of h](t1){};
				\node[scale=0.4,circle,fill=black,label=right:{$t_2$},below right =of h](t2){};
				\draw[->,thick] (o1) to node[above]{$2x^2+1$} (t1);
				\draw[->,thick] (o1) to node[left]{$x^2+1$}
				(s);
				\draw[->,thick] (o2) to node[left]{$x+1$}
				(s);
				\draw[->,thick] (s) to node[above]{$x+1$} (h);
				\draw[->,thick] (h) to node[right]{$3x^2+1$} (t1);
				\draw[->,thick] (h) to node[right]{$x+1$} (t2);
				\draw[->,thick] (o2) to node[above]{$2x+1$}
				(t2);
			\end{tikzpicture}
			\subcaption{$\Gamma$ with  $d^{(n)}=(d_1^{(n)},d_2^{(n)})$}
		\end{subfigure}
	\begin{subfigure}{0.3\textwidth}
		\centering 
		\begin{tikzpicture}[
		>=latex
		]
		\node[scale=0.4,circle,fill=black,label=left:{$o_1$}](o1){};
		\node[scale=0.4,circle,fill=black,label=left:$H$,below right =of o1](s){};
		\node[scale=0.4,circle,fill=black,label=left:{$o_2$},below left =of s](o2){};
		\node[scale=0.4,circle,fill=black,label=right:$F$,right =of s](h){};
		\node[scale=0.4,circle,fill=black,label=right:{$t_1$},above right =of h](t1){};
		\node[scale=0.4,circle,fill=black,label=right:{$t_2$},below right =of h](t2){};
		\draw[->,thick] (o1) to node[above]{$2x^2+1$} (t1);
		\draw[->,thick] (o1) to node[left]{$x^2+1$}
		(s);
		\draw[dashed,->,thick] (o2) to node[left]{$x+1$}
		(s);
		\draw[->,thick] (s) to node[above]{$x+1$} (h);
		\draw[->,thick] (h) to node[right]{$3x^2+1$} (t1);
		\draw[dashed,->,thick] (h) to node[right]{$x+1$} (t2);
		\draw[dashed,->,thick] (o2) to node[above]{$2x+1$}
		(t2);
		\end{tikzpicture}
		\subcaption{$\Gamma_{|\{1\}}$ with 
		$d_{|\{1\}}^{(n)}=(d_1^{(n)})$}
	\end{subfigure}
\begin{subfigure}{0.3\textwidth}
	\centering
	\begin{tikzpicture}[
	>=latex
	]
	\node[scale=0.4,circle,fill=black,label=left:{$o_1$}](o1){};
	\node[scale=0.4,circle,fill=black,label=left:$H$,below right =of o1](s){};
	\node[scale=0.4,circle,fill=black,label=left:{$o_2$},below left =of s](o2){};
	\node[scale=0.4,circle,fill=black,label=right:$F$,right =of s](h){};
	\node[scale=0.4,circle,fill=black,label=right:{$t_1$},above right =of h](t1){};
	\node[scale=0.4,circle,fill=black,label=right:{$t_2$},below right =of h](t2){};
	\draw[dashed,->,thick] (o1) to node[above]{$2x^2+1$} (t1);
	\draw[dashed,->,thick] (o1) to node[left]{$x^2+1$}
	(s);
	\draw[->,thick] (o2) to node[left]{$x+1$}
	(s);
	\draw[->,thick] (s) to node[above]{$x+1$} (h);
	\draw[dashed,->,thick] (h) to node[right]{$3x^2+1$} (t1);
	\draw[->,thick] (h) to node[right]{$x+1$} (t2);
	\draw[->,thick] (o2) to node[above]{$2x+1$}
	(t2);
	\end{tikzpicture}
	  \subcaption{$\Gamma_{|\{2\}}$ with 
	  	$d_{|\{2\}}^{(n)}=(d_2^{(n)})$}   
\end{subfigure}
		\caption{A non-scalable game and its decomposition}
		\label{fig:NoScalableSubsequences}
\end{figure}

\Wuuu{\Wuuuu{For this
	regular sequence $(d^{(n)})_{n\in \N}$, we consider} the two O/D pairs
	$(o_1,t_1)$ and $(o_2,t_2)$ as independent ``subgames'' $\Gamma_{|\{1\}}$ and $\Gamma_{|\{2\}}$ of the game 
	$\Gamma$ in Figure~\ref{fig:NoScalableSubsequences}(a), (i.e., the
	two independent games formed only by the solid arcs in 
	Figure~\ref{fig:NoScalableSubsequences} (b)--(c), respectively), 
	and obtain their limits
	$\Gamma^{(\infty)}_{|\{1\}}$ and $\Gamma_{|\{2\}}^{(\infty)}$
	in Figure~\ref{fig:MiniExample_LimitGame} (a)--(b)
	w.r.t. the two scaling sequences 
	$(g_n^{(1)}=n^{2/3})_{n\in \N}$ 
	and $(g_n^{(2)}=n)_{n\in \N}$, respectively.
	Herein, we identify the ``restricted vector'' 
	$(d_k^{(n)})$ of the demand vector $d^{(n)}=(d_1^{(n)},d_2^{(n)})$ as 
	the demand vector of $\Gamma_{|\{k\}}$
	and ignore the other subgame
	completely in  subgame $\Gamma_{|\{k\}},$ $k=1,2.$
    This forms an {\em asymptotic decomposition} of $\Gamma$ w.r.t.
    the regular sequence $(d^{(n)})_{n\in \N}.$
   }

   \Wuuuu{  
   	We first sort the two subgames $\Gamma_{|\{1\}}$ and $\Gamma_{|\{2\}}$
   	decreasingly according to the asymptotic sizes of their demands
   	$d_1^{(n)}$ and $d_2^{(n)}$, and
   	write the decomposition of game 
   	$\Gamma$ w.r.t. this regular 
   	sequence $(d^{(n)})_{n\in \N}$ as $\Gamma\asymp_{d^{(n)}}\Gamma_{|\{2\}}\oplus\Gamma_{|\{1\}}.$
   	We then use that the subgames
   	$\Gamma_{|\{1\}}$ and $\Gamma_{|\{2\}}$ are scalable and
   	apply Lemma~\ref{lma:PriceConvergenceOfStrategies} independently 
   	to the two subgames. 
   	This yields an {\em asymptotic} upper bound $\overline{\text{PoA}}(d^{(n)})$
   	such that $\varlimsup_{n\to \infty}\overline{\text{PoA}}(d^{(n)})\ge
   	\varlimsup_{n \to \infty}\text{PoA}(d^{(n)}).$
   	This upper bound $\overline{\text{PoA}}(d^{(n)})$
   	equals the {\em ratio} of the NE cost of game 
   	$\Gamma$ over
    the sum of the NE costs for the two subgames $\Gamma_{|\{1\}}$ and
    $\Gamma_{|\{2\}}$,
    see \eqref{eq:MiniProof_PoA_UpperBound} below. 
    Using again the scaling properties in Lemma~\ref{lma:PriceConvergenceOfStrategies} for the two subgames and their limit games, we show that this
   upper bound converges to $1,$ which implies
   that $\lim_{n\to\infty}$PoA$(d^{(n)})=1$
   for the regular sequence $(d^{(n)})_{n\in \N}$.}
   \Wuuuu{
   	The distinction of the asymptotic sizes of the demands of the subgames is crucial in this analysis, see the remarks  after the \Wuu{definitions} of ``subgame'' (Definition~\ref{def:MarginalGame}) and 
   	``decomposition'' (Definition~\ref{def:EssentialDecomposition}) and
   	the general proof in Appendix~\ref{proof:AsymptoticDecomposition}.}
\end{example}

\begin{figure}[!htb]
	\centering
	\begin{subfigure}{0.49\textwidth}
		\centering
		\begin{tikzpicture}[
		>=latex
		]
		\node[scale=0.4,circle,fill=black,label=left:{$o_1$}](o1){};
		\node[scale=0.4,circle,fill=black,label=left:$H$,below right =of o1](s){};
		\node[scale=0.4,circle,fill=black,label=right:$F$,right =of s](h){};
		\node[scale=0.4,circle,fill=black,label=right:{$t_1$},above right =of h](t1){};
		\draw[->,thick] (o1) to node[above]{$2x^2$} (t1);
		\draw[->,thick] (o1) to node[left]{$x^2$}
		(s);
		\draw[->,thick] (s) to node[above]{$0$} (h);
		\draw[->,thick] (h) to node[right]{$3x^2$} (t1);
		\end{tikzpicture}
		\caption{$\Gamma_{|\{1\}}^{(\infty)}$}
	\end{subfigure}
	\begin{subfigure}{0.49\textwidth}
		\centering
		\begin{tikzpicture}[
		>=latex
		]
		\node[scale=0.4,circle,fill=black,label=left:{$o_2$}](o2){};
		\node[scale=0.4,circle,fill=black,label=left:$H$,above right =of o1](s){};
		\node[scale=0.4,circle,fill=black,label=right:$F$,right =of s](h){};
		\node[scale=0.4,circle,fill=black,label=right:{$t_2$},below right =of h](t2){};
		\draw[->,thick] (o2) to node[left]{$x$}
		(s);
		\draw[->,thick] (s) to node[above]{$x$} (h);
		\draw[->,thick] (h) to node[right]{$x$} (t2);
		\draw[->,thick] (o2) to node[above]{$2x$}
		(t2);
		\end{tikzpicture}
		\caption{$\Gamma_{|\{2\}}^{(\infty)}$}
	\end{subfigure}
	\caption{The limit games}
	\label{fig:MiniExample_LimitGame}
\end{figure}

\begin{definition}[Subgame]
	\label{def:MarginalGame}
Consider a game $
	\Gamma=\Big(A,\K,\S=\bigcup_{k\in \K}\S_k,\tau=\big(\tau_a(\cdot)\big)_{a\in A},d=(d_k)_{k\in\K}\Big)
	$
	and
	a non-empty subset $\K'$ of $\K.$ We  call 
	the game
	\[
	\Gamma_{|\K'}=\Big(A_{|\K'}=\{a\in A:\ \exists s\in \S_{|\K'} \ s.t.\ a\in s\},\K',\S_{|\K'}:=\bigcup_{k\in \K'}\S_k,\tau_{|\K'}:=\big(\tau_a(\cdot)\big)_{a\in A_{|\K'}},d_{|\K'}:=(d_k)_{k\in \K'}\Big)
	\] 
	the \emph{subgame} 
	of $\Gamma$ \emph{induced by} $\K'.$ 
\end{definition}

\Wuuuu{We also need some helpful notation relevant to subgames.} \Wuu{For a subgame}
$\Gamma_{|\K'},$ we call resources
$a\in A_{|\K'}$ \emph{relevant}, and the others \emph{irrelevant},
%
\Wuu{see, e.g., these solid and dashed arcs in 
Figure~\ref{fig:NoScalableSubsequences}(b)--(c), respectively.
Clearly, subgame $\Gamma_{|\K'}$ uses only its relevant resources.}
\Wuuuu{The restricted vector $d_{|\K'}=
	(d_k)_{k\in \K'}$ denotes the demand vector
	of subgame $\Gamma_{|\K'}.$
Note that groups $k\in \K\setminus\K'$
are not considered in that subgame.}
\Wuuuu{For a profile $f=(f_s)_{s\in \S}$  of $\Gamma$
w.r.t.\ demand vector $d,$  we call its restriction $f_{|\K'}:=(f_s)_{k\in \K',s\in \S_k}$ 
the {\em subprofile}  
w.r.t.\ the restricted demand vector $d_{|\K'}.$} \Wuu{Trivially, $f_{|\K'}$
fulfills all demands from subgame $\Gamma_{|\K'},$
and is thus a profile of $\Gamma_{|\K'}$ w.r.t.
demand vector $d_{|\K'}.$} \Wuuu{For a relevant resource $a\in A_{|\K'},$ 
we call $f_{a|\K'}:=\sum_{k \in \K}\sum_{s\in S_k : a \in s} f_s$
and $\tau_a(f_{a|\K'})$
the {\em independent consumption} and {\em independent cost} of resource $a$ from
subgame $\Gamma_{|\K'}$ w.r.t. profile $f,$ respectively. Note that
this independent consumption is not larger than
\Wuu{its} joint consumption,  i.e.,
$f_{a|\K'}\le f_a$ for each $a\in A_{|\K'},$
and that the resulting independent cost $\tau_a(f_{a|\K'})$
is also not larger than its
{\em joint} cost $\tau_a(f_a),$ i.e., 
$\tau_a(f_{a|\K'})\le \tau_a(f_a)$ because 
$\tau_a(\cdot)$ is non-decreasing.}

\Wuuuu{The independent cost of resources induces 
an independent cost of strategies.}
For each strategy $s\in \S_{|\K'},$ we  call  
$\tau_{s}(f_{|\K'}):=\sum_{a\in s} \tau_{a}(f_{a|\K'})$ the {\em independent cost} of strategy $s$ for
\Wuu{subgame} $\Gamma_{|\K'}$, 
$\sum_{k  \in \K'}\sum_{s \in S_k} f_s\cdot
\tau_{s} (f_{|\K'})$ 
the \Wuu{\em independent total cost}
of $\Gamma_{|\K'},$ and  $\sum_{k  \in \K'}\sum_{s\in S_k}
f_s\cdot \tau_s(f)$ the \Wuu{\em joint total cost}
of $\Gamma_{|\K'}$ \Wuu{w.r.t. profile $f,$ respectively.}
 \Wuuuu{Clearly, the independent total cost is not larger than
\Wuu{its} joint total cost.
For instance,} if $f$ is a profile in Example~\ref{example:NoScalableSubsequences}(a) that routes 1 unit of demand on every $(o_k,t_k)$-path, $k=1,2$, then the \Wuu{independent total cost} of $\Gamma_{|\{1\}}$ is 7, while its {\em joint total cost} is 8. 
\Wuu{Obviously, the independent total cost of $\Gamma_{|\K'}$ w.r.t
the profile}
$f$ of game $\Gamma$
is just
the \Wuu{social} cost $C_{\Gamma_{|\K'}}(f_{|\K'})=\sum_{k \in \K'}\sum_{s\in S_k} f_s
\cdot \tau_{s}(f_{|\K'})$ of 
\Wuuuu{the profile $f_{|\K'}$ of subgame $\Gamma_{|\K'},$
while the joint total cost is the {\em total contribution}
of subgame $\Gamma_{|\K'}$ to the social cost $C(f)$ of the
profile $f$.}
\Wuuu{Note that
we put}
``$\Gamma_{|\K'}$" in the subscript of $C(\cdot)$
\Wuuu{to explicitly refer} to the \Wuu{social} cost of
$\Gamma_{|\K'}.$

\Wuuuu{With the above notation, we can now formally
define the decomposition. As illustrated in Example~\ref{example:NoScalableSubsequences}, the decomposition for a particular
regular sequence $(d^{(n)})_{n\in \N}$ needs the property  that 
the sizes $d_k^{(n)}$ are mutually comparable
in the limit, i.e., 
$\lim_{n \to \infty}d_k^{(n)}/d_{k'}^{(n)}\in [0,\infty]$
exists for each pair $(k,k')\in \K\times\K$.} \Wuuuu{We call a regular sequence} $\big(d^{(n)}\big)_{n\in \N}$
{\em decomposable}, 
if $\lim_{n\to\infty}d_k^{(n)}=\infty$
for each $k\in \K,$ and if
$\lim_{n\to\infty}d_k^{(n)}/d_{k'}^{(n)}
\in [0,\infty]$ for each $k,k'\in \K.$
\Wuuuu{Note that every regular sequence $(d^{(n)})_{n\in \N}$
with $\varlimsup_{n \to \infty}d_k^{(n)}=\infty$ for all
$k\in \K$ has a decomposable subsequence.} 
\Wuuuu{Given a decomposable sequence
	$(d^{(n)})_{n\in \N}$,} \Wuu{we} can partition $\K$ 
by the asymptotic growth rates of the demand sizes $d_k^{(n)}$
\Wuuuu{as in Example~\ref{example:NoScalableSubsequences}.}
We write
$k\preceq k'$ if $\lim_{n\to\infty}
\frac{d_k^{(n)}}{d_{k'}^{(n)}}<\infty,$ $k\not\preceq k'$ if  
$\lim_{n\to\infty}\frac{d_k^{(n)}}{d_{k'}^{(n)}}=\infty$, and $k\prec k'$ if $k\preceq k'$ and $k'\not\preceq k$.
Two groups $k, k'$ are \emph{equivalent}, written as $k \sim k',$ if $k\preceq k'$ and $k'\preceq k$. 
The partition of $\K$ is defined by putting equivalent groups $k, k'$ into one class $\K_u$ of the partition. The \Wuu{subgames} $\Gamma_{|\K_u}$ induced by these classes and ordered {\em decreasingly} by their demand sizes form an {\em asymptotic decomposition} of $\Gamma,$ if every \Wuu{subgame} $\Gamma_{|\K_u}$ is scalable w.r.t\ $(d_{|\K_u}^{(n)})_{n\in \N}$.

\Wuu{A formal definition follows}. \Wuu{The notation
$\asymp_{d^{(n)}}$ indicates} that the asymptotic decomposition
depends on $(d^{(n)})_{n\in \N},$
\Wuuu{and thus different demand sequences may lead to different
decompositions of $\Gamma$}.

\begin{definition}[Asymptotic decomposition of games]
	\label{def:EssentialDecomposition}
	Consider a game $\Gamma$ and a decomposable sequence $(d^{(n)})_{n\in \N}.$
	We call \Wuu{subgames} $\Gamma_{|\K_1},\ldots,\Gamma_{|\K_m}$  
	 of $\Gamma$
	an \emph{asymptotic decomposition} of $\Gamma$ w.r.t.\ $(d^{(n)})_{n\in \N},$ denoted by 
	$
	\Gamma\asymp_{d^{(n)}} \Gamma_{|\K_1}\oplus\cdots\oplus\Gamma_{|\K_m},
	$
	if the following conditions hold. 
	\begin{itemize}
		\item[(AD1)] 
		If $k,k'\in \K_u,$ then $k\sim k'.$
		
		\item[(AD2)] For each
		$u,v\in \{1,\ldots,m\},$
		if $u<v,$ then $k'\prec k$ for all
		$k\in \K_u$ and all $k'\in \K_v.$ 
		
		\item[(AD3)] For each $u\in \{1,\ldots,m\},$ 
		$\big(d^{(n)}_{|\K_u }=(d_k^{(n)})_{k\in \K_u}\big)_{n\in \N}$
		is scalable 
		w.r.t.\ $\Gamma_{|\K_u}$, i.e.,
		there is a scaling sequence $(g_{n}^{(u)})_{n\in \N}$
		such that $\Gamma^{(\infty)}_{|\K_u}=
		\lim_{ d^{(n)}_{|\K_u}\to \infty}\Gamma_{|\K_u}^{[g_n^{(u)}]}$ 
		for a limit game $\Gamma^{(\infty)}_{|\K_u}$.
	\end{itemize}
\end{definition}


\Wuuuu{Given an asymptotic decomposition $
	\Gamma\asymp_{d^{(n)}} \Gamma_{|\K_1}\oplus\cdots\oplus\Gamma_{|\K_m},
	$ it follows easily that}
\begin{equation}\label{eq:MarginalSOCost}
\sum_{k  \in \K_u}\sum_{s\in S_k}
f_s^{*(n)}\cdot \tau_s(f^{*(n)})
\ge C_{\Gamma_{|\K_u}}(f_{|\K_u}^{*(n)})\ge C_{\Gamma_{|\K_u}}(f^{*(\K_u,n)}),
\quad \forall u=1,\ldots,m,
\end{equation}
\Wuuu{where $f^{*(n)}$ is an SO profile
of $\Gamma$ w.r.t. $d^{(n)},$
and $f^{*(\K_u,n)}$ is an SO profile of subgame 
$\Gamma_{|\K_u}$ w.r.t. its demand vector 
$d^{(n)}_{|\K_u}.$ \Wuuuu{Just observe} that 
the joint total cost of 
subgame $\Gamma_{|\K_u}$ in the SO profile 
$f^{*(n)}$ of game $\Gamma$ is not less than
its independent total cost 
$C_{\Gamma_{|\K_{u}}}(f^{*(n)}_{|\K_u}),$
and that the social cost $C_{\Gamma_{|\K_{u}}}(f^{*(n)}_{|\K_u})$
of the profile $f_{|\K_u}^{*(n)}$ of game
$\Gamma_{|\K_u}$
is not less than the SO cost $C_{\Gamma_{|\K_{u}}}(f^{*(\K_u,n)})$
of game
$\Gamma_{|\K_u}.$} 

Lemma~\ref{lma:Prop_AD}
\Wuuuu{below uses \eqref{eq:MarginalSOCost} and 
	Lemma~\ref{lma:PriceConvergenceOfStrategies}, and shows some basic properties of} the decomposition $\Gamma\asymp_{d^{(n)}}\Gamma_{|\K_1}\oplus
\cdots\oplus\Gamma_{|\K_m}.$  \Wuuu{Herein, we put $\M:=\{1,\ldots,m\}.$
The proof is trivial and omitted.}

\begin{lemma}[Elementary properties of the asymptotic decomposition]\label{lma:Prop_AD}
	Consider a game $\Gamma$ and a decomposable sequence  
	$(d^{(n)})_{n\in \N}$ \Wuu{s.t.}
	$\Gamma\asymp_{d^{(n)}}\Gamma_{|\K_1}\oplus
	\cdots\oplus\Gamma_{|\K_m}$
	w.r.t. scaling sequences 
	$(g_n^{(u)})_{n\in \N},$ $u\in \M.$ Let $(\tilde{f}^{(n)})_{n\in \N}$ and $(f^{*(n)})_{n\in \N}$ be sequences of \Wuu{NE profiles and SO profiles} w.r.t.\ $d^{(n)},$ respectively. Let 
	$\tilde{f}^{(\K_u,n)}$ and $f^{*(\K_u,n)}$ be an
	NE profile and an SO profile \Wuu{of subgame $\Gamma_{|\K_u}$} w.r.t.\
	$d^{(n)}_{|\K_u}$ for each $u\in \M$ and $n \in \N$, respectively.
	\Wuu{Then:}
	\begin{itemize}
		\item[a)] $\lim_{n\to\infty}
		\frac{C_{\Gamma_{|\K_u}}(\tilde{f}^{(\K_u,n)})}{C_{\Gamma_{|\K_u}}(f^{*(\K_u,n)})}=1$ and 
		$\lim_{n\to\infty}
		\frac{C_{\Gamma_{|\K_u}}(\tilde{f}^{(\K_u,n)})}{
			T(d^{(n)}_{|\K_u})\cdot g_n^{(u)}
		}=C_{\Gamma^{(\infty)}_{|\K_u}}(\tilde{f}^{(\K_u,\infty)})
		$ for each $u\in \M,$
		where $\Gamma^{(\infty)}_{|\K_u}$ is the limit
		of $\Gamma_{|\K_u}$ w.r.t.\ $(d^{(n)}_{|\K_u})_{n\in \N}$
		under scaling sequence
		$(g_n^{(u)})_{n\in \N},$ \Wuu{and $\tilde{f}^{(\K_u,\infty)}$ 
		is an NE profile of subgame $\Gamma^{(\infty)}_{|\K_u}.$}
		
		
		\item[b)] $\sum_{k\in \K_u} \sum_{s\in \S_k}
		f_{s}^{*(n)}\cdot \tau_s(f^{*(n)})\ge C_{\Gamma_{|\K_u}}(f^{*(n)}_{|\K_u})
		\ge C_{\Gamma_{|\K_u}}(f^{*(\K_u,n)})$ for each $u\in \K_u$ and thus
	\Wuuu{\begin{equation*}
		\begin{split}
		\varlimsup_{n\to\infty}\text{PoA}(d^{(n)})&=
		\varlimsup_{n\to\infty}\frac{C(\tilde{f}^{(n)})}{
			\sum_{u\in \M} \big[\sum_{k\in \K_u} \sum_{s\in \S_k}
			f_{s}^{*(n)}\cdot \tau_s(f^{*(n)})\big]}\le \varlimsup_{n\to\infty}\frac{C(\tilde{f}^{(n)})}{
			\sum_{u\in \M} C_{\Gamma_{|\K_u}}(f^{*(\K_u,n)})}\\
		&= \varlimsup_{n\to\infty}\frac{C(\tilde{f}^{(n)})}{
			\sum_{u\in \M} C_{\Gamma_{|\K_u}}(\tilde{f}^{(\K_u,n)})}=:
		\varlimsup_{n \to \infty}\overline{\text{PoA}}(d^{(n)}).
		\end{split}
	\end{equation*}}
	\end{itemize}
\end{lemma}

\Wuuu{Theorem~\ref{lma:Prop_AD}b) provides an asymptotic upper bound 
	$\overline{\text{PoA}}(d^{(n)})=\frac{C(\tilde{f}^{(n)})}{
		\sum_{u\in \M} C_{\Gamma_{|\K_u}}(\tilde{f}^{(\K_u,n)})}$
	for the PoA$(d^{(n)}).$ We can thus prove the convergence of
	$\text{PoA}(d^{(n)})$ to $1$ by \Wuuuu{showing that} $\lim_{n \to \infty}\overline{\text{PoA}}(d^{(n)})=1$
	when the decomposition exists.}

\Wuuu{For \Wuuuu{the game} $\Gamma$ in Example~\ref{example:NoScalableSubsequences}, 
this upper bound equals
\begin{equation}\label{eq:MiniProof_PoA_UpperBound}
\overline{\text{PoA}}(d^{(n)})=\frac{C(\tilde{f}^{(n)})}{\sum_{k=1}^{2}C_{\Gamma_{|\{k\}}}(\tilde{f}^{(\{k\},n)})}=\frac{\sum_{k=1}^{2}\big[\sum_{s\in \S_k}\tilde{f}_s^{(n)}\cdot\tau_s(\tilde{f}^{(n)})\big]}{\sum_{k=1}^{2}C_{\Gamma_{|\{k\}}}(\tilde{f}^{(\{k\},n)})}
\end{equation} 
for the \Wuuuu{decomposable} sequence $(d^{(n)})_{n\in \N}$
with $d_1^{(n)}=n^{1/3}$ and $d_2^{(n)}=n$ and its resulting
asymptotic decomposition $\Gamma\asymp_{d^{(n)}}\Gamma_{|\{2\}}\oplus\Gamma_{|\{1\}}.$
Using
Lemma~\ref{lma:PriceConvergenceOfStrategies},
we prove the convergence of $\overline{\text{PoA}}(d^{(n)})$ to $1$ by \Wuuuu{independently} comparing
the NE cost $C_{\Gamma_{|\{k\}}}(\tilde{f}^{(\{k\},n)})$ of subgame $\Gamma_{|\{k\}}$ with its total contribution (i.e., the joint total cost) $\sum_{s\in \S_k}\tilde{f}_s^{(n)}\cdot\tau_s(\tilde{f}^{(n)})$ 
in the NE cost $C(\tilde{f}^{(n)})$, 
$k=1,2.$ \Wuuuu{The asymptotic growth rates
of the total demands $T(d_{|\{1\}}^{(n)})=d_1^{(n)}$ and $T(d_{|\{2\}}^{(n)})=d_2^{(n)}$
determine the order for the comparisons, i.e., first for subgame $\Gamma_{|\{2\}}$ and then
for subgame $\Gamma_{|\{1\}}.$}}

\Wuuu{\Wuuuu{These comparisons exploit the fact that the subgames are scalable 
		and thus have limits, see Figure~\ref{fig:MiniExample_LimitGame}(a)--(b).} 
 Using Lemma~\ref{lma:PriceConvergenceOfStrategies}, the comparison 
 for subgame $\Gamma_{|\{2\}}$ \Wuuuu{results in} the limit
\begin{equation}\label{eq:Marg1}
\lim_{n\to \infty}\frac{\sum_{s\in \S_2}
	\tilde{f}_{s}^{(n)}\!\cdot\! \tau_{s}
	(\tilde{f}^{(n)})}{C_{\Gamma_{|\{2\}}}(\tilde{f}^{(\{2\},n)})}\!=\!\lim_{n\to \infty}\frac{\sum_{s\in \S_2}
	\tilde{f}_{s}^{(n)}\!\cdot\! \tau_{s}
	(\tilde{f}^{(n)})/\big(T(d_{|\{1\}}^{(n)})\!\cdot\! g_n^{(2)}\big)}{C_{\Gamma_{|\{2\}}}(\tilde{f}^{(\{2\},n)})/\big(T(d_{|\{1\}}^{(n)})\!\cdot\! g_n^{(2)}\big)}\!=\!\frac{C_{\Gamma_{|\Gamma^{(\infty)}}}(\tilde{f}_{|\{2\}}^{(\infty)})}{C_{\Gamma_{|\Gamma^{(\infty)}}}(\tilde{f}^{(\{2\},\infty)})}\!=\!1.
\end{equation}
Herein, $\tilde{f}_{|\{2\}}^{(\infty)}$ is \Wuuuu{the} limit
of the profile distributions $\tilde{f}_{|\{2\}}^{(n)}/T(d_{|\{2\}}^{(n)}),$
while $\tilde{f}^{(\{2\},\infty)})$ is \Wuuuu{the} limit 
of the profile distributions $\tilde{f}^{(\{2\},n)}/T(d_{|\{2\}}^{(n)}).$ Since
$\lim_{n \to \infty}d_1^{(n)}/d_2^{(n)}=\lim_{n \to \infty}n^{1/3}/n=0,$ 
the scaled joint cost $\tau_a(\tilde{f}^{(n)}_a)/g_n^{(2)}$ of the unique
common arc $a=(H,F)$
is asymptotically equal to its scaled independent cost $\tau_a(\tilde{f}^{(n)}_{a|\{2\}})/g_n^{(2)}$. 
Since $\tilde{f}^{(n)}$ and $\tilde{f}^{(\{2\},n)}$
are NE profiles,
we obtain by 
Lemma~\ref{lma:PriceConvergenceOfStrategies} that $\tilde{f}_{|\{2\}}^{(\infty)}$ and $\tilde{f}^{(\{2\},\infty)}$ are NE profiles of $\Gamma_{|\{2\}}^{(\infty)}$
and thus have equal cost in $\Gamma_{|\{2\}}^{(\infty)}$.
This shows \eqref{eq:Marg1}.}

\Wuuu{The analysis for subgame $\Gamma_{|\{2\}}$ implies
	that the joint cost of the common arc $(H,F)$
	in the NE profile $\tilde{f}^{(n)}$ is $\Theta(g_n^{(2)}).$ 
   \Wuuuu{Observing that} 
$g_n^{(1)}=n^{2/3}\in o(g_n^{(2)})=o(n),$ we obtain that $\tau_a(\tilde{f}^{(n)}_{a})\in O(g_n^{(2)})$
for each relevant arc $a\in A_{|\{1\}}$ of subgame 
$\Gamma_{|\{1\}}.$ This means that the total contribution
$\sum_{s\in \S_1}\tilde{f}^{(n)}_s\cdot \tau_s(\tilde{f}_s^{(n)})$
of subgame $\Gamma_{|\{1\}}$ \Wuuuu{to} the NE cost
of profile $\tilde{f}^{(n)}$ is 
$O\big(T(d_{|\{1\}}^{(n)})\cdot g_n^{(2)}\big)
\subseteq o\big(T(d_{|\{2\}}^{(n)})\cdot g_n^{(2)}\big).$
By Lemma~\ref{lma:Prop_AD}a), we \Wuuuu{obtain} also \Wuuuu{that}
$C_{\Gamma_{|\{1\}}}(\tilde{f}^{(\{1\},n)})
\in \Theta\big(T(d_{|\{1\}}^{(n)})\cdot g_n^{(1)}\big)
\subseteq o\big(T(d_{|\{2\}}^{(n)})\cdot g_n^{(2)}\big).$
\Wuuuu{Combining this} with \eqref{eq:Marg1}
\Wuuuu{implies} $\lim_{n\to \infty}\overline{\text{PoA}}(d^{(n)})=1.$}


\Wuuuu{Generalizing these proof ideas yields} Theorem~\ref{thm:AsymptoticDecomposition}.
It
shows that $\lim_{n \to \infty}\overline{\text{PoA}}(d^{(n)}) = 1$ 
for an arbitrary game \Wuuuu{$\Gamma$}
when $(d^{(n)})_{n\in \N}$
is decomposable and \Wuuu{$\Gamma$ has} an asymptotic
decomposition, 
see the detailed proof in Appendix~\ref{proof:AsymptoticDecomposition}.}

\begin{theorem}[Asymptotic decomposition theorem]\label{thm:AsymptoticDecomposition}
	Let $\Gamma$ be a game and $(d^{(n)})_{n\in \N}$ be a decomposable sequence.
	If $\Gamma\asymp_{d^{(n)}} \Gamma_{|\K_1}\oplus\cdots\oplus\Gamma_{|\K_m},$
	then $\overline{\text{PoA}}(d^{n}) = 1,$
	and $\lim_{n\to \infty}\text{PoA}(d^{(n)})=1.$
\end{theorem}

\Wuuuu{The proof for Example~\ref{example:NoScalableSubsequences}
	considers only the case that 
	one subgame in the decomposition, i.e., $\Gamma_{|\{2\}},$ completely 
	determines the limit of the upper bound $\overline{\text{PoA}}(d^{(n)})$. 
	In general, several subgames may determine the 
	limit together, although
	some of them have a negligible total demand compared
	to the total demand $T(d^{(n)})$ of $\Gamma.$  
	The proof in the Appendix~\ref{proof:AsymptoticDecomposition} shows that $\lim_{n\to \infty}\overline{\text{PoA}}(d^{(n)})=1$
	for this general case
	by proving inductively over  
	$u\in \M$ that $\lim_{n\to \infty}\sum_{v=1}^u
	\sum_{k\in \K_v}\sum_{s\in \S_k}
	\tilde{f}_s^{(n)}\cdot \tau_a(\tilde{f}^{(n)})/\sum_{v=1}^{u}
	C_{\Gamma_{|\K_v}}(\tilde{f}^{(\K_v,n)})=1,$ i.e., 
	the total contribution of the ``joint'' subgame 
	$\Gamma_{|\bigcup_{v=1}^u\K_v}$
	to the NE cost $C(\tilde{f}^{(n)})$
	of game $\Gamma$
	asymptotically equals 
	the sum of the NE costs $C_{\Gamma_{|\K_v}}(\tilde{f}^{(\K_v,n)})$
	of the individual subgames $\Gamma_{|\K_v},$ 
	$v=1,\ldots,u.$ }
 
 \Wuuu{Note that there are only two possible cases for each 
	step $u\in \M$ in the induction. Either
	$g_n^{(u)}\in O(\max_{v=1}^{u-1}\ g_n^{(v)})$ or
	$g_n^{(u)}\in \omega(\max_{v=1}^{u-1}\ g_n^{(v)}).$
	If  $g_n^{(u)}\in O(\max_{v=1}^{u-1}\ g_n^{(v)}),$ then the contribution of the subgame $\Gamma_{|\K_u}$
	to the NE cost $C(\tilde{f}^{(n)})$ is completely negligible compared to that
	of the joint subgame $\Gamma_{|\bigcup_{v=1}^{u-1}\K_v},$
	and an argument similar to that for subgame $\Gamma_{|\{1\}}$
	of Example~\ref{example:NoScalableSubsequences} applies.
	If $g_n^{(u)}\in \omega(\max_{v=1}^{u-1}\ g_n^{(v)}),$ then
	the scaled joint cost 
	$\tau_a(\tilde{f}^{(n)})/g_n^{(u)}$ of an overlapping resource
	$a\in A_{|\K_u}\cap A_{|\bigcup_{v=1}^{u-1}\K_v}$
	asymptotically equals its independent cost 
	$\tau_a(f_{|\K_u}^{(n)})/g_n^{(n)}$ and a slight adaptation of 
	the above argument for subgame 
	$\Gamma_{|\{2\}}$ of Example~\ref{example:NoScalableSubsequences}
	applies.
}

\Wuuu{With the upper bound $\overline{\text{PoA}}(d^{(n)}),$
	we need no longer \Wuuuu{consider SO profiles}
	$f^{*(n)}$ of game $\Gamma$ in the convergence analysis, \Wuuuu{and so the proof does not use any} 
 particular properties other than our standard assumptions \Wuuuu{on} the cost functions.} \Wuuu{In particular, the cost functions 
need not be differentiable, which is usually a premise in the worst-case analysis of the PoA, \Wuu{see,
	e.g., } \cite{Roughgarden2000How,Roughgarden2001Designing,Roughgarden2002The,Roughgarden2015Intrinsic}.}
\Wuuuu{Instead, the proof builds} essentially on the existence of scaling sequences
$(g_n^{(u)})_{n\in \N}$ for \Wuu{subgames} 
$\K_u$.
%
In the sequel, we thus only need to \Wuu{verify} the existence of
scaling sequences \Wuuu{that make all} \Wuu{subgames}
scalable.

. 

\Wuuuu{An arbitrary unbounded 
sequence need not have a decomposable subsequence, since
$\lim_{n \to \infty} d_k^{(n)}\in [0,\infty)$ 
may hold
for some $k\in \K$.}
However, we will see in Corollary~\ref{thm:ADGame_Results} \Wuuu{below} that 
the convergence of the PoA for decomposable sequences still
carries over to all unbounded sequences by \Wuu{Lemma~\ref{lma:Convergence_iff_RegularSequence}c)} and
a slight adaptation of the
decomposition.

\subsection{Games scalable by decomposition}\label{sec:scalableByDecompostion}

We now refine the above asymptotic
decomposition technique to extend the convergence result \Wuu{in
	Theorem~\ref{thm:AsymptoticDecomposition}} \Wuuuu{from}
decomposable sequences to
arbitrary \Wuu{regular} sequences. 
Given a regular sequence $(d^{(n)})_{n\in \N},$
we call a group $k\in \K$ {\em regular} w.r.t.\
$(d^{(n)})_{n\in \N},$ if  $\lim_{n\to\infty}d_k^{(n)}=\infty$, and denote 
by $\K_{reg}$ the set of all regular groups w.r.t.\ 
$(d^{(n)})_{n\in \N}.$ 
%
Let $d^{(n)}_{|\K_{reg}}=(d_k^{(n)})_{k\in \K_{reg}}$ be the demand sequence of the regular groups and 
$T(d^{(n)}_{|\K_{reg}})=\sum_{k \in \K_{reg}}d_k^{(n)}$ their total demand.
Clearly, $\lim_{n\to\infty}
\frac{T(d^{(n)}_{|\K_{reg}})}{T(d^{(n)})}=1$ and 
$\sum_{k  \in \K\setminus\\K_{reg}}d_k^{(n)}\in O(1).$
%
\Wuuu{Corollary~\ref{thm:ADGame_Results} below shows that the PoA still converges to $1$} when 
$(d^{(n)}_{|\K_{reg}})_{n\in\N}$ is decomposable and
$\Gamma_{|\K_{reg}}\asymp_{d^{(n)}_{|\K_{reg}}}=\Gamma_{|\K_1}
\oplus\cdots\oplus\Gamma_{|\K_m}$ for
\Wuu{a partition} $\K_1,\ldots,\K_m$ of
$\K_{reg}.$ 
\Wuu{It} follows immediately 
 \Wuuu{from the fact that 
subgame $\Gamma_{|\K\setminus\K_{reg}}$ contributes only
a negligible part
to the NE cost
in comparison to the NE cost of subgame $\Gamma_{|\K_{reg}}.$} 
We move the simple proof to Appendix~\ref{proof:ADGame_Results}.
\begin{corollary}\label{thm:ADGame_Results}
	Let $\Gamma$ be a game and $(d^{(n)})_{n\in \N}$ be a regular
	sequence. Then $\lim_{n\to\infty}$PoA$(d^{(n)})=1$ if the demand sequence
	$(d^{(n)}_{|\K_{reg}})_{n\in \N}$ of the set $\K_{reg}$ of all regular groups is decomposable
	and $\Gamma_{|\K_{reg}}\asymp_{d^{(n)}_{|\K_{reg}}}\Gamma_{|\K_1}
	\oplus\cdots\oplus\Gamma_{|\K_m}$ for some subsets $\K_1,\ldots,\K_m$ of $\K_{reg}$.
\end{corollary}

We use the assumptions of Corollary~\ref{thm:ADGame_Results} to define the class of games \Wuu{scalable by decomposition}.

\begin{definition}[Games scalable by decomposition]\label{def:ScalableGames}
	We call a game $\Gamma$  \emph{scalable by decomposition}, if 
	each regular sequence $(d^{(n)})_{n\in \N}$
	has an infinite subsequence $(n_i)_{i\in \N}$ such that 
	$(d^{(n_i)}_{|\K_{reg}})_{i\in \N}:=((d_k^{(n_i)})_{k\in \K_{reg}})_{i\in \N}$ is decomposable
	and $\Gamma_{|\K_{reg}}\asymp_{d^{(n_i)}_{|\K_{reg}}}=\Gamma_{|\K_1}
	\oplus\cdots\oplus\Gamma_{|\K_m},$
	where $\K_{reg}$ is the set of regular groups
	w.r.t.\ $(d^{(n)})_{n\in \N}$ and
	$\K_1,\ldots,\K_m$ are non-empty subsets of $\K_{reg}.$ 
\end{definition}

\Wuu{For a regular sequence $(d^{(n)})_{n\in \N}$, its} regular \Wuu{groups}  are uniquely determined and will
not change for different subsequences $(d^{(n_i)})_{i\in\N}.$
\Wuuu{This means that} $\K_{reg}$ \Wuu{is} the same for different subsequences
$(d^{(n_i)})_{i\in\N}.$ Thus Definition~\ref{def:ScalableGames}
is not ambiguous, although it defines
an asymptotic decomposition of $\Gamma_{|\K_{reg}}$
on an arbitrary decomposable subsequence $(d_{|\K_{reg}}^{(n_i)})_{i\in\N}.$
Each regular sequence
$(d^{(n)})_{n\in \N}$ has of course a 
subsequence $(d^{(n_i)})_{i\in \N}$ such that 
$(d_{|\K_{reg}}^{(n_i)})_{i\in \N}$ is decomposable.
\Wuu{Lemma~\ref{lma:Convergence_iff_RegularSequence}c)}
and Corollary~\ref{thm:ADGame_Results} \Wuuu{then imply that every game} $\Gamma$ scalable by decomposition is \aw designed.
We summarize this trivial result in Corollary~\ref{thm:Scalability}.

\begin{corollary}\label{thm:Scalability}
	Games that are scalable by decomposition are \aw designed.
\end{corollary}

%

%

\subsection{An extensive class of \awd games and a conjecture}
\label{sec:Rechness}


Theorem~\ref{thm:Results_RVcost} below
demonstrates that games 
scalable by decomposition form 
an extensive class of asymptotically well designed games. 
We move the proof to Appendix~\ref{proof:Results_RVcost}.
\begin{theorem}\label{thm:Results_RVcost}
	Games with
	regularly varying cost functions are scalable by decomposition and thus asymptotically well designed.
\end{theorem}

A game $\Gamma$ has the {\em trivial asymptotic decomposition} into itself if the decomposition has just one class, i.e., $k \sim k'$ for all $k,k' \in \K$
for all decomposable sequence $(d^{(n)})_{n\in \N}$. Obviously, such a game \Wuuu{has exactly} one group and is
by definition scalable. On the other hand, every \Wuu{singleton subgame} $\Gamma_{|\{k\}}$, $k\in \K,$ of a game $\Gamma$ 
that is scalable  by decomposition is scalable. 
This holds because each regular 
sequence $\big(d^{(n)}=(0,\ldots,0,d_k^{(n)},0,\ldots,0)\big)_{n\in \N}$
has $\K_{reg}=\{k\},$ and so
there is a scalable subsequence  
$\big(d^{(n_i)}_{|\K_{reg}}=(d_k^{(n_i)})\big)_{i\in \N}$ for \Wuu{subgame}
$\Gamma_{|\{k\}}$ when $\Gamma$ is scalable by
decomposition.
We summarize this in Corollary~\ref{thm:CharacterizationSG} below.
\begin{corollary}\label{thm:CharacterizationSG}
	A game $\Gamma$ scalable by decomposition
	has scalable \Wuuu{subgames} $\Gamma_{|\{k\}}$ for each $k\in \K$.
\end{corollary}

This observation leads to the natural question whether the converse also holds, i.e., whether a game $\Gamma$ is scalable
by decomposition if all its singleton \Wuuu{subgames} $\Gamma_{|\{k\}}$ are scalable. 
Obviously, this is true for games with regularly varying
	cost functions, and holds also when the cost functions $\tau_a(\cdot)$
	are mutually comparable, i.e.,
	$\lim_{x\to\infty}\frac{\tau_a(x)}{\tau_b(x)}\in (0,\infty)$
	for each pair $(a,b)\in A\times A$.
We thus believe that this should be true in general,  but have so far not been able to prove it. We pose it as a conjecture.

\begin{conjecture}\label{conj:CharacterizationSG}
	 A game $\Gamma$ is scalable by decomposition if and only if each singleton \Wuu{subgame}
	$\Gamma_{|\{k\}}$, $k\in \K,$ is scalable.
\end{conjecture}

If this conjecture \Wuuu{were} true, then it would provide an easy way to check that a game is scalable by decomposition.

\subsection{Relationship between classes of games and more conjectures}
\label{sec:ClassRelations}

\begin{figure}[!htb]
	\centering
	\includegraphics[width=0.325\linewidth,height=0.18\textheight]{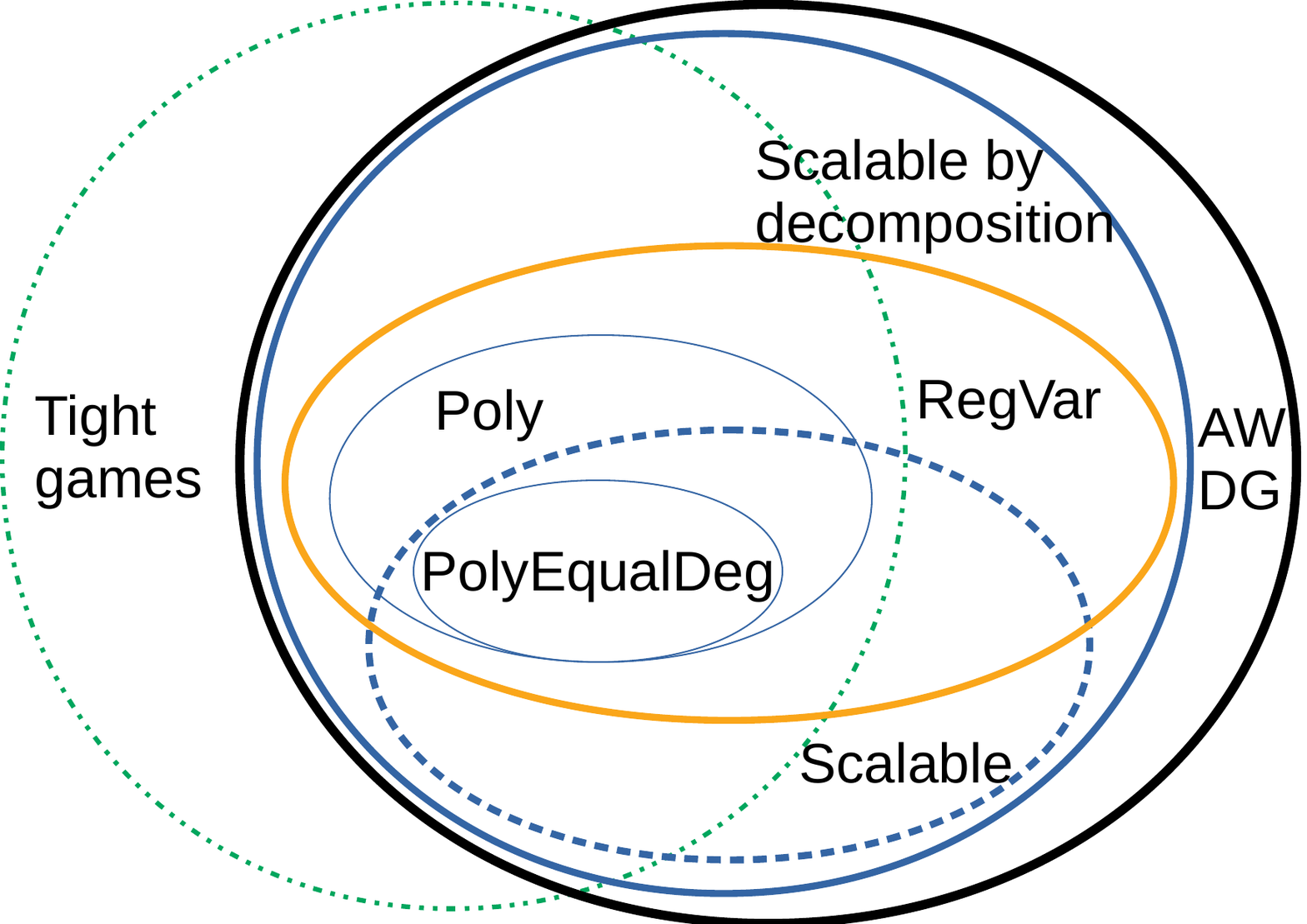}
	\caption{Relationship between different classes of games: ``AWDG"
		denotes the class of \awd games; ``RegVar" denotes
the class of games with arbitrary regularly varying cost functions; ``Poly" denotes 
the class of games with arbitrary polynomial cost functions;
``PolyEqualDeg" denotes the class of games with polynomial cost functions
of the same degree.}
\label{fig:GamesRelation}
\end{figure}

Figure~\ref{fig:GamesRelation} summarizes the
relationship between different classes of games  that
we have investigated so far. Games scalable by decomposition form a subclass of 
\awd games. The relationship between tight games and \awd games
is rather clear. None is included in the other, but they have
an intersection containing, e.g., \Wuuuu{all} games with arbitrary polynomial
cost functions.

%
%
Games scalable by decomposition
need not be scalable, see Example~\ref{example:NoScalableSubsequences}.
Lemma~\ref{lma:Scalable<=ScalableByDecomposition} below
shows that 
scalable games are also scalable by decomposition, and thus form a {\em proper} subclass of
games scalable by decomposition. We move the proof
of Lemma~\ref{lma:Scalable<=ScalableByDecomposition} to
Appendix~\ref{proof:Scalable<=ScalableByDecomposition}.
\begin{lemma}
	\label{lma:Scalable<=ScalableByDecomposition}
	Every scalable game is scalable by decomposition.
\end{lemma}

Tight games need not be scalable by decomposition, since they need not be
\aw designed, see Example~\ref{exam:TightNeedNotAWDG}. 
Neither games scalable by decomposition 
nor scalable games need be tight,  see Example~\ref{example:SSDDivergentSequence} and Example~\ref{example:NonRegularlyVaryingBenchmark}.
However, these three classes overlap and 
include all games with polynomial cost functions of the same degree.

The cost functions of a tight game need not be regularly
varying, see Example~\ref{exam:TightNeedNotAWDG}.
However, we do not know whether a game with regularly
varying cost functions is tight. This is related to the existence of two regularly varying functions
$h_1(x)$ and $h_2(x)$ s.t. 
$0\le \varliminf_{x\to\infty}\frac{h_1(x)}{h_2(x)}
<\varlimsup_{n\to \infty}\frac{h_1(x)}{h_2(X)}\le \infty.$
We believe that there are such regularly varying functions,
but are not able to prove this at present. 
We leave it as Conjecture~\ref{conj:RV_Comparability} below.
Under Conjecture~\ref{conj:RV_Comparability}, there is
a game that is not tight, but has regularly varying  cost functions, see \Wuu{the discussion in}
Section~\ref{sec:conclusion}.
\begin{conjecture}\label{conj:RV_Comparability}
	There are two non-decreasing, non-negative and
	continuous regularly varying functions 
	$h_1(x)$ and $h_2(x)$ s.t. 
	$0\le \varliminf_{x\to\infty}\frac{h_1(x)}{h_2(x)}
	<\varlimsup_{n\to \infty}\frac{h_1(x)}{h_2(X)}\le \infty.$
\end{conjecture}

Theorem~\ref{thm:Results_RVcost} shows that games with regularly varying cost functions
are scalable by decomposition.
Example~\ref{example:SSDDivergentSequence}
shows that cost functions of a scalable game
need not be regularly varying, and neither
need the cost functions of a game scalable by decomposition. 
So, the class of games with regularly varying cost functions
is a {\em proper} subclass of games scalable by decomposition,
and overlaps with the class of scalable games.

Finally, we do not know whether each asymptotically well designed
game is scalable by decomposition. 
We guess that this is not true, and leave it as Conjecture~\ref{conj:AWDG_Decomposition} below.
\begin{conjecture}\label{conj:AWDG_Decomposition}
	There are asymptotically well designed games that are not
	scalable by decomposition.
\end{conjecture}

\section{Results for routing games with BPR \Wuu{cost} functions}
\label{sec:RoutingPoA}

\subsection{Approximation and convergence results for routing games with BPR \Wuu{cost} functions}

BPR \Wuu{cost} functions are popular in static traffic models, see \cite{BPR}. They are polynomials of the form 
$
h(x)=h(0)\cdot \big(1+\alpha\cdot \big(\frac{x}{u}\big)^{\beta}\big),\ h(0)>0,
\alpha> 0, u>0, \beta> 0,
$
and are used to model the \Wuu{flow dependent} \Wuu{cost} (latency) of a street. The constant 
$h(0)$ is the \Wuu{free flow} \Wuu{cost}, 
$u$ is the ``practical'' capacity of that street, and
$\alpha,\beta$ are constants reflecting the latency. 
Typical values in practice are $\alpha=0.15$ and $\beta=4.$

\Wuu{We study} them in their general form 
$\tau_{a}(x)=\gamma_a\cdot x^{\beta}+\eta_a$
for some constants $\beta> 0,$ $\gamma_a>0$ and $\eta_a\ge 0.$ 
%
\Wuu{By Theorem~\ref{thm:Results_RVcost}, routing games $\Gamma$ with BPR cost functions are  asymptotically well designed.} But their cost
functions  are polynomials of the same degree $\beta$, and that enables stronger results.
Our first stronger result shows that 
every SO profile is \Wuuu{an}
{\em $\epsilon$-approximate NE profile}, see Theorem~\ref{eq:Limit_PoA_epsilon}.

\begin{definition}[\Wuu{see, e.g., } \cite{Roughgarden2000How}]
	\label{def:eNE}
	Let $\Gamma$ be a game and $\epsilon>0$ be a constant. A strategy profile $f$ \Wuu{of $\Gamma$} is an \emph{$\epsilon$-approximate NE profile} if
	\Wuu{$
	\tau_s(f) \le (1+\epsilon)\cdot \tau_{s'}(f)
	$}
	for each $k\in \K,$ and
		each $s,s'\in \S_k$ with $f_s>0.$
\end{definition}

\begin{theorem}\label{eq:Limit_PoA_epsilon}
	Consider a game $\Gamma$ with BPR cost functions $\tau_a(x)=\gamma_a\cdot x^\beta +\eta_a$ for all $a \in A$.
	Let $d$ be a demand vector for $\Gamma$ and let $d_{\min}=\min\{d_k: k\in\K\}.$ Then
	every SO profile of $\Gamma$ is an $O(d_{\min}^{-\beta})$-approximate NE profile of $\Gamma$.
\end{theorem}

The proof has been moved to Appendix~\ref{proof:Limit_PoA_epsilon}. It actually shows the stronger property that 
$
\tau_s(f^*) \le \big(1+O(d_k^{-\beta})\big)\cdot \tau_{s'}(f^*)
$
for every SO profile $f^*=(f_s^{*})_{s\in \S}$ and every $k \in \K$ with 
	$s, s' \in \S_k$ and $f_s^{*} > 0$.
Users of O/D pairs with large  demands $d_k$ will \Wuu{thus} approximately follow paths
of an SO profile, and their choices
will be independent of the choices of other users.
In particular, when all OD pairs have large travel demands,  
an SO profile is an $O(T(d)^{-\beta})$-approximate NE profile. 

Our second result obtains the convergence rate PoA$(d)=1+o(T(d)^{-\beta})$ for the PoA 
 and gives a detailed answer to
the conjecture by \cite{O2016Mechanisms} in 
Theorem~\ref{theo:PoA_Limit_BPR} below. Their conjecture may hold when
the game has only one OD pair with parallel links, but does not in general. 
In fact, the PoA of a game may have largely
different convergence rates for \Wuuu{different} unbounded sequences.

\begin{theorem}\label{theo:PoA_Limit_BPR}
	Let $\Gamma$ be a game with BPR cost functions $\tau_a(x)=\gamma_a\cdot x^\beta+\eta_a$ with 
	$\beta>0,\ \gamma_a>0$ and $\eta_a\ge 0$ for all $a \in A,$ and let $d=(d_k)_{k\in \K}$ be an arbitrary demand vector. Then:
	\begin{itemize}
		\item[a)] $\text{PoA}(d)=1+o(T(d)^{-\beta}).$
		
		\item[b)] For each $\beta\in (0,1),$
		there is an instance such that,  for
		each $\theta\in (2\cdot \beta,\beta+1],$ there is an unbounded sequence 
		$(d^{(n)})_{n\in \N}$ for which  
		PoA$(d^{(n)})=1+\Theta(T(d^{(n)})^{-\theta}).$
		
		\item[c)] For each $\beta\ge 1,$
		there is an instance such that,  for
		each $\theta\in [\beta+1,2\cdot \beta),$ there is an unbounded sequence 
		$(d^{(n)})_{n\in \N}$ for which  
		PoA$(d^{(n)})=1+\Theta(T(d^{(n)})^{-\theta}).$
		
		\item[d)] For each $\beta>0,$
		there is an instance with a single OD pair and 
		parallel links, such that  
		PoA$(d)=1+\Theta(T(d)^{-2\cdot \beta}).$
	\end{itemize}
\end{theorem}


The proof has been moved to Appendix~\ref{proof:PoA_Limit_BPR}.

The next theorem gives additional insight into the convergence of the PoA
and provides a method to estimate the cost of both, NE profiles $\tilde{f}$ and 
SO profiles $f^*$, when the total demand $T(d)$ is large.
Their cost is almost the same and depends essentially only on the distribution $\frac{d}{T(d)}$.  Moreover, $\big(\frac{\tilde{f}_a}{T(d)}\big)_{a\in A}$ and $\big(\frac{f^*_a}{T(d)}\big)_{a\in A}$
are almost identical.

\begin{theorem}
\label{theo:ImprovedPoABound}
Let $\Gamma$ be a game with BPR cost functions and let  $(d^{(n)})_{n\in \N}$ be a regular demand sequence
with limit distribution $d^{(\infty)}=(d_k^{(\infty)})_{k\in \K}.$ For each $n\in \N,$ let $\fne^{(n)}$ and $f^{*(n)}$ be an NE profile and an SO profile of $\Gamma$ for the demand vector $d^{(n)}$, respectively.
Then:
\begin{itemize}
\item[a)] 	$
	\lim_{n\to \infty}\frac{C(f^{*(n)})}{\big(T(\D^{(n)})\big)^{\beta+1}}
	=C_{\Gamma^{(\infty)}}(\tilde{f}^{(\infty)})=\lim_{n\to \infty}\frac{C(\fne^{(n)})}{\big(T(\D^{(n)})\big)^{\beta+1}}>0,
	$
	where $C_{\Gamma^{(\infty)}}(\tilde{f}^{(\infty)})$ denotes the NE cost of the limit game 
	$\Gamma^{(\infty)}$ under scaling factors $g_n=T(d^{(n)})^{\beta}.$ 
\item[b)] For $n$ large enough, the consumption distributions
	$\big(\frac{f_a^{*(n)}}{T(d^{(n)})}\big)_{a\in A}$ and $\big(\frac{\tilde{f}_a^{(n)}}{T(d^{(n)})}\big)_{a\in A}$
    are almost identical, i.e., for each $\epsilon>0,$ there is
    $N\in \N$ such that
    $
    \max_{s\in \S} \big|\frac{f_a^{*(n)}}{T(d^{(n)})}-\frac{\tilde{f}_a^{(n)}}{T(d^{(n)})}\big|<\epsilon
    \ \text{ for all }n\ge N.
    $
\end{itemize}
\end{theorem}

The proof of Theorem \ref{theo:ImprovedPoABound} follows directly
from the scaling properties of Lemma~\ref{lma:PriceConvergenceOfStrategies}. 

Theorem~\ref{theo:ImprovedPoABound}a) is particularly interesting. It states that
$C_{\Gamma^{(\infty)}}(\tilde{f}^{(\infty)})\cdot T(d)^{\beta+1}$
approximates the cost of both, NE profiles
and SO profiles, for an arbitrary demand vector $d=(d_k)_{k\in \K}$ with
	sufficiently large total demand $T(d).$
	Note that the NE cost $C_{\Gamma^{(\infty)}}(\tilde{f}^{(\infty)})$
	of limit game $\Gamma^{(\infty)}$ depends mainly on the
	demand distribution $\frac{d}{T(d)},$ when road conditions
	are given.
So, it further implies that, asymptotically, the demand distribution
$\frac{d}{T(d)}$ is the crucial factor to optimize a traffic system with a large total demand $T(d)$. As $\frac{d}{T(d)}$ depends essentially on the location of resources such as working
and living places, hospitals, shopping malls, schools,
government offices, and others, it means that good urban planning is the key to cope with heavy traffic.

\subsection{An experimental study}\label{sec:empirical}

\begin{figure}[!htb]
\centering
\includegraphics[scale=0.15, angle=180]{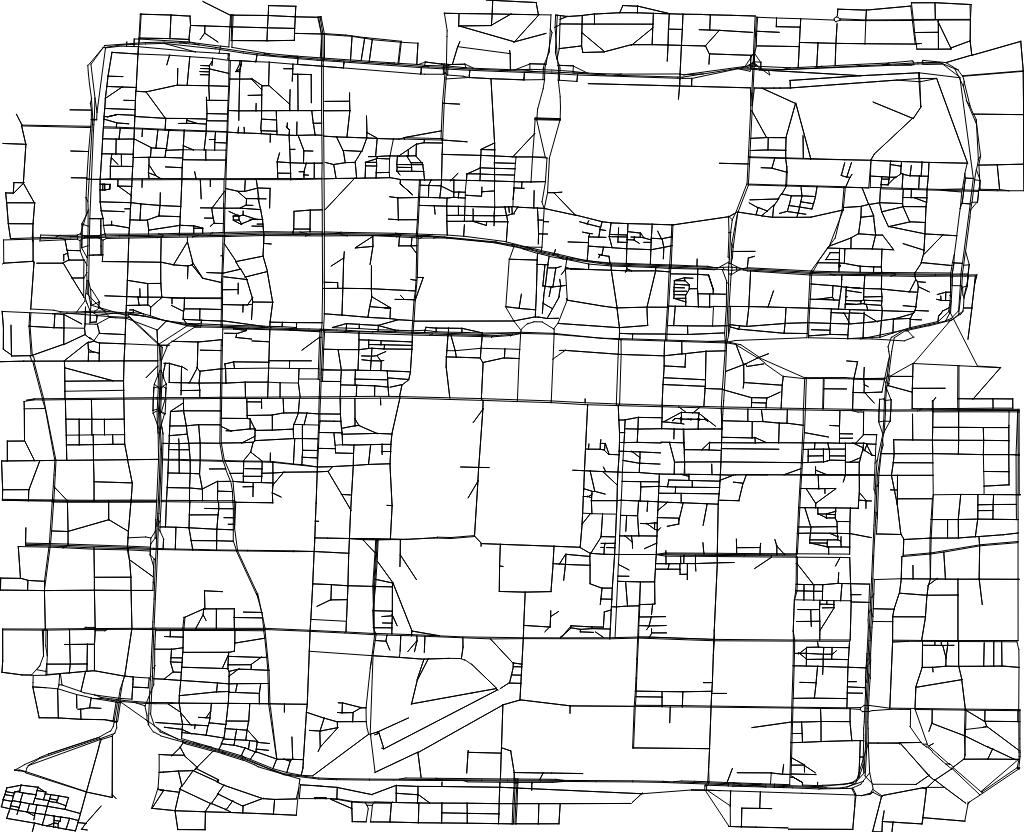}
\vspace{0.5cm}
\caption{The street network within the 2nd ring road of Beijing}
\label{fig:BeijingNet}
\end{figure}

This subsection empirically verifies our theoretical findings. We analyzed real traffic data during
rush hour (7:00 a.m.--9:00 a.m.) within the second ring road of Beijing. The O/D pairs and travel demand were gathered
from GPS data. After a suitable
calibration, we obtained
$|\K|=K=33,\!426$ different O/D pairs with total demand $T(d)=\sum_{k\in \K}d_k=101,\!074.$
Figure \ref{fig:BeijingNet} displays the \Wuuu{street} network
$G=(V,A)$ within that area of
Beijing, which was taken from OpenStreetMap. It has $|V|=4,\!716$ nodes and $|A|=10,\!267$ arcs.

Calculations were done with the program ``CMCF'' developed by 
the COGA group at Berlin University of Technology.
It has been applied before in \cite{Jahn2004System}
and \cite{Harks2015Computing} to compute SO profiles,
NE profiles, and tolls for congestion pricing. For computing SO and NE profiles,
it uses a variant of the Frank-Wolfe algorithm  \cite{Fukushima1984A}
together with Dijkstra's algorithm  \cite{Dijksra1959} for shortest path computations in each iteration.
Our implementation was done under Mac OS Sierra on a Laptop with a 2.7
GHz Intel Core i7 CPU. We stopped each run of the program once the current solution 
had an objective value within $1\%$ of the optimum. 

The experiment has actually been carried out in two separate phases. The first phase
had already been done before conceiving
this paper without knowing the results of \cite{Colini2016On,Colini2017WINE,Colini2017arxiv}, and only computed the empirical PoA.  
Table \ref{tab:PoA_Beijing} below reports the result
from the first phase. It shows that
the PoA within that area of Beijing is very close to $1.$
\begin{table}[!htb]
\centering
\begin{tabular}{c|c|c|c|c}
\hline
\hline
PoA & SO cost & NE cost & CPU\_SO (s) & CPU\_NE (s)\\
\hline
1.0   & 1.23093000E+15 & 1.23083000E+15 & 29287.245 & 29307.265\\
\hline\hline
\end{tabular}
\vspace{0.5cm}
\caption{\footnotesize{The PoA within the 2nd ring road of Beijing.
Column ``PoA'' reports the price of anarchy, column ``SO cost'' reports
the cost of SO profiles, column ``NE cost'' reports the
cost of NE profiles, and the last two columns report the CPU time
for computing the cost of the SO and NE, respectively.}}
\label{tab:PoA_Beijing}
\end{table}

The second phase was done after obtaining the theoretical results with the aim to empirically verify the convergence of the PoA.
To this end, we took
$65$ different subsets of the \Wuuu{entire} $33,\!426$ O/D pairs, and ran the algorithm
\Wuu{for every one of them.}  To save space, we only report
	the results for some of the $65$ subsets in
Table \ref{tab:Conver_PoA}. 
Column ``Perc.'' lists the percentage of the $33,\!426$ O/D pairs contained
in a subset, column ``$K$'' lists the corresponding number of O/D
pairs of that subset, and column ``$T$'' lists the corresponding
total demands. For instance, 
for the first row in Table \ref{tab:Conver_PoA}, we took $0.01\%$ of the $33,\!426$
O/D pairs, which results in $K=\lceil 33,\!426\times 0.01\%\rceil=4$ O/D pairs with total travel demand $T=15$.

\begin{longtable}[!htb]{c||c|c|c|c|c}
\hline
\hline
Perc. &    SO cost          &         NE cost       &           PoA     &      $K$ & $T$\\
\hline
\hline
\endhead
\hline\hline
\multicolumn{6}{c}{Table \ref{tab:Conver_PoA}  Convergence of the PoA (To be continued on the next page) }
  \endfoot
 \endlastfoot
 0.01\%	&5.92E+03	&5.92E+03	&1.00 	&4	&15\\
0.05\%	&1.45E+04	&1.61E+04	&1.11 	&17	&51\\
0.10\%	&2.91E+04	&3.30E+04	&1.13 	&34	&90\\
0.15\%	&3.76E+04	&4.16E+04	&1.11 	&51	&116\\
0.20\%	&4.65E+04	&5.14E+04	&1.10 	&67	&146\\
0.30\%	&7.56E+04	&8.32E+04	&1.10 	&101 &216\\
0.35\%	&1.39E+05	&1.51E+05	&1.08 	&117	 &264\\
0.45\%	&1.73E+05	&1.89E+05	&1.09 	&151	&392\\
0.50\%	&2.62E+05	&2.90E+05	&1.11 	&168	&483\\
0.60\%	&3.12E+05	&3.48E+05	&1.12 	&201	&550\\
0.65\%	&3.37E+05	&3.75E+05	&1.11 	&218	&626\\
0.95\%	&3.75E+06	&3.85E+06	&1.03 	&318	&1111\\
1.00\%	&3.84E+06	&3.94E+06	&1.03 	&335	&1149\\
1.50\%	&5.12E+06	&5.22E+06	&1.02 	&502	&1531\\
2.00\%	&7.73E+06	&7.82E+06	&1.01 	&669	&1938\\
2.50\%	&1.43E+07	&1.44E+07	&1.01 	&836	&2276\\
3.00\%	&3.81E+07	&3.81E+07	&1.00 	&1003	&2726\\
3.50\%	&6.65E+07	&6.65E+07	&1.00 	&1170	&3280\\
20.00\%	&4.00E+11	&4.00E+11	&1.00 	&6686	&20098\\
90.00\%	&7.18E+14	&7.18E+14	&1.00 	&30084	&90302\\
100.00\%	&1.23E+15	&1.23E+15	&1.00 	&33426	&101074\\
\hline
\hline
\caption{Convergence of the PoA}
\label{tab:Conver_PoA}
\end{longtable}

\begin{figure}[!htb]
\centering
\begin{subfigure}{0.49\textwidth}
\centering
\includegraphics[scale=0.5]{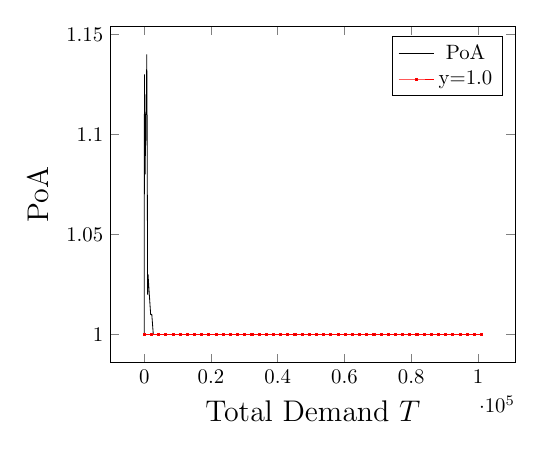}
\caption{}
\end{subfigure}
\begin{subfigure}{0.49\textwidth}
\centering
\includegraphics[scale=0.5]{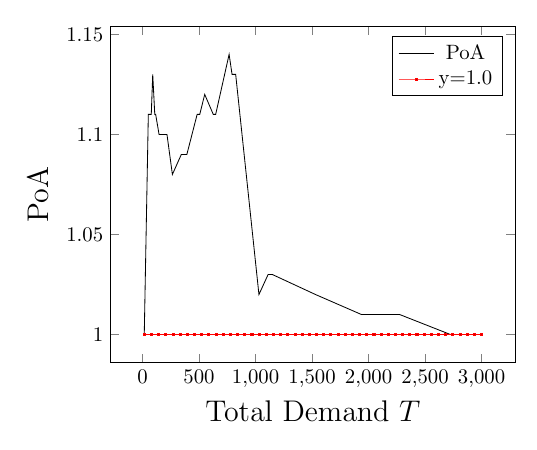}
\caption{}
\end{subfigure}
\caption{Convergence of PoA}
\label{fig:Conver_PoA}
\end{figure}

Table \ref{tab:Conver_PoA} shows that the PoA has
	already converged to 1 when $K \ge 1,\!003$ (which accounts 
for only $3\%$ of the $33,\!426$ O/D pairs). 
Figure~\ref{fig:Conver_PoA} plots the PoA as a function of the total demand $T$ for the data
from Table~\ref{tab:Conver_PoA}. 
Part (a) displays the PoA 
with $T$ up to
$101,\!074$ and shows that it quickly converges to 
$1$ as $T$ increases.

We observe a very sudden and steep decline to $1$.
This empirically verifies Theorem~\ref{theo:PoA_Limit_BPR}.
Part (b) of Figure \ref{fig:Conver_PoA} 
takes a closer look at that decline of the PoA by keeping $T$ below $3,\!000$.
The PoA increases quickly with growing but still small total demand $T\le 100$.
 However,
when $T$ gets moderately large, i.e., $100 \le T\le 1,\!200,$ the PoA becomes
choppy with several oscillations.
After $T\ge 1,\!200$, the PoA decreases very fast to $1.0.$

\begin{figure}[!htb]
\centering
\begin{subfigure}{0.32\textwidth}
\centering
\includegraphics[scale=0.5]{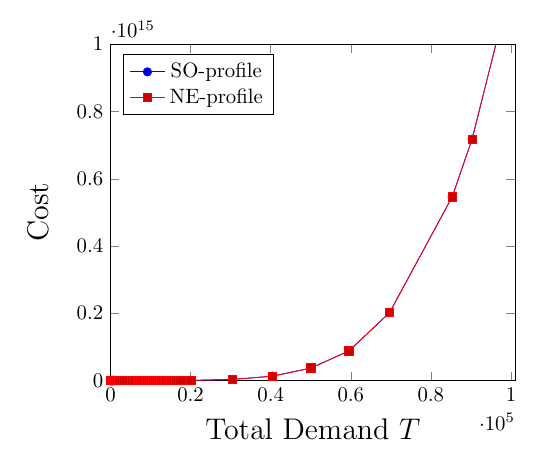}
\caption{}
\end{subfigure}
\begin{subfigure}{0.32\textwidth}
\centering
\includegraphics[scale=0.5]{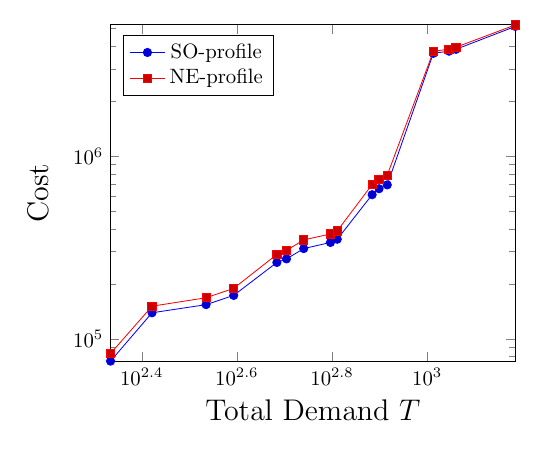}
\caption{}
\end{subfigure}
\begin{subfigure}{0.32\textwidth}
	\centering
	\includegraphics[scale=0.5]{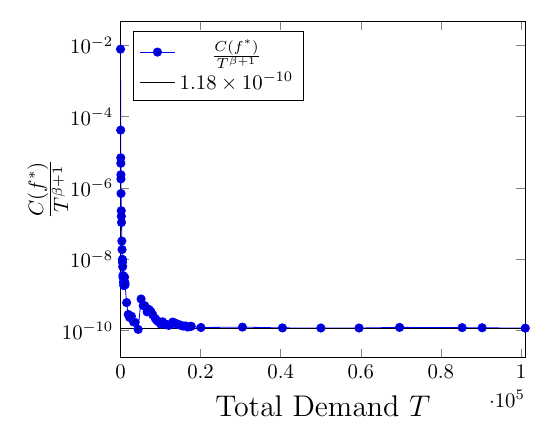}
	\caption{}
\end{subfigure}
\caption{Curves of SO cost, NE cost and $\frac{C(f^*)}{T^{\beta+1}}$}
\label{fig:Cost_SONE}
\end{figure}

Figure~\ref{fig:Cost_SONE}(a) displays the cost
	curves of SO profiles and NE profiles for $T\le 101,\!074$. The cost differences are so small 
on that scale that the two curves seem to coincide. Figure~\ref{fig:Cost_SONE}(b) changes the scale to 
$T\in (200, 2,\!000)$ and shows that the two curves  gradually become
identical as $T$ increases. This empirically
verifies Theorem~\ref{theo:ImprovedPoABound}~$(b).$

Finally, Figure~\ref{fig:Cost_SONE}~(c) shows 
the ratio of the SO cost over $T^{\beta +1}$ for the data from
the $65$ subsets for $\beta=4$. 
This ratio converges quickly to a constant as
$T$ increases and empirically
verifies Theorem~\ref{theo:ImprovedPoABound}(a).
When $T$ has reached
$2\cdot 10^4,$ the ratio has already converged
to the constant $1.18\cdot 10^{-10}$ and is, by Theorem~\ref{theo:ImprovedPoABound}(a),
an estimator of the NE cost $C_{\Gamma^{(\infty)}}(\tilde{f}^{(\infty)})$ of the limit game $\Gamma^{(\infty)}$. 

It took about 29,307 seconds CPU time to
compute the SO cost for $T=101,\!074$, and about $5,\!967$
seconds CPU time to compute
the SO cost for $T=20,\!098$. Thus, when we use the approximation method
of Theorem \ref{theo:ImprovedPoABound}(a),
we save about $\frac{29,307-5,967}{29,307}\approx 
79.6\%$ of the time to
compute the SO cost within the second ring road of Beijing.

All these empirical results show that the PoA$(d)$ converges very fast to $1$ as $T(d)$ tends to
infinity, and that there is a moderately large threshold value (a \emph{saturation point} $T_{sat}$) for 
$T(d)$ at which the PoA has already decreased to 1 and stays at 1. The NE and SO cost can then be  computed efficiently for $T \ge T_{sat}$. In our computations, this saturation point seems to be at about $T_{sat}=2,\!726,$
which is far below the current total travel demand of $101,\!074$. 

Moreover, the PoA$(d)$ deviates from $1$ only in a small interval of the total demand $T(d)$. 
So selfishness is actually good for most values of $T(d)$.

\section{Summary}\label{sec:conclusion}

Our limit analysis of the PoA has identified several classes of non-atomic congestion games for which the PoA converges to 1 regardless of the growth of the demand sequence.
To that end, we have developed a new framework that is based on limit games and the asymptotic decomposition of games. This framework shows that the convergence of the PoA to 1 does not depend on a ``regular'' growth of the total \Wuuu{demand} $T(d)$, but on the existence of limit games in a decomposition.
When they exist, the PoA is 1 in the limit and user selfish behavior in these games already leads to the social optimum in the limit for every sequence of growing demands. In particular, with our framework, we are able to prove
\Wuuu{this} convergence for a large class of \Wuuu{games, e.g.,  those} with arbitrary regularly varying cost functions (Theorem~\ref{thm:Results_RVcost}).

Our results can be generalized
	by using constants $r(a,s)\ge 0$ instead of
	the membership relation ``$a\in s$".
	Then
	each strategy $s\in \S$ has
	a general {\em consumption pattern} $\big(r(a,s)\big)_{a\in A}$
	s.t. $r(a,s)\ge 0$ represents
	the amount of resource $a\in A$ demanded by
	strategy $s.$
	The joint consumption of $a\in A$ is then $f_a=\sum_{a\in A} r(a,s)\cdot f_s,$ and 
	the cost of strategy $s\in \S$ is $\tau_s(f)=\sum_{a\in A}r(a,s)\cdot \tau_a(f_a).$ Non-atomic congestion
	games form
	a particular case in which $r(a,s)=\1_{s}(a)$ for
	each $(a,s)\in A\times \S,$ where 
	$\1_s(a)$ is the indicator function of the relation ``$a\in s$". Moreover, letting $r(a,s)=\{0,w_k\}$ with
$w_k>0$ for each $k\in \K$ and all $(a,s)\in A\times \S_k$  yields the class of weighted non-atomic congestion games.

Some of our results have been strengthened for routing games with BPR \Wuu{cost} functions. 
Socially optimal strategy profiles in such games are $\epsilon$-approximate \Wuu{equilibria} for a small $\epsilon>0$ tending to $0$ as the total travel demand $T(d)$ increases. Also, the PoA follows the power law $1+o(T(d)^{-\beta})$, where  $\beta$ is the degree of the \Wuu{BPR} functions. But it has
largely different convergence rates for different unbounded sequences and so the above-mentioned conjecture by \cite{O2016Mechanisms} does not hold in general.

Finally, to empirically verify our theoretical findings, we have analyzed real traffic data within the 2nd ring road of Beijing in an experimental study. Our empirical results definitely validate our findings. They show that the current traffic in that area of Beijing is already far beyond the point at which the PoA is 1, and so no route guidance policy can reduce the total \Wuu{cost} without significantly reducing the current huge total travel demand.

There are still some open questions. One is Conjecture~\ref{conj:CharacterizationSG}, which might be hard to prove but would 
provide an easy method to check if a game is scalable by
decomposition.
The second is Conjecture~\ref{conj:RV_Comparability}, which postulates
the existence of two regularly varying functions $h_1(\cdot)$
and $h_2(\cdot)$ s.t. $\varliminf_{x \to \infty}\frac{h_1(x)}{h_2(x)}\ne \varlimsup_{x \to \infty}\frac{h_1(x)}{h_2(x)}.$
Conjecture~\ref{conj:RV_Comparability} is crucial to analyze the relationship between tight games and games
with regularly varying cost functions. If Conjecture~\ref{conj:RV_Comparability} holds,  
we \Wuuu{can easily} construct a routing game that is not
tight, but has regularly varying  cost functions,  see Figure~\ref{fig:CTGNeqTight}. 

Our last open problem is Conjecture~\ref{conj:AWDG_Decomposition},
which postulates that not all \awd games are
scalable by decomposition. This could \Wuuu{also be}  difficult to prove.
\begin{figure}[!htb]
	\centering
	\begin{tikzpicture}[
	>=latex
	]
	\node[scale=0.4,circle,fill=black,label=left:$o$](o){};
	\node[above right =of o](h1){$h_1(x)$};
	\node[right =of o](h2){$h_2(x)$};
	\node[scale=0.4,circle,fill=black,label=right:$t$,right =of h2](t){};
	\draw[-,thick] (o) to [out=90,in=180] (h1);
	\draw[->,thick] (h1) to [out=0,in=90] (t);
	\draw[-,thick] (o) to (h2);
	\draw[->,thick] (h2) to (t);
	\end{tikzpicture}
	\caption{The relation between tight games and
		games with regularly varying cost functions}
	\label{fig:CTGNeqTight}
\end{figure}

\Wuuuu{Other interesting topics are the
	application of our approach to
	more general settings, e.g., to {\em atomic congestion games} (\cite{Nisan2007}), or to 
	{\em non-atomic congestion games with non-separable cost for strategies} (\cite{Perakis2007}).}
\Wuuuu{For instance, our approach applies directly to the special non-separable function $\tau_s(f)=\max_{a\in s}\tau_a(f_a)$ for each strategy $s\in \S.$}

\Wuuu{A further} topic for future research is the convergence
rate of the PoA for games with certain classes of cost functions, e.g.,
arbitrary polynomials. We have shown such a result for BPR functions, see Theorem~\ref{theo:PoA_Limit_BPR}, and 
\cite{Colini2017WINE} have
 shown an inspiring result
for games with polynomial cost functions under gaugeable sequences. Results for arbitrary demand sequences and arbitrary polynomial cost functions are still missing.

\section*{Acknowledgement}
The first author acknowledges support from the National Science Foundation of China
	with grant No.~61906062, support from the Science Foundation of Anhui Science and Technology Department with grant No.~1908085QF262, and support from
	the Talent Foundation of Hefei University with grant No.~1819RC29;
    The first and second authors acknowledge  support from the Science Foundation of the Anhui Education Department with grant No.~KJ2019A0834.

\Wuuuu{Moreover,} the authors would like to thank Christoph Hansknecht from TU Braunschweig for his support in installing the package CMCF and
Xin Sun for computing the results of Table~\ref{tab:Conver_PoA} for small subsets of the O/D pairs. \Wuu{We also thank} Marc Uetz for bringing the work of \cite{Colini2016On} to our attention and Max Klimm for pointing us to the conjecture by \cite{O2016Mechanisms}. Finally, we would like to thank the referees for their many hints that have led to an improved presentation of our results.

\bibliographystyle{plain}
\bibliography{TO.bib}


\appendix

\section{Appendix: mathematical Proofs}

We will use the standard asymptotic notation. Let $h(x)$ be an arbitrary
non-negative real function. Then $O(h)$ denotes the set of non-negative real functions
$u(x)$ such that $\varlimsup_{x\to \infty}\frac{u(x)}{h(x)}<\infty$, while  $o(h)$ denotes
the set of non-negative real functions $u(x)$ such that $\varlimsup_{x\to \infty}\frac{u(x)}{h(x)}=0$. Similarly, $\Omega(h)$ is the set of non-negative real functions $u(x)$ such that $h\in O(u)$, and 
$\omega(h)$ is the set of non-negative real functions $u(x)$ with $h\in o(u).$
Moreover, we write $h\in \Theta(u)$ if $u\in \Omega(h)\bigcap O(h).$

\subsection{Proof of Lemma~\ref{lma:PriceConvergenceOfStrategies}}\label{proof:PriceConvergenceOfStrategies}
	Consider a game $\Gamma$ and
		a scalable sequence $(d^{(n)})_{n\in \N}$
		with limit distribution $d^{(\infty)}.$
		Let $(g_n)_{n\in \N}$ be a scaling sequence 
		s.t. $\lim_{d^{(n)}\to \infty} \Gamma^{[g_n]}
		=\Gamma^{(\infty)}$ for a limit game
		$\Gamma^{(\infty)}$ as defined in 
		Definition~\ref{def:LimitGameAndStrongScalable}. 
		Let $f^{(n)},\tilde{f}^{(n)}$ and $f^{*(n)}$
		be an arbitrary strategy profile, an NE profile
		and an SO profile for $d^{(n)}$, respectively.
	
	\textbf{Proof of $a):$} 
	Assume w.l.o.g.\ that 
	$\lim_{n\to \infty}\frac{f^{(n)}}{T(d^{(n)})}=f^{(\infty)}.$
	Trivially, $f^{(\infty)}$ is a strategy
	profile of $\Gamma^{(\infty)}$ for the limit distribution $d^{(\infty)}$ of $(d^{(n)})_{n\in \N}.$ Moreover, 
	each $f_a^{(\infty)}=\sum_{s\in S : s \ni a} 
	f_s^{(\infty)}=\lim_{n\to \infty} \sum_{s\in S : s \ni a}\frac{f_s^{(n)}}{T(d^{(n)})}=\lim_{n \to \infty}
	\frac{f_a^{(n)}}{T(d^{(n)})}\in I_a,$ since each domain $I_a$ is non-empty and closed, and
	$\frac{f_a^{(n)}}{T(d^{(n)})}\in I_a$ is 
	the consumption rate of resource $a \in A$
	for each $n\in \N$ and each $a\in A.$
	
	Consider now an arbitrary tight strategy $s\in \S$
	with $f_s^{(\infty)}>0.$
	By (L2), the limit cost function $\tau^{(\infty)}_a(\cdot)$ is finite and
	continuous on $I_a$ for each $a \in s.$
	We will prove that 
	$\lim_{n\to \infty} \frac{\tau_a(f_a^{(n)})}{g_n}=\tau^{(\infty)}_a(f_a^{(\infty)})$
	for each $a\in s$, which in turn implies that $\lim_{n\to\infty}
	\frac{\tau_s(f^{(n)})}{g_n}=\sum_{a\in s}
	\tau^{(\infty)}_a(f_a^{(\infty)})=\tau^{(\infty)}_s(f^{(\infty)})<\infty.$  Trivially, $f_a^{(\infty)}>0$
	for each $a\in s$.
	
	Consider now an arbitrary resource $a\in s$. We distinguish two cases w.r.t.\ its domain $I_a$.
	
	\textbf{(Case I: $I_a$ is \Wuu{a} singleton)} Then  $\frac{f_a^{(n)}}{T(d^{(n)})}\equiv f_a^{(\infty)} > 0$ for each
	$n\in \N.$ 
	By the definition of
	$\tau^{(\infty)}_a(\cdot)$, we obtain trivially that $\lim_{n\to \infty} \frac{\tau_a(f_a^{(n)})}{g_n}=\lim_{n\to \infty} \frac{\tau_a\big(T(d^{(n)})\cdot f_a^{(\infty)}\big)}{g_n}=\tau^{(\infty)}_a(f_a^{(\infty)}).$
	
	\textbf{(Case II: $I_a$ is not a singleton)} Then
	$I_a$ is a non-empty closed interval. 
	%
	 Let $\epsilon>0$ be an arbitrary small constant. Since $f_a^{(\infty)}=\lim_{n\to \infty} \frac{f_a^{(n)}}{T(d^{(n)})},$ we obtain 
	that $\frac{f_a^{(n)}}{T(d^{(n)})}\in [f_a^{(\infty)}-\epsilon, f_a^{(\infty)}+\epsilon]\bigcap (I_a\setminus\{0\}),$ when $n$ is large enough, since $f_a^{(\infty)}>0.$
	We assume w.l.o.g.\ that $[f_a^{(\infty)}-\epsilon, f_a^{(\infty)}+\epsilon]
	\subseteq I_a\setminus\{0\}.$ 
	Since $\tau_a(x)$ is non-decreasing, we obtain that 
	\[
	\frac{\tau_a\big(T(d^{(n)})\cdot (f_a^{(\infty)}-\epsilon)\big)}{g_n}\le \frac{\tau_a(f_a^{(n)})}{g_n} \le \frac{\tau_a\big(T(d^{(n)})\cdot (f_a^{(\infty)}+\epsilon)\big)}{g_n}
	\]
	for $n$ large enough.
	So, letting $n\to \infty$ yields by (L1) that
	\[
	\tau^{(\infty)}_a(f_a^{(\infty)}-\epsilon)\le \varliminf_{n\to\infty}\frac{\tau_a(f_a^{(n)})}{g_n}\le  \varlimsup_{n\to\infty}\frac{\tau_a(f_a^{(n)})}{g_n}\le \tau^{(\infty)}_a(f_a^{(\infty)}+\epsilon).
	\]
	Since $\tau^{(\infty)}_a(\cdot)$ is continuous on $I_a\setminus\{0\}$, we obtain  
	 $\lim_{n\to \infty}\frac{\tau_a(f_a^{(n)})}{g_n}
	=\tau^{(\infty)}_a(f_a^{(\infty)})$ when $\epsilon\to 0$.
	
	So  
	$\lim_{n\to\infty}\frac{\tau_s(f^{(n)})}{g_n}
	=\tau^{(\infty)}_s(f^{(\infty)})$ for each tight strategy $s\in \S$ with
	$f_s^{(\infty)}>0$.
	
	We now consider an arbitrary
	tight strategy $s'\in \S$ with 
	$f_{s'}^{(\infty)}=0$ and \Wuu{prove that} $\varlimsup_{n\to\infty}\frac{\tau_{s'}(f^{(n)})}{T(d^{(n)})}
	\le \tau^{(\infty)}_{s'}(f^{(\infty)})<\infty$. To this end, we consider
	$\varlimsup_{n\to\infty}\frac{\tau_a(f_a^{(n)})}{g_n}$
	for an arbitrary resource $a\in s'$.
	If $f_a^{(\infty)}>0,$ \Wuu{we obtain}
	as before that $\varlimsup_{n\to\infty}\frac{\tau_{a}(f_a^{(n)})}{g_n}
	=\lim_{n\to\infty}\frac{\tau_{a}(f_a^{(n)})}{g_n}=\tau^{(\infty)}_{a}(f_a^{(\infty)})<\infty$.
	
	So, we assume now that $f_a^{(\infty)}=0.$ Then $I_a=[0,b]$ for some $b>0$,
	since each resource is used by some
	strategy (see~\eqref{eq:StrategiesScarce}) and 
	$I_a$ is either a singleton
	or a closed interval. Let $\eta\in (0,b)$
	be an arbitrary small constant.
	Trivially, 
	$\varlimsup_{n\to\infty}\frac{\tau_a(f_a^{(n)})}{g_n}
	\le \lim_{n\to\infty}\frac{\tau_a(f_a^{(n)}+\eta
		\cdot T(d^{(n)}))}{g_n}=
	\tau^{(\infty)}_a(\eta)<\infty.$  Letting $\eta\to 0^+,$
	we obtain that $\varlimsup_{n\to\infty}\frac{\tau_a(f_a^{(n)})}{g_n}
	\le 
	\tau^{(\infty)}_a(f^{(\infty)}_a)=\tau^{(\infty)}_a(0)=
	\lim_{\eta \to 0^+} \tau^{(\infty)}_a(\eta)<\infty.$
	Therefore,
	\[
	\varlimsup_{n\to\infty}
	\frac{\tau_{s'}(f^{(n)})}{g_n}
	=\lim_{n\to\infty}\sum_{f_a^{(\infty)}>0,a\in s'}\frac{
		\tau_a(f_a^{(n)})}{g_n}+\varlimsup_{n\to\infty}
	\sum_{f_a^{(\infty)}=0,a\in s'}\frac{ 
		\tau_a(f_a^{(n)})}{g_n}
	\le \tau^{(\infty)}_{s'}(f^{(\infty)})<\infty.
	\]
	
	Altogether, we easily obtain that
	$
	\lim_{n\to\infty}\frac{f_s^{(n)}\cdot \tau_s(f^{(n)})}{T(d^{(n)})\cdot g_n}=
	f_s^{(\infty)}\cdot \tau^{(\infty)}_s(f^{(\infty)})
	$
	for each tight strategy $s\in \S.$
	Herein, 
	$\lim_{n\to\infty}\frac{f_s^{(n)}\cdot \tau_s(f^{(n)})}{T(d^{(n)})\cdot g_n}=0=f_s^{(\infty)}\cdot \tau^{(\infty)}_s(f^{(\infty)})$
	for each tight
	strategy $s$ with $f_s^{(\infty)}=0.$

	\textbf{Proof of $b):$}  
	Let $k\in \K$ be an arbitrary group, and
	let $s'\in \S_k$ be a non-tight
	strategy (if applicable). 
	By (L2), there is a tight strategy
	$s\in \S_k.$ Let $(n_i)_{i\in \N}$
	be an infinite subsequence such that 
	$\tilde{f}_{s'}^{(\infty)}=\lim_{i\to \infty}
	\frac{\tilde{f}_{s'}^{(n_i)}}{T(d^{(n_i)})}
	=\varlimsup_{n\to\infty}\frac{\tilde{f}_{s'}^{(n)}}{T(d^{(n)})}.$
	To simplify notation, assume w.l.o.g.\ that
	$(n_i)_{i\in \N}=(n)_{n\in \N}.$
	To apply $a),$ we assume further that 
	the strategy distribution 
	$\frac{\tilde{f}^{(n)}}{T(d^{(n)})}$
	converges to a limit $\tilde{f}^{(\infty)}=(\tilde{f}^{(\infty)}_{s''})_{s''\in \S}.$
	Otherwise, one can take an infinite subsequence with that property
	and continue the discussion with this subsequence.
	
	Using $a),$ we obtain that 
	$\varlimsup_{n\to\infty} \frac{\tau_{s}(\tilde{f}^{(n)})}{g_n}<\infty,$ since
	$s$ is tight.
	We will now prove by contradiction that $\tilde{f}_{s'}^{(\infty)}=\lim_{n\to \infty} \frac{\tilde{f}_{s'}^{(n)}}{T(d^{(n)})}=0$.   
	So assume  that $\tilde{f}_{s'}^{(\infty)}>0.$ 
	Then $\tilde{f}_{a}^{(\infty)}=\sum_{s''\in S: s'' \ni a} \tilde{f}_{s''}^{(\infty)}
	\ge \tilde{f}_{s'}^{(\infty)}>0$ for
	each $a\in s'$. 
	Thus $\tilde{f}_a^{(\infty)}\in I_a\setminus\{0\}$
	for each $a\in s'$. 	
	
	The definition of the limit functions $\tau^{(\infty)}_a(\cdot)$ then yields
	$\tau^{(\infty)}_{s'}(\tilde{f}^{(\infty)})=\lim_{n\to\infty} \frac{\tau_{s'}(\tilde{f}^{(n)})}{g_n}
	=\sum_{a\in s'} \tau^{(\infty)}_a(\tilde{f}_a^{(\infty)})= \infty > \varlimsup_{n\to\infty} \frac{\tau_{s}(\tilde{f}^{(n)})}{g_n}$ 
	since $s'$ is not tight and
	there is an $a\in s'$ such that 
	\Wuu{$\tau^{(x)}_a(\cdot)\equiv \infty$} on 
	$I_a\setminus\{0\}.$
	This, in turn, implies that $\tau_{s}(\tilde{f}^{(n)})<\tau_{s'}(\tilde{f}^{(n)})$
	when $n$ is large enough. So, $\tilde{f}_{s'}^{(n)}\equiv 0$ for large enough $n,$ and thus $\tilde{f}_{s'}^{(\infty)}=0,$ a contradiction.
		
	So, 
	$\tilde{f}_{s'}^{(\infty)}=\varlimsup_{n\to\infty}
	\frac{\tilde{f}_{s'}^{(n)}}{T(d^{(n)})}=0$. 
	
	We will now show that 
	$\varlimsup_{n\to \infty} \frac{\tilde{f}_{s'}^{(n)}\cdot \tau_{s'}(\tilde{f}^{(n)})}{T(d^{(n)})
		\cdot g_n}=0.$ As above, we assume w.l.o.g.\ that 
	$\varlimsup_{n\to \infty} \frac{\tilde{f}_{s'}^{(n)}\cdot \tau_{s'}(\tilde{f}^{(n)})}{T(d^{(n)})
		\cdot g_n}=\lim_{n\to \infty} \frac{\tilde{f}_{s'}^{(n)}\cdot \tau_{s'}(\tilde{f}^{(n)})}{T(d^{(n)})
		\cdot g_n}$ and that the  distribution $\frac{\tilde{f}^{(n)}}{T(d^{(n)})}$
	converges to $\tilde{f}^{(\infty)}.$
	
	We assume again by contradiction that  
	$\lim_{n\to \infty} \frac{\tilde{f}_{s'}^{(n)}\cdot \tau_{s'}(\tilde{f}^{(n)})}{T(d^{(n)})
		\cdot g_n}>0.$
	Then we obtain immediately that $\lim_{n\to \infty} \frac{ \tau_{s'}(\tilde{f}^{(n)})}{g_{n}}=\infty,$
	since $\tilde{f}_{s'}^{(\infty)}=\varlimsup_{n\to\infty}
	\frac{\tilde{f}_{s'}^{(n)}}{T(d^{(n)})}=0.$ So (L2) yields
	that 
	$\tilde{f}_{s'}^{(n)}\equiv 0$ for $n$ 
	large enough, which in turn implies that $\frac{\tilde{f}_{s'}^{(n)}\cdot \tau_{s'}(\tilde{f}^{(n)})}{T(d^{(n)})
		\cdot g_n}\equiv 0$
	for $n$ large enough, a contradiction.
	
	So, $\varlimsup_{n\to \infty} \frac{\tilde{f}_{s'}^{(n)}\cdot \tau_{s'}(\tilde{f}^{(n)})}{T(d^{(n)})
		\cdot g_n}=0.$

\textbf{Proof of $c):$} 
	Assume again w.l.o.g.\ that the distribution 
	$\frac{\tilde{f}^{(n)}}{T(d^{(n)})}$
	converges  to the limit $\tilde{f}^{(\infty)}.$
	Trivially, $\tilde{f}^{(\infty)}$ is a strategy
	profile of $\Gamma^{(\infty)}.$
	We will now show that $\tilde{f}^{(\infty)}$
	is an NE profile of $\Gamma^{(\infty)}.$
	
	Consider $k\in \K$ and two strategies $s,s'\in \S_k$
	with $\tilde{f}_s^{(\infty)}>0.$ We need to show
	that $\tau^{(\infty)}_s(\tilde{f}^{(\infty)})\le 
	\tau^{(\infty)}_{s'}(\tilde{f}^{(\infty)}).$
	By $b),$ $s$ is tight. If 
	$s'$ is not tight, then 
	$\tau^{(\infty)}_s(\tilde{f}^{(\infty)})\le 
	\tau^{(\infty)}_{s'}(\tilde{f}^{(\infty)})=\infty$.
		
	So assume that $s'$ is also tight. 
	Since $\tilde{f}_s^{(\infty)}>0$ and
	each $\tilde{f}^{(n)}$ is an NE profile, 
	we obtain 
	$\tilde{f}_s^{(n)}>0$
	and 
	$\frac{\tau_s(\tilde{f}^{(n)})}{g_n}
	\le \frac{\tau_{s'}(\tilde{f}^{(n)})}{g_n}$
	for $n$ large enough. Thus
	$\tau^{(\infty)}_s(\tilde{f}^{(\infty)})
	\le \varlimsup_{n\to \infty }\frac{\tau_{s'}
		(\tilde{f}^{(n)})}{g_n}\le \tau^{(\infty)}_{s'}(\tilde{f}^{(\infty)})$ by a).
	
	Altogether, this shows that $\tilde{f}^{(\infty)}$
	is an NE profile of $\Gamma^{(\infty)}.$
	Since the PoA$(d^{(\infty)})$ of $\Gamma^{(\infty)}$
	equals $1,$ $\tilde{f}^{(\infty)}$ is also an SO profile of $\Gamma^{(\infty)}.$
	
	\textbf{Proof of d) and e):} Let $\tilde{f}^{(\infty)}$
	be an NE profile of $\Gamma^{(\infty)}.$
	By a)--c) and (L4),
	we obtain that 
	\[
	\lim_{n\to \infty} \frac{C(\tilde{f}^{(n)})}{T(d^{(n)})\cdot g_n}
	=\sum_{s\in S:\ s\text{ is tight}}
	\tilde{f}_s^{(\infty)}\cdot 
	\tau^{(\infty)}_s(\tilde{f}^{(\infty)})=C_{\Gamma^{(\infty)}}(\tilde{f}^{(\infty)})\in (0,\infty),
	\]
	and that 
	$\tilde{f}^{(\infty)}$ is an SO 
	profile of $\Gamma^{(\infty)}.$ Here we also used 
	that the NE cost of $\Gamma^{(\infty)}$ is unique.
	
	d) follows directly from the optimality of profiles $f^{*(n)}.$ In fact,
	$
	\varlimsup_{n\to \infty} 
	\frac{C(f^{*(n)})}{T(d^{(n)})\cdot g_n}
	\le \lim_{n\to \infty} \frac{C(\tilde{f}^{(n)})}{T(d^{(n)})\cdot g_n}
	=C_{\Gamma^{(\infty)}}(\tilde{f}^{(\infty)})<\infty
	$
implies that $\lim_{n\to \infty}\frac{f_s^{*(n)}}{T(d^{(n)})}=0$ for non-tight strategies $s\in \S$.
	Otherwise,
	$
	\varlimsup_{n\to \infty} 
	\frac{C(f^{*(n)})}{T(d^{(n)})\cdot g_n}
	\ge \varlimsup_{n\to \infty} \frac{
		f_s^{*(n)}\cdot \tau_s(f^{*(n)})}{T(d^{(n)})\cdot g_n}=\infty
	$
	for a non-tight strategy $s$ with
	$\varlimsup_{n\to \infty}\frac{f_s^{*(n)}}{T(d^{(n)})}>0.$
	
	To prove e), we assume again w.l.o.g.\ that 
	$\lim_{n\to\infty}\frac{f^{*(n)}}{T(d^{(n)})}
	=f^{*(\infty)}$ for a limit distribution
	$f^{*(\infty)}.$ Otherwise, we can take
	an infinite subsequence $(n_i)$
	fulfilling this condition and 
	$\lim_{i\to \infty}$PoA$(d^{(n_i)})
	=\varlimsup_{n\to\infty}$PoA$(d^{(n)}).$
	Then
	we obtain from a)--c) that 
	\[
	\begin{split}
	\lim_{n\to\infty}
	\frac{C(\tilde{f}^{(n)})}{T(d^{(n)})\cdot g_n}&= C_{\Gamma^{(\infty)}}(\tilde{f}^{(\infty)})
	\le C_{\Gamma^{(\infty)}}(f^{*(\infty)})
	=\sum_{s\in S: s\text{ is tight}}
	f_s^{*(\infty)}\cdot \tau^{(\infty)}_s(f^{*(\infty)})\\
	&=\lim_{n\to\infty}\frac{\sum_{s\in S: s\text{ is tight}}f_s^{*(n)}\cdot \tau_s(f^{*(n)})}{T(d^{(n)})\cdot g_n}
	\le \varliminf_{n\to \infty} 
	\frac{C(f^{*(n)})}{T(d^{(n)})\cdot g_n},
	\end{split}
	\] 
	which in turn implies that 
	$
	\varlimsup_{n\to\infty}\text{PoA}(d^{(n)})
	\le 1.
	$
	So e) holds.
	\hfill$\square$
		
\subsection{Proof of Example~\ref{example:SSDDivergentSequence}}\label{sec:proofExampleSSDDivergentSequence}
We need to show that 
each regular sequence of $\Gamma$ has a scalable subsequence.	
Let $(\D^{(n)})_{n\in \N}$
be an arbitrary regular sequence. We assume w.l.o.g.\ that
$T(d^{(n)})\in [b_n,b_{n+1})$ for each $n\in \N.$
Otherwise, we can continue the discussion with an infinite subsequence 
$(n_j)_{j\in \N}$ of $\N$ that satisfies 
$T(d^{(n_j)})\in [b_{n_j},b_{n_{j}+1})$ for
all $j\in \N.$ Such an infinite subsequence must exist, since 
$\bigcup_{n\in \N}[b_n,b_{n+1})=[0,\infty).$

By induction \Wuu{over} $n\in \N,$ we easily obtain that 
$\tau(b_{n+1})\le \theta_{n}\cdot b_{n+1}$ for all 
$n\in \N,$ since both $(b_n)_{n\in \N}$ and 
$(\theta_n)_{n\in \N}$ are strictly increasing.
So, $\lim_{n \to \infty}\frac{\tau(b_n)}{b_{n+1}}=0,$
since $\lim_{n \to \infty}\frac{\theta_{n-1}\cdot b_n}{b_{n+1}}=0.$

Let 
$
\kappa :=\varlimsup_{n\to \infty}\frac{T(\D^{(n)})}{b_{n}}\in [1,\infty].
$
We assume further that
$
\lim_{n\to \infty}\frac{T(\D^{(n)})}{b_{n}}=\varlimsup_{n\to \infty}\frac{T(\D^{(n)})}{b_{n}}=\kappa.
$
Otherwise, we can take another subsequence of the current sequence fulfilling
this condition.
We set the scaling factor $g_n$ to
	$
	g_n:=\tau\big(T(d^{(n)})\big)=\theta_n\cdot(T(d^{(n)})-b_n)
	+\tau(b_n)
	=\theta_n\cdot(T(d^{(n)})-b_n)
	+\theta_{n-1}\cdot(b_n-b_{n-1})+\tau(b_{n-1})
	$
for each $n\in \N.$ There are three cases: $\kappa=1,\ \kappa=\infty,$ and $\kappa\in (1,\infty)$. The following analysis shows that the limit game w.r.t.\
$(g_n)_{n\in \N}$ exists in all cases.

\textbf{(Case I: $\kappa=1$)}
In this case, we obtain for any $x\in (0,1)$ that $b_{n-1}<x\cdot T(\D^{(n)})<b_{n}\le T(d^{(n)})<b_{n+1}$
for $n$ large enough, since $\frac{b_{n}}{b_{n-1}}\to \infty$ 
and $\frac{T(\D^{(n)})}{b_{n}}\to\kappa=1$ as $n\to \infty.$ 
So, for every $x\in (0,1),$ 
\[
\begin{split}
\tau^{(\infty)}(x):=\lim_{n\to\infty}\frac{\tau\big(T(\D^{(n)})\cdot x\big)}{g_n}
=\lim_{n\to\infty}\frac{
\theta_{n-1} \cdot \big(T(d^{(n)})x-b_{n-1}\big)+\tau(b_{n-1})}{
\theta_n\cdot(T(d^{(n)})-b_n)
+\theta_{n-1}\cdot(b_n-b_{n-1})+\tau(b_{n-1})
	}=x,
\end{split}
\]
where we observe that $\lim_{n \to \infty}\frac{b_n}{T(d^{(n)})}=\kappa=1,$
$\lim_{n \to \infty}\frac{\tau(b_{n-1})}{b_n}=0,$
$\lim_{n \to \infty}\frac{b_n-b_{n-1}}{b_n}=1$
and $\lim_{n \to \infty}\frac{\theta_n}{\theta_{n-1}}=\epsilon>1.$ 
So, $\lim_{d^{(n)}\to \infty}\Gamma^{[g_n]}=\Gamma^{(\infty)}$ 
for the limit game
$\Gamma^{(\infty)}$
with the \Wuu{cost} function 
$\tau^{(\infty)}(x)=x$ 
for both arcs. Trivially, PoA$(d^{(\infty)})=1.$
Hence, $\big(d^{(n)}\big)_{n\in \N}$
is scalable.

\textbf{(Case II: $\kappa=\infty$)}
In this case, for each $x\in (0,1],$ 
$b_{n}<x\cdot T(d^{(n)})\le T(d^{(n)})<b_{n+1}$ holds
for large enough
$n,$ since $\lim_{n\to \infty}\frac{T(d^{(n)})}{b_{n}}=\infty.$ The definition of $\tau(\cdot)$ 
yields directly that $\tau^{(\infty)}(x)
:=\lim_{n\to \infty}\frac{\tau\big(x\cdot T(d^{(n)})\big)}{g_n}=x$
for each $x\in (0,1].$
So $\big(d^{(n)}\big)_{n\in \N}$
is also scalable in this case.

\textbf{(Case III: $\kappa\in (1,\infty)$)} 
We distinguish between different values of $x \in (0,1]$.
If $x\in (\frac{1}{\kappa},1],$ then $b_{n}<T(\D^{(n)})\cdot x\le T(\D^{(n)})
<b_{n+1}$ for $n$ large enough. If $x\in (0,\frac{1}{\kappa}),$ then
$b_{n-1}<T(\D^{(n)})\cdot x<b_{n}
\le T(d^{(n)})$ for $n$
large enough.   
The definition of $\tau(\cdot)$
yields that the limit cost function
\[
\tau^{(\infty)}(x):=
\lim_{n \to \infty}\frac{\tau(T(d^{(n)})x)}{
g_n}=
\begin{cases}
\frac{x/\epsilon }{1-1/\kappa+1/(\kappa\cdot \epsilon)}&\text{if }x\in (0,1/\kappa)\\
\frac{x-1/\kappa+1/(\kappa\cdot\epsilon)}{1-1/\kappa+1/(\kappa\cdot \epsilon)}&\text{if }
x\in [1/\kappa,1]
\end{cases}
\] is continuous, convex, strictly increasing
and non-negative, although it is a piece-wise function. 
Herein, the convexity follows because $\epsilon>1.$
So, both NE profiles and SO profiles of the limit game
$\Gamma^{(\infty)}$ are unique.
The unique NE profile is obviously $
\tilde{f}^{(\infty)}=(\tilde{f}^{(\infty)}_u,\tilde{f}^{(\infty)}_\ell)=(0.5,0.5),$
where we denote by $u$ and $\ell$ the upper and the lower
arc, respectively. \Wuu{Note that $\tilde{f}^{(\infty)}$ is also the unique SO profile, since $\tau^{(\infty)}(\cdot)$ is strictly convex and}
$C_{\Gamma^{(\infty)}}(f)=C_{\Gamma^{(\infty)}}(f')$ for any two profiles $f=(f_u,f_\ell)$ and $f'=(f'_u,f'_\ell)$
of $\Gamma^{(\infty)}$ when $f+f'=(1,1)$.
So, PoA$(d^{(\infty)})=1,$ and 
$\big(d^{(n)}\big)_{n\in \N}$ is  scalable.

So \Wuu{$\Gamma$} is scalable. The limit game
in Case III is essentially different \Wuu{to}
those in Cases I and II. Hence,
limit games for a scalable game need not be
essentially unique, and
an arbitrary regular demand sequence of a scalable game itself
need not be scalable, but contains an scalable 
subsequence.
\hfill$\square$

\subsection{Proof of Theorem~\ref{thm:AsymptoticDecomposition}}\label{proof:AsymptoticDecomposition}
	Let $\Gamma\asymp_{d^{(n)}} \Gamma_{|\K_1}\oplus\cdots\oplus\Gamma_{|\K_m}.$
	Definition~\ref{def:EssentialDecomposition}~(AD3) yields that 
	there is a scaling sequence
	$(g^{(u)}_{n})_{n\in \N}$
	for each $u\in\M= \{1,\ldots,m\}$
	such that the limit game $\Gamma_{|\K_u}^{(\infty)}$
	of the scaled games $\Gamma_{|\K_u}^{[g^{(u)}]}$ w.r.t.\ 
	$(d^{(n)}_{|\K_u})_{n\in \N}=\big((d_k^{(n)})_{k\in \K_u}\big)_{n\in \N}$ exists.
	For each $u\in \M$  we put
	\[
	\Gamma^{(\infty)}_{|\K_u} := \Big(A_{|\K_u},\K_u,\S_{|\K_u}:=\bigcup_{k\in \K_u}\S_k,\big(\tau^{(\infty)}_{a,u}(\cdot)\big)_{a\in A_{|\K_u}},d_{|\K_u}^{(\infty)}=(d_k^{(u,\infty)})_{k\in \K_u}\Big),
	\]
	where each limit cost function
	$\tau^{(\infty)}_{a,u}(x)=\lim_{n\to\infty}\frac{\tau_a(T(d^{(n)}_{|\K_u})\cdot x)}{g_n^{(u)}}$ is either the ``constant"
	$\infty$ on $I_a\setminus\{0\},$ or 
	finite and continuous on $I_a\setminus\{0\}$. 
	Recall that we consider only
	relevant resources $a\in A_{|\K_u}$ in subgames
	$\Gamma_{|\K_u}$.
	
	Consider now a sequence $(\tilde{f}^{(n)})_{n\in \N}$ of NE profiles 
	of $\Gamma$ w.r.t.\ the decomposable sequence $(d^{(n)})_{n\in \N}.$
	For each $u\in \M,$ let $(\tilde{f}^{(\K_u,n)})_{n\in \N}$ and $(f^{*(\K_u,n)})_{n\in \N}$
	be a sequence of NE profiles and SO profiles of $\Gamma_{|\K_u}$ w.r.t.
	$(d^{(n)}_{|\K_u})_{n\in \N},$ respectively.
	Let $(n_i)_{i\in \N}$ be an infinite subsequence
	such that $\lim_{i\to \infty}\overline{\text{PoA}}(d^{(n_i)})$=$\varlimsup_{n\to \infty}\overline{\text{PoA}}(d^{(n)}).$
	We again assume w.l.o.g.\ that 
	$(n_i)_{i\in \N}=(n)_{n\in \N},$ i.e.,
	$\lim_{n\to \infty}\overline{\text{PoA}}(d^{(n)})\in [0,\infty]$
	exists.
	
	By taking again an appropriate subsequence,  we may assume w.l.o.g.
	that: 
	\begin{itemize}
		\item For each $u\in \M,$ the limit distribution
		$\lim_{n\to \infty}
		\frac{\tilde{f}_{|\K_u}^{(n)}}{T(d^{(n)}_{|\K_u})}=:\tilde{f}^{(u,\infty)}=(\tilde{f}^{(u,\infty)}_s)_{s\in \S_k,k\in\K_u}$
		exists.
		\item For each $u\in \M,$ the limit distribution
		$\lim_{n\to \infty}
		\frac{\tilde{f}^{(\K_u,n)}}{T(d^{(n)}_{|\K_u})}=:\tilde{f}^{(\K_u,\infty)}=(\tilde{f}^{(\K_u,\infty)}_s)_{s\in \S_k,k\in\K_u}$
		exists.
		\item Scaling sequences are {\em mutually comparable},
		\Wuu{i.e.,} $\lim_{n\to\infty}\frac{g_n^{(u)}}{g_n^{(v)}}\in [0,\infty]$ for $(u,v)\in \M\times\M.$
	\end{itemize}

	With these preparations, we will now show inductively over $u\in \M$ that
	\begin{equation}\label{eq:AD_NE_Induction_Obj}
	\lim_{n\to \infty}
	\frac{\sum_{v=1}^{u}\sum_{k  \!\in\! \K_v}
		\sum_{s\!\in\! S_k}\tilde{f}_s^{(n)}
		\cdot \tau_s(\tilde{f}^{(n)})}{\sum_{v=1}^{u}
		C_{\Gamma_{|\K_v}}(\tilde{f}^{(\K_v,n)})}=1,
	\end{equation}
	 which yields
	that $\lim_{n \to \infty}\overline{\text{PoA}}(d^{(n)})=1.$ 
	
	\textbf{(Case $u=1$)} Conditions (AD1)--(AD2) from Definition~\ref{def:EssentialDecomposition} yield
	$d_k^{(1,\infty)}=\lim_{n\to \infty}
	\frac{d_k^{(n)}}{T(d^{(n)}_{|\K_1})}>0$
	for each $k\in \K_1$ and 
	$\lim_{n\to\infty}\frac{T(d^{(n)}_{|\K_1})}{T(d^{(n)})}=1.$ 
	To prove \eqref{eq:AD_NE_Induction_Obj} for
	$u=1,$ we will show that
	\begin{equation}\label{eq:AD_Step1_Obj}
	\begin{split}
	\lim_{n \to \infty} 
	\frac{\sum_{k \in \K_1}\sum_{s \in S_k}
		\tilde{f}_s^{(n)} \cdot 
		\tau_s(\tilde{f}^{(n)})
	}{T(d^{(n)}_{|\K_1}) \cdot g_n^{(1)}}
	 & = \sum_{k \in \K_1} \sum_{s \in S_k: s \; \text{tight}}
	\tilde{f}_s^{(1,\infty)}
	 \cdot 
	\tau^{(\infty)}_{s,1}(\tilde{f}^{(1,\infty)}) \\
	 & = C_{\Gamma^{(\infty)}_{|\K_1}}(\tilde{f}^{(1,\infty)})
	 \in (0,\infty)
	\end{split}
	\end{equation}
	and that $\tilde{f}^{(1,\infty)}$ is an NE profile
	of $\Gamma^{(\infty)}_{|\K_1}.$ 
	This directly proves \eqref{eq:AD_NE_Induction_Obj}
	for $u=1$ with Lemma~\ref{lma:Prop_AD}~a).
	To prove \eqref{eq:AD_Step1_Obj}, we will
	use a similar argument as in the proof of 
	Lemma~\ref{lma:PriceConvergenceOfStrategies}~a)--c)
	for the convergence of scaled prices of relevant resources. 	
	
	Consider now resource $a\in A_{|\K_1}.$ Trivially, $\lim_{n\to\infty}\frac{\tau_a(\tilde{f}_a^{(n)})}{g_n^{(1)}}
	=\lim_{n\to \infty}\frac{\tau_a(\tilde{f}_{a|\K_1}^{(n)})}{g_n^{(1)}}
	=\tau^{(\infty)}_{a,1}(\tilde{f}_{a}^{(1,\infty)}),$
	when $\tilde{f}_{a}^{(1,\infty)}=
	\sum_{k \in \K_1}\sum_{s\in S_k : s \ni a}
	\tilde{f}_s^{(1,\infty)}>0$ and
	$\tau^{(\infty)}_{a,1}(\cdot)$ is finite and continuous
	on $I_a\setminus\{0\}.$
	Herein we used the facts that each
	$\tilde{f}_a^{(n)}=\tilde{f}_{a|\K_1}^{(n)}+
	\tilde{f}_{a|\bigcup_{v=2}^{m}\K_v}^{(n)},$
	$\lim_{n\to\infty}\frac{\tilde{f}_{a|\bigcup_{v=2}^{m}\K_v}^{(n)}}{T(d^{(n)}_{|\K_1})}=0,$
	and every $\tau^{(\infty)}_{a,1}(x)=\lim_{n\to\infty}\frac{\tau_a(T(d^{(n)}_{|\K_1})\cdot x)}{g_n^{(1)}}$ for each consumption rate $x\in I_a\setminus\{0\}.$ 
Similar to the proof of Lemma~\ref{lma:PriceConvergenceOfStrategies}$a),$
	we obtain that $\varlimsup_{n\to\infty}\frac{\tau_a(\tilde{f}_a^{(n)})}{g_n^{(1)}}\le \tau^{(\infty)}_{a,1}(\tilde{f}^{(1,\infty)}_a)=\tau^{(\infty)}_{a,1}(0)$
	when $\tilde{f}_{a}^{(1,\infty)}=0$ and 
	$\tau^{(\infty)}_{a,1}(\cdot)$ is finite and continuous on
	$I_a\setminus\{0\}.$ 
	
	So, 
	each {\em tight} strategy 
	$s\in \S_{|\K_1}$ fullfils 
	\begin{equation}\label{eq:AD_Step1_UserTotalPrice}
	\lim_{n\to\infty}\frac{\tilde{f}_s^{(n)}\cdot
		\tau_s(\tilde{f}^{(n)})}{T(d^{(n)}_{|\K_1})\cdot g_n^{(1)}}	=\tilde{f}_s^{(1,\infty)}\cdot \tau^{(\infty)}_{s,1}(\tilde{f}^{(1,\infty)})
	=\tilde{f}_s^{(1,\infty)}\cdot\sum_{a\in s} \tau^{(\infty)}_{a,1}(\tilde{f}_a^{(1,\infty)})
	<\infty,
	\end{equation} 
	and
	\begin{equation}\label{eq:AD_Step1_UserPrice}
	\varlimsup_{n\to\infty}\frac{
		\tau_s(\tilde{f}^{(n)})}{g_n^{(1)}}	\le \tau^{(\infty)}_{s,1}(\tilde{f}^{(1,\infty)})
	=\sum_{a\in s} \tau^{(\infty)}_{a,1}(\tilde{f}_a^{(1,\infty)})
	<\infty.
	\end{equation}
	Moreover, if $\tilde{f}_s^{(1,\infty)}>0$ then the limit $\lim_{n\to\infty}\frac{\tau_s(\tilde{f}^{(n)})}{g_n^{(1)}}$
	exists and 
	\begin{equation}\label{eq:AD_Step1_UserPrice_Positive}
	\lim_{n\to\infty}\frac{
		\tau_s(\tilde{f}^{(n)})}{g_n^{(1)}}	= \tau^{(\infty)}_{s,1}(\tilde{f}^{(1,\infty)})
	=\sum_{a\in s} \tau^{(\infty)}_{a,1}(\tilde{f}_a^{(1,\infty)})
	<\infty.
	\end{equation}
	Therefore, for each $k\in \K_1,$ its user equilibrium cost is $\tilde{L}_k^{(n)}:=\tau_s(\tilde{f}^{(n)})
	\in O(g_n^{(1)})$ for each 
	$s\in \S_k$ with $\tilde{f}_s^{(n)}>0.$
	
	Using \eqref{eq:AD_Step1_UserPrice}--\eqref{eq:AD_Step1_UserPrice_Positive}, 
	and arguments similar to those
	in the proof of Lemma~\ref{lma:PriceConvergenceOfStrategies}b),
	we obtain 
	for each non-tight strategy $s'\in \S_{|\K_1}$ that 
	\begin{equation}\label{eq:AD_Step1_Nontight}
	\tilde{f}_{s'}^{(1,\infty)}=\lim_{n\to\infty}
	\frac{\tilde{f}_{s'}^{(n)}}{T(d^{(n)}_{|\K_1})}=0
	\quad\text{and }
	\lim_{n\to\infty}\frac{\tilde{f}_{s'}^{(n)}\cdot
		\tau_{s'}(\tilde{f}^{(n)})}{T(d^{(n)}_{|\K_1})\cdot g_n^{(1)}}=0.
	\end{equation}
	
	Finally, \eqref{eq:AD_Step1_UserPrice}--\eqref{eq:AD_Step1_UserPrice_Positive}, 
	and arguments similar to those
	in the proof of Lemma~\ref{lma:PriceConvergenceOfStrategies}c),
	yield 
	that $\tilde{f}^{(1,\infty)}$ is an NE profile
	of $\Gamma^{(\infty)}_{|\K_1}.$
	Then
	\eqref{eq:AD_Step1_Obj} follows
	immediately from \eqref{eq:AD_Step1_UserTotalPrice} and
	\eqref{eq:AD_Step1_Nontight}.
	So, \eqref{eq:AD_NE_Induction_Obj} holds for $u=1.$

\textbf{(Case $u=t\in \{2,\ldots,m\}>1$)} 
	We first make the following inductive assumptions for $u=t-1$: 
	\begin{itemize}
		\item[(IA1)] The joint total cost
		$
		\sum_{v=1}^{t-1}\sum_{k  \in \K_v}
		\sum_{s\in S_k} \tilde{f}_s^{(n)}
		\cdot \tau_s(\tilde{f}^{(n)})
		=\Theta\Big(\max_{v=1}^{t-1}\ T(d^{(n)}_{|\K_v})\cdot g_n^{(v)}\Big).
		$
		
		\item[(IA2)] The user equilibrium cost $\tilde{L}_k^{(n)}\in O\big(\max_{v=1}^{t-1}\ g_n^{(v)}\big)$
		for $k\in \bigcup_{v=1}^{t-1}\K_v.$ 
		
		\item[(IA3)] \eqref{eq:AD_NE_Induction_Obj} holds for $u=t-1.$
	\end{itemize}
	We need to show that (IA1)--(IA3) also hold for $u=t$. 
	
	Consider a resource $a\in A_{|\K_t}$. Trivially, the joint consumption
	$
	\tilde{f}^{(n)}_a=\sum_{v=1}^{m} \tilde{f}_{a|\K_v}^{(n)}
	=\sum_{v=1}^{t-1}\tilde{f}_{a|\K_v}^{(n)}
	+\sum_{v=t}^{m}\tilde{f}_{a|\K_v}^{(n)}.
	$
	So if $\sum_{v=1}^{t-1}\tilde{f}_{a|\K_v}^{(n)}
	=\sum_{v=1}^{t-1}\sum_{k \in \K_v}\sum_{s \in S_k : s \ni a}
	\tilde{f}_s^{(n)}>0,$ then
	there are a group $k\in \bigcup_{v=1}^{t-1}\K_v$
	and a strategy $s'\in \S_k$
	such that $a \in s'$ and $\tilde{f}_{s'}^{(n)}>0,$ i.e.,
	$s'$ is adopted by some users from group 
	$k.$
	In this case we obtain from the inductive assumption (IA2) for $u=t-1$ that
	\[
	\tau_{a}\Big(\sum_{v=1}^{t-1}\tilde{f}_{a|\K_v}^{(n)}\Big)
	\in O\Big(\sum_{b\in s'}  
	\tau_{b} (\tilde{f}_b^{(n)})
	\Big)=O\Big(
	\tau_{s'} (\tilde{f}^{(n)})
	\Big)\subseteq O\big(\max_{v=1}^{t-1}\ g_n^{(v)}\big),
	\]
	since $\tilde{f}^{(n)}$ is an NE profile of
	$\Gamma$  and  
	$\tau_{s'}(\tilde{f}^{(n)})=\tilde{L}_k^{(n)}.$
	
	Therefore
	\begin{equation}\label{eq:PriceFunction_AD_Explicit}
	\tau_{a}(\tilde{f}_a^{(n)})
	=
	\begin{cases}
	O\big(\max_{v=1}^{t-1}\ g_n^{(v)}\big)&
	\text{if } \sum_{v=1}^{t-1}\tilde{f}^{(n)}_{a|\K_v}>0,\\
	\tau_a\big(\sum_{v=t}^{m}\tilde{f}^{(n)}_{a|\K_v}\big)&\text{otherwise}
	\end{cases}
	\text{\quad for each }a \in A.
	\end{equation}
	By \eqref{eq:PriceFunction_AD_Explicit} we can analyze 
    the \Wuuuu{subgame} $\Gamma_{|\K_t}$ independently from the others. 
	
	Since $(g_n^{(1)})_{n\in \N},\ldots,(g_n^{(m)})_{n\in \N}$
	are mutually comparable, we obtain that 
	either $g_n^{(t)}\in O(\max_{v=1}^{t-1}\ g_n^{(v)}),$
	or $g_n^{(t)}\in \omega(\max_{v=1}^{t-1}\ g_n^{(v)}).$ We now 
	discuss $\Gamma_{|\K_t}$ in these two subcases.
	
	\textbf{(Subcase I: $g_n^{(t)}\in O(\max_{v=1}^{t-1} \ g_n^{(v)})$)}
	We will show that $\Gamma_{|\K_t}$ is
	negligible w.r.t.\ the \Wuuuu{subgame} $\Gamma_{|\bigcup_{v=1}^{t-1}\K_v}.$ 
	Equation \eqref{eq:PriceFunction_AD_Explicit} implies for each $a\in A$ that
	\begin{equation}\label{eq:Price_Important}
	\begin{split}
	\tau_a(\tilde{f}_a^{(n)})\in 
	O\Bigg(\max\Big\{\max_{v=1}^{t-1}\ g_n^{(v)},\ 
	\tau_a\big(\sum_{v=t}^{m}\tilde{f}^{(n)}_{a|\K_v}\big) \Big\}\Bigg).
	\end{split}
	\end{equation}
	Since $\tilde{f}_{|\K_t}^{(n)}=\big(\tilde{f}_{s}^{(n)})_{s\in \S_{k},k\in \K_t}$ is a strategy profile of \Wuuuu{subgame}
	$\Gamma_{|\K_t}$ w.r.t.\ $(d^{(n)}_{|\K_t}=(d_k^{(n)})_{k\in \K_t}\big)_{n\in \N},$ and since $\Gamma^{(\infty)}_{|\K_t}$
	is the limit game of $\Gamma_{|\K_t}^{[g_n^{(t)}]}$ w.r.t.\ 
	$\big(d^{(n)}_{|\K_t}\big)_{n\in \N},$
	we obtain with Lemma~\ref{lma:PriceConvergenceOfStrategies}a)
	that
	\begin{equation}\label{eq:TightStrategy_Step_t}
	\varlimsup_{n\to\infty}
	\frac{\tau_a\big(\tilde{f}_{a|\K_t}^{(n)}\big)}{g_n^{(t)}}
	=\varlimsup_{n\to\infty}\frac{\tau_a\big(T(d^{(n)}_{|\K_t})\cdot 
		\frac{\tilde{f}_{a|\K_t}^{(n)}}{T(d^{(n)}_{|\K_t})}\big)}{g_n^{(t)}}
	\le \tau^{(\infty)}_{a,t}(\tilde{f}_{a}^{(t,\infty)})<\infty
	\end{equation}
	for each tight strategy $s\in \S_{|\K_t}$ and each $a\in s$. 
	Herein we have used  that $\tilde{f}^{(t,\infty)}=\lim_{n\to\infty}
	\frac{\tilde{f}_{|\K_t}^{(n)}}{T(d^{(n)}_{|\K_t})}$
	and $\tilde{f}_{a}^{(t,\infty)}=\sum_{k \in \K_t}\sum_{s'\in S_k: s'\ni a} \tilde{f}_{s'}^{(t,\infty)}.$
	Equation~\eqref{eq:TightStrategy_Step_t} and
	the fact that $\lim_{n\to\infty}\frac{\sum_{v=t+1}^{m}\sum_{k  \!\in\! \K_v}d_k^{(n)}}{T(d^{n}_{|\K_t})}=\lim_{n\to\infty}\frac{\sum_{v=t+1}^{m}\sum_{k  \!\in\! \K_v}d_k^{(n)}}{\sum_{k\in \K_t}d_k^{(n)}}=0$ (see Definition~\ref{def:EssentialDecomposition}~(AD2)) yield that 
	\begin{equation}\label{eq:TightStrategy_Step_t_II}
	\begin{split}
	\varlimsup_{n\to\infty}
	\frac{\tau_a\big(\sum_{v=t}^{m}\tilde{f}_{a|\K_v}^{(n)}\big)}{g_n^{(t)}}
	&=\varlimsup_{n\to\infty}\frac{\tau_a\big(T(d^{(n)}_{|\K_t})\cdot 
		\frac{\sum_{v=t}^{m}\tilde{f}_{a|\K_v}^{(n)}}{T(d^{(n)}_{|\K_t})}\big)}{g_n^{(t)}}\\
	&=\varlimsup_{n\to\infty}\frac{\tau_a\big(T(d^{(n)}_{|\K_t})\cdot 
		\frac{\tilde{f}_{a|\K_t}^{(n)}}{T(d^{(n)}_{|\K_t})}\big)}{g_n^{(t)}}
	\le \tau^{(\infty)}_{a,t}(\tilde{f}_{a|\K_t}^{(t,\infty)})<\infty
	\end{split}
	\end{equation}
	for each relevant resource $a$ with $a\in s$ 
	for some tight strategy $s\in \bigcup_{k\in \K_t}\S_k.$
	
	Equations~\eqref{eq:Price_Important} and 
	\eqref{eq:TightStrategy_Step_t_II} and the fact that $g_n^{(t)}\in O(\max_{v=1}^{t-1}g_n^{(v)})$ imply that 
	\[
	\tau_s(\tilde{f}^{(n)})=\sum_{a\in s}
	\tau_a(\tilde{f}_a^{(n)})
	\in O\Bigg(\max\Big\{\max_{v=1}^{t-1}\ g_n^{(v)},\ 
	\tau_a\Big(\sum_{v=t}^{m}\tilde{f}^{(n)}_{a|\K_v}\Big) \Big\}\Bigg)
	\subseteq O\Bigg(\max_{v=1}^{t-1}\ g_n^{(v)}\Bigg),
	\]
	for each tight strategy $s\in \bigcup_{k\in \K_t}\S_k$. This, in turn, yields
	$\tilde{L}_k^{(n)}\in O\big(\max_{v=1}^{t-1}g_n^{(v)}\big)= O\big(\max_{v=1}^{t}g_n^{(v)}\big)$
	for each $k\in \K_t,$ since each 
	group $k\in \K_t$ has a tight strategy
	and $\tilde{f}^{(n)}$ is an NE profile.
	So, the inductive assumption (IA2) holds
	for $u=t.$ 
	
	Note that 
	$
	\sum_{s\!\in\! S_k}\tilde{f}_s^{(n)}\cdot
	\tau_s(\tilde{f}^{(n)})=\tilde{L}_k^{(n)}\cdot 
	d_k^{(n)}
	$
	for each $k\in \K_t$. 
	So condition (AD2) from
	Definition~\ref{def:EssentialDecomposition} implies 
	\begin{equation}\label{eq:AD_Subcase_PoA_Marginal}
	\sum_{k  \!\in\! \K_t}\sum_{s\!\in\! S_k}\tilde{f}_s^{(n)}\cdot
	\tau_s(\tilde{f}^{(n)})
	\in  O\Big(T(d^{(n)}_{|\K_t})\cdot
	\max_{v=1}^{t-1}\ g_n^{(v)}\Big)
	\subseteq o\Big(
	\max_{v=1}^{t-1}\ T(d^{(n)}_{|\K_v})\cdot g_n^{(v)}
	\Big),
	\end{equation}
	since $T(d^{(n)}_{|\K_t})\in o\big(T(d^{(n)}_{|\K_v})\big)$
	for each $v<t.$ 
	
	This means that $\Gamma_{|\K_t}$ is negligible w.r.t.
	$\Gamma_{|\bigcup_{v=1}^{t-1}\K_v}.$
	The inductive assumption (IA1) for
	$u=t-1$ then gives
	$
	\sum_{v=1}^{t}\sum_{k  \!\in\! \K_v}\sum_{s\!\in\! S_k}\tilde{f}_s^{(n)}\cdot
	\tau_s(\tilde{f}^{(n)})
	\in \Theta\Big(\sum_{v=1}^{t-1}\sum_{k  \!\in\! \K_v}\sum_{s\!\in\! S_k}\tilde{f}_s^{(n)}\cdot
	\tau_s(\tilde{f}^{(n)})\Big)
	=\Theta\Big(\max_{v=1}^{t}\ T(d^{(n)}_{|\K_v})\cdot g_n^{(v)}\Big).
	$
	Hence, (IA1) holds also for $u=t.$
		
	By \eqref{eq:AD_Subcase_PoA_Marginal} and
	the induction assumption (IA1) for $u=t-1$, we obtain
	immediately that 
	\begin{equation}\label{eq:SubcaseI_IA3_a}
	\lim_{n\to\infty} \frac{\sum_{v=1}^{t-1}\sum_{k  \!\in\! \K_v}\sum_{s\!\in\! S_k}\tilde{f}_s^{(n)}\cdot
		\tau_s(\tilde{f}^{(n)})}{\sum_{v=1}^{t}\sum_{k  \!\in\! \K_v}\sum_{s\!\in\! S_k}\tilde{f}_s^{(n)}\cdot
		\tau_s(\tilde{f}^{(n)})}=1.
	\end{equation}
	Since 
	$
	\sum_{k  \!\in\! \K_t}\sum_{s\!\in\! S_k}\tilde{f}_s^{(n)}\cdot
	\tau_s(\tilde{f}^{(n)})
	\ge \sum_{k  \!\in\! \K_t}\sum_{s\!\in\! S_k}\tilde{f}_s^{(n)}\cdot
	\tau_{s}(\tilde{f}_{|\K_t}^{(n)})
	=C_{\Gamma_{|\K_t}}(\tilde{f}^{(n)}_{|\K_t})
	\ge C_{\Gamma_{|\K_t}}(f^{*(\K_t,n)}),
	$
	\eqref{eq:AD_Subcase_PoA_Marginal}, Lemma~\ref{lma:Prop_AD}a) 
	and the inductive assumption (IA3) for
	$u=t-1$ yield
	\[
	\begin{split}
	C_{\Gamma_{|\K_t}}(\tilde{f}^{(\K_t,n)}) & \in \Theta\Big(C_{\Gamma_{|\K_t}}(f^{*(\K_t,n)})\Big) 
	\subseteq O\Big(\sum_{k \in \K_t}\sum_{s \in S_k}\tilde{f}_s^{(n)}\cdot \tau_s(\tilde{f}^{(n)}) \Big) \\
	& \subseteq o\Big(\sum_{v=1}^{t-1}\sum_{k  \!\in\! \K_v}\sum_{s\!\in\! S_k}\tilde{f}_s^{(n)}\cdot
	\tau_s(\tilde{f}^{(n)})\Big)
	= o\Big(\sum_{v=1}^{t-1}C_{\Gamma_{|\K_v}}(\tilde{f}^{(\K_v,n)})\Big).
	\end{split}
	\] 
	This implies that
	\begin{equation}\label{eq:SubcaseII_IA3_b}
	\lim_{n\to\infty}\frac{\sum_{v=1}^{t-1}C_{\Gamma_{|\K_v}}(\tilde{f}^{(\K_v,n)})}{\sum_{v=1}^{t}C_{\Gamma_{|\K_v}}(\tilde{f}^{(\K_v,n)})}=1.
	\end{equation}
	Therefore, using \eqref{eq:SubcaseI_IA3_a}, \eqref{eq:SubcaseII_IA3_b}, and
	the inductive assumption (IA3) for $u=t-1,$
	we obtain \eqref{eq:AD_NE_Induction_Obj}
	for $u=t$. 		
	
	\textbf{(Subcase II: $g_n^{(t)}\in \omega\big(
		\max_{v=1}^{t-1} g_n^{(v)}
		\big)$)} Equation~\eqref{eq:PriceFunction_AD_Explicit} yields that 
	\begin{equation}\label{eq:SubcaseII_PriceFunction}
	\varlimsup_{n\to\infty}
	\frac{\tau_a(\tilde{f}^{(n)}_a)}{g_n^{(t)}}
	=\varlimsup_{n\to\infty}\frac{\tau_{a}\Big(
		\sum_{v=t}^{m}\tilde{f}_{a|\K_v}^{(n)}\Big)}{g_n^{(t)}}
	\end{equation}
	for $a\in A_{|\K_t},$
	since $\lim_{n\to\infty}\frac{\max_{v=1}^{t-1}g_n^{(v)}}{g_n^{(t)}}=0,$ and thus  
	$\sum_{v=1}^{t-1}\tilde{f}_{a|\K_v}^{(n)}$
	is asymptotically negligible in this case. Because of \eqref{eq:SubcaseII_PriceFunction},
	we can ignore these groups $k$ in the union $ \bigcup_{v=1}^{t-1}\K_v$ and concentrate on 
	the remaining groups. These, in turn, can be treated
	in a similar way as in the proof for $u=1.$ We observe the following facts:
	\begin{itemize}
		\item $\lim_{n\to\infty}\frac{T(d^{(n)}_{|\K_v})}{T(d^{(n)}_{|\K_t})}=0$
		for each $v\in \{t+1,\ldots,m\},$ and
		$\lim_{n\to\infty}T(d^{(n)}_{|\K_t})=\infty.$
		
		\item $\Gamma^{(\infty)}_{|\K_t}$ is the limit of 
		$\Gamma_{|\K_t}^{[g_n^{(t)}]}$ w.r.t.\ $(d^{(n)}_{|\K_t})_{n\in \N}$
		and scaling sequence $(g_n^{(t)})_{n\in \N}.$
		
		\item \eqref{eq:SubcaseII_PriceFunction} implies that  
		$\tilde{f}_{|\bigcup_{v=t}^{m}\K_v}^{(n)}=(\tilde{f}_s^{(n)})_{s\in \S_k, k\in \bigcup_{v=t}^{m}\K_v}$ 
		behaves in the limit as an NE profile of the game 
		$\Gamma_{|\bigcup_{v=t}^{m}\K_v}$ for large $n$.
	\end{itemize} 
	Combining these facts with the arguments used in the proof of Lemma~\ref{lma:PriceConvergenceOfStrategies}a)--c),
	we obtain that
	\begin{itemize}
		\item 
		\begin{equation}\label{eq:SubcaseII_StrategyCost}
		\varlimsup_{n\to\infty}\frac{\tau_s(\tilde{f}^{(n)})}{g_n^{(t)}}
		=\varlimsup_{n\to\infty}\frac{\tau_{s}(\tilde{f}_{|\K_t}^{(n)})}{g_n^{(t)}}
		\le \tau_{s,t}^{(\infty)}(\tilde{f}^{(t,\infty)})=\sum_{a\in s}  \tau^{(\infty)}_{a,t}(\tilde{f}_{a}^{(t,\infty)})<\infty,
		\end{equation}
		for each tight strategy $s\in \K_t.$
		
		\item \begin{equation}\label{eq:SubcaseII_TotalCost_t}
		\begin{split}
		\lim_{n \to \infty}\!
		\frac{\sum_{k \in \K_t}\sum_{s\in \S_k}\tilde{f}_s^{(n)}
				\cdot \tau_s(\tilde{f}^{(n)})}{T(d_{|\K_t}^{(n)})\cdot g_n^{(t)}}
		& =\lim_{n\to \infty}
		\frac{\sum_{k \in \K_t}\sum_{s\in \S_k}\tilde{f}_s^{(n)}
			\cdot \tau_{s}(\tilde{f}_{|\K_t}^{(n)})}{T(d_{|\K_t}^{(n)})\cdot g_n^{(t)}} \\
			& = C_{\Gamma^{(\infty)}_{|\K_t}}(\tilde{f}^{(t,\infty)})\in (0,\infty),
		\end{split}
		\end{equation}
		where we recall that $C_{\Gamma^{(\infty)}_{|\K_t}}(\cdot)$
		is the cost function of the limit game $\Gamma^{(\infty)}_{|\K_t}.$
		
		\item $\tilde{f}^{(t,\infty)}=(\tilde{f}^{(t,\infty)}_s)_{s\in \S_k, k\in \K_t}$ is an NE profile of
		$\Gamma^{(\infty)}_{|\K_t},$ since each $k\in\K_t$
		has a tight strategy.
		%
	\end{itemize} 
So
\begin{equation}\label{eq:NE_PartialCost_LimitII}
\lim_{n\to \infty} 
\frac{\sum_{k  \!\in\! \K_t}\sum_{s\!\in\! S_k}
	\tilde{f}_s^{(n)}\cdot \tau_s(\tilde{f}^{(n)})}{C_{\Gamma_{|\K_t}}(\tilde{f}^{(\K_t,n)})}=1.
\end{equation}
	Combining \eqref{eq:SubcaseII_StrategyCost}--\eqref{eq:NE_PartialCost_LimitII} with 
	the inductive assumptions (IA1)--(IA3)
	for $u=t-1$  then yields that (IA1)--(IA3)
	also hold for $u=t$ in this subcase.
	
	This completes the proof of
	Theorem~\ref{thm:AsymptoticDecomposition}.
	\hfill$\square$
	
	\subsection{Proof of Corollary~\ref{thm:ADGame_Results}}
	\label{proof:ADGame_Results}
	If $\K_{reg}=\K,$ then
	the proof is completely the same as that
	in Appendix~\ref{proof:AsymptoticDecomposition}.
If $\K\setminus\K_{reg}\ne \emptyset,$
	we first obtain \eqref{eq:AD_NE_Induction_Obj} for $\Gamma_{|\K_{reg}}$
	by an induction similar to that in Appendix~\ref{proof:AsymptoticDecomposition},
	since $\Gamma_{|\K_{reg}}\asymp_{d^{(n)}_{|\K_{reg}}}
		\Gamma_{|K_1}\oplus\cdots \oplus\Gamma_{|\K_m}.$
		
    Since $\tau_a(x)\not\equiv 0$ for
    each $a\in A$ and
    $T(d^{(n)}_{|\K\setminus\K_{reg}})\in O(1),$ the \Wuuuu{subgame} $\Gamma_{|\K\setminus\K_{reg}}$
	is negligible w.r.t. 
	$\Gamma_{|\K_{reg}}$ when we compare their
	total cost. 
	Then, by a discussion similar to subcase I in the
	proof of Theorem~\ref{thm:AsymptoticDecomposition},
	we obtain that 
	$\lim_{n \to \infty}\overline{\text{PoA}}(d^{(n)})=1.$\hfill$\square$

\subsection{Proof of Theorem~\ref{thm:Results_RVcost}}\label{proof:Results_RVcost}
	

	We only need to show the existence of suitable scaling
	sequences for an asymptotic decomposition of $\Gamma_{|\K_{reg}}$. 
	
	Consider a game $\Gamma$ with
	regularly varying cost functions $\tau_a(\cdot).$ 
	Let $(d^{(n)})_{n\in \N}$ be an arbitrary regular demand sequence. 
	By Appendix~\ref{proof:ADGame_Results},
	we assume that
	$(d^{(n)})_{n\in \N}$ is decomposable
	and $\K_{reg}=\K.$ 
	
	Let $\K_1,\ldots,\K_m$ be a partition of 
	$\K$ satisfying (AD1)--(AD2) of Definition~\ref{def:EssentialDecomposition}.
	We will show that
	there exist an infinite subsequence $(n_i)_{i\in \N}$
	and scaling sequences $(g_i^{(u)})_{i\in \N}$ for each 
	$u\in \M$ s.t. 
	$(d_{|\K_u}^{(n_i)})_{i\in \N}$ is a scalable demand sequence
	of $\Gamma_{|\K_u}$ w.r.t. $(g_i^{(u)})_{i\in \N}$ for
	each $u\in \M.$
	
	Trivially, there is an infinite subsequence $(n_i)_{i\in \N}$ s.t.
	the limit
	\[
	\lambda_u(a,b):=\lim_{i \to \infty}\frac{\tau_a\big(T(d^{(n_i)}_{|\K_u})\big)}{\tau_b\big(T(d^{(n_i)}_{|\K_u})\big)}\in [0,\infty]
	\]
	exists for each $u\in \M$ and for all $(a,b)\in A_{|\K_u}\times A_{|\K_u}$. 
	For such a subsequence $(n_i)_{i\in \N},$  we define for each 
	$u\in \M$
	an ordering $\preceq_u$ on $A_{|\K_u}$ as follows.

    For each pair $(a,b)\in A_{|\K_u}\times A_{|\K_u},$ we set $a\preceq_u b$ if $\lambda_u(a,b)<\infty,$ $a\sim_u b$ if $\lambda_u(a,b)\in (0,\infty),$
    and $b\prec_u a$ if $\lambda_u(a,b)= \infty.$
    
   For each $u\in \M,$ we then define 
   $
   a_u:=\max_{k\in \K_u}\min_{s\in \S_k}
   \max_{a\in s}\ a.
   $
   Herein, both the maximization and minimization are
   taken w.r.t. the ordering $\preceq_u.$
   If there are multiple maxima or minima, we pick an arbitrary one.  
   We then set the scaling factor $g_i^{(u)}$ to $\tau_{a_u}\big(T(d^{(n_i)}_{|\K_u})\big)$ for each $u\in \M$ and $i\in \N.$ 
   
   Consider now $u\in \M$ and a resource $a\in A_{|\K_u}.$
  The limit cost functions $\tau_{a,u}^{(\infty)}(\cdot)$ exist
   because 
   \[
   \begin{split}
   \tau_{a,u}^{(\infty)}(x)=
   \lim_{i\to \infty}
   \frac{\tau_a(T(d^{(n_i)}_{|\K_u})\cdot x)}{g_i^{(u)}}
   &=\lim_{i\to \infty}
   \frac{\tau_a(T(d^{(n_i)}_{|\K_u})\cdot x)}{
   	\tau_a(T(d^{(n_i)}_{|\K_u}))}\cdot 
   \frac{\tau_a(T(d^{(n_i)}_{|\K_u}))}{\tau_{a_u}(T(d^{(n_i)}_{|\K_u}))}
   =x^{\rho_{a}}\cdot \lambda_{u}(a,a_u)\\
   &=\begin{cases}
   0 &\text{if } \lambda_{u}(a,a_u)=0,\\
  \lambda_{u}(a,a_u)\cdot  x^{\rho_{a_u}} &\text{if }\lambda_{u}(a,a_u)\in (0,\infty),\\
   \infty&\text{if }\lambda_{u}(a,a_u)=\infty,
   \end{cases}
   \end{split}
   \]
   for each $x\in (0,1]$.
   Herein, we used that the functions $\tau_a(\cdot)$ are regularly varying with regular variation index $\rho_a\ge 0.$
   So, (L1)--(L2) hold. (L3) follows, since  
   $\min_{s\in \S_k}\max_{a\in s}\ a\preceq a_u$
   for each $k\in \K_u,$ and thus there is a strategy $s\in \S_k$
    for 
   each $k\in \K_u$ s.t. $\lambda_{u}(a,a_u)<\infty$ for each $a\in s.$
   (L4) follows since there is at least one $k\in \K_u$ s.t. 
   $\min_{s\in \S_k}\max_{a\in s}\ a=a_u,$ and since 
   $(d_{|\K_u}^{(n_i)})_{i\in\N}$ is decomposable and
   all $\tau_{a,u}^{(\infty)}$ are
   monomials of the same degree $\rho_{a_u}$.
   Altogether, we obtain that 
   $(d^{(n_i)}_{|\K_u})_{i\in\N}$
   is a scalable sequence of $\Gamma_{|\K_u}$ for
   each $u\in \M.$
   This completes the proof. 
	\hfill$\square$
	
	\subsection{Proof of Lemma~\ref{lma:Scalable<=ScalableByDecomposition}}
	\label{proof:Scalable<=ScalableByDecomposition}
	
	Consider a scalable game $\Gamma.$ 
	Let $(d^{(n)})_{n\in \N}$ be an
	arbitrary decomposable demand sequence and let
	$\K_1,\ldots,\K_m$ be the partition of $\K$
	satisfying (AD1)--(AD2). We will show that
	$\Gamma_{|\K_u}$ is scalable for each $u\in \M=\{1,\ldots,m\}.$
	
	
	For each $u\in \M,$ let $D_{|\K_u}^{(n)}=(D_k^{(n)})_{k\in \K}$ 
	be a demand vector of the game $\Gamma$ by extending $d_{|\K_u}^{(n)}$
	as follows. $D_k^{(n)}=0$ if $k\in \K\setminus\K_u,$ and
	$D_k^{(n)}=d_k^{(n)}$ if $k\in \K_u.$
	
	Obviously, $(D_{|\K_u}^{(n)})_{n\in \N}$
	is a regular demand sequence of $\Gamma,$ since $(d^{(n)})_{n\in \N}$ is
	decomposable and $(d_{|\K_u}^{(n)})_{n\in \N}$
	is a regular sequence of \Wuuuu{subgame} $\Gamma_{|\K_u}$. Since $\Gamma$ is scalable,
	there are an infinite subsequence $(n_i)_{i\in \N}$
	and a scaling sequence $(g_i^{(u)})_{i\in \N}$ for
	each $u\in \M$ s.t. 
	$(D_{|\K_u}^{(n_i)})_{i\in \N}$ is a scalable sequence
	of $\Gamma$ w.r.t. $(g_i^{(u)})_{i\in \N}.$
	It follows that $(d^{(n_i)}_{|\K_u})_{i\in \N}$ is a scalable sequence of $\Gamma_{|\K_u}$ w.r.t. $(g_i^{(u)})_{i\in \N}$ for
	each $u\in \M,$ since $\Gamma$ with demands $D^{(n_i)}_{|\K_u}$
	coincides with the subgame $\Gamma_{|\K_u}$
	with demands $d_{|\K_u}^{(n_i)}$ for each $u\in \M.$
	Hence, $\Gamma\asymp_{d^{(n_i)}}\Gamma_{|\K_1}
	\oplus \cdots \oplus \Gamma_{|\K_m}.$\hfill$\square$
	
\subsection{Proof of Theorem~\ref{eq:Limit_PoA_epsilon}}\label{proof:Limit_PoA_epsilon}
	Consider a game $\Gamma$ with cost functions
	$\tau_a(x)=\gamma_a\cdot x^{\beta}+\eta_a$.
	Let $d=(d_k)_{k\in \K}$ be a demand vector
	with a large enough total demand $T(d).$ 
	Let $d_{\min}=\min\{d_k:\ k\in \K\}>0$
	and $f^{*}$ be an SO profile of $\Gamma$ w.r.t.
	$d.$ 
	
	The optimality conditions for the SO (\Wuu{see, e.g.,}~\cite{Roughgarden2000How}) yield that 
	$
	\sum_{a\in s} \big(\tau_a(f^{*}_a)+
	f_a^*\cdot \tau'_a(f^{*}_a)
	\big)\le \sum_{a\in s'}\big(\tau_a(f^{*}_a)+
	f_a^*\cdot \tau'_a(f^{*}_a)
	\big)
	$
	for each $k\in \K$ and every $s,s'\in \S_k$ with $f_{s}^*>0,$ where $\tau'_a(\cdot)$ denotes the derivative
	of $\tau_a(\cdot)$.  
	
	So
	$
	\sum_{a\in s}  \Big((\beta
	+1)\cdot \tau_a(f^{*}_a)-
	\beta\cdot \eta_a
	\Big)\le \sum_{a\in s'} \Big((\beta
	+1)\cdot \tau_a(f^{*}_a)-
	\beta\cdot \eta_a
	\Big).
	$
	Therefore, 
	\begin{equation}\label{eq:BOUND_EPSILON}
	(1+\beta)\cdot\tau_s(f^*)\le (1+\beta)\cdot 
	\tau_{s'}(f^*)+\beta\cdot\big(\tau_s(\boldsymbol{0})-\tau_{s'}(\boldsymbol{0})\big),
	\end{equation}
	where $\boldsymbol{0}=(0)_{s''\in \S}$ denotes the 
	zero flow. 
	Hence,
	$
	\tau_s(f^*)\le \tau_{s'}(f^*)\cdot
	\Big(1+\frac{\beta\cdot |\tau_s(\boldsymbol{0})-\tau_{s'}(\boldsymbol{0})|}{(1+\beta)\cdot \tau_{s'}(f^*)}\Big).
	$
We now show that $\tau_{s'}(f^*)\in \Omega\big(d_{\min}^{\beta}),$ which directly implies
Theorem~\ref{eq:Limit_PoA_epsilon}.

Let $s_k\in \S_k$ be a strategy such that 
$f^{*}_{s_k}\ge \frac{d_k}{|\S|}\ge \frac{d_{\min}}{|\S|}>0.$
Clearly, \eqref{eq:BOUND_EPSILON} holds for
$s=s_k.$ Hence,
$
\tau_{s'}(f^*)\ge \tau_{s_k}(f^*)- 
\frac{\beta\cdot\big(\tau_{s'}(\boldsymbol{0})-\tau_{s_k}(\boldsymbol{0})\big)}{1+\beta}\in \Theta\big(\tau_{s_k}(f^*)\big).
$
Note that
$
\tau_{s_k}(f^*) =\sum_{a\in s_k} \tau_a(f_a^*)
\ge \sum_{a\in s_k} \tau_a\big(f_{s_k}^*\big)
\ge \sum_{a\in s_k} \tau_a\Big( \frac{d_{\min}}{|\S|}\Big),
$
where we recall Assumption~\eqref{eq:StrategiesScarce}, i.e., $a \in s_k$ for some $a\in A.$ 

Then $\tau_{s_k}(f^*)\in \Omega\big(d_{\min}^{\beta}\big),$
since each $\tau_a(\cdot)$ has degree $\beta.$
Therefore $\tau_{s'}(f^*)\in \Omega\big(d_{\min}^{\beta}\big).$
This completes the proof.
	\hfill$\square$

\subsection{Proof of Theorem~\ref{theo:PoA_Limit_BPR}}\label{proof:PoA_Limit_BPR}
\textbf{Proof of a):}
Consider a game $\Gamma$ as in Theorem~\ref{theo:PoA_Limit_BPR} and a demand vector
$d.$ Let $\tilde{f}$ and $f^*$ be an NE profile and an
SO profile, respectively.
Then 
\[
	\begin{split}
	C(\tilde{f})&=(\beta+1)\cdot \sum_{a\in A}
	\int_{0}^{\tilde{f}_a}\tau_a(t)dt-\beta\cdot 
	\sum_{a\in A}\eta_a\cdot \tilde{f}_a
	\le (\beta+1)\cdot \sum_{a\in A}
	\int_{0}^{f^*_a}\tau_a(t)dt-\beta\cdot 
	\sum_{a\in A}\eta_a\cdot \tilde{f}_a\\
	&=C(f^*)+\beta\cdot \sum_{a\in A}\eta_a\cdot \big(f^*_a-\tilde{f}_a\big).
	\end{split}
\]



Clearly, $C(f^*)=\Theta(T(d)^{\beta+1}).$ 
Lemma~\ref{lma:PriceConvergenceOfStrategies} yields that 
$\frac{f_a^*-\tilde{f}_a}{T(d)}\to 0$ as 
$T(d)\to \infty,$ since $\Gamma$ is tight and has strictly
increasing limit cost functions
that are essentially unique
for all regular demand sequences, and since every unbounded demand sequence $(d^{(n)})_{n\in \N}$ has a scalable subsequence $(d^{(n_i)})_{i\in \N}$ with
$\lim_{n \to \infty}\frac{f^{*(n)}_a-\tilde{f}_a^{(n)}}{T(d^{(n)})}=0.$
Hence
$f^{*}_a-\tilde{f}_a\in o(T(d))$ for each $a\in A$
and PoA$(d)=1+o(T(d)^{-\beta}).$

\textbf{Proof of b)--d): 
Consider the routing game $\Gamma$ in Figure~\ref{fig:BPR-Rate-Counterexample}(a).}
$\Gamma$ has two OD pairs $(o_k,t_k),$ $k=1,2.$
All \Wuu{cost} functions are displayed
next to the arcs and are BPR functions with the same
degree $\beta>0$. Suppose that 
$\eta_1> \eta_2>0.$

\begin{figure}[!htb]
	\centering
	\begin{subfigure}{0.30\textwidth}
		\centering
		\begin{tikzpicture}[
		>=latex
		]
		\node[scale=0.4,circle,fill=black,label=left:$o_1$](o1){};
		\node[scale=0.4,circle,fill=black,label=left:$C$,below right= of o1](C){};
		\node[scale=0.4,circle,fill=black,label=right:$D$,right =of C](D){};
		\node[scale=0.4,circle,fill=black,label=right:$t_1$,above right =of D](t1){};
		\node[scale=0.4,circle,fill=black,label=left:$o_2$,below left =of C](o2){};
		\node[scale=0.4,circle,fill=black,label=right:$t_2$,below right =of D](t2){};
		\draw[->,thick] (o1) to node[scale=0.8,above]{$\gamma_2x^{\beta}+\eta_2$} (t1);
		\draw[->,thick] (o1) to [out=90,in=90] node[scale=0.8,above]{$\gamma_1x^{\beta}+\eta_1$} (t1);
		\draw[->,thick] (C) to  node[scale=0.8,above]{$\gamma_3x^{\beta}+\eta_3$} (D);
		\draw[->,thick] (o2) to node[scale=0.8,above]{$\gamma_8x^{\beta}+\eta_8$} (t2);
		\draw[->,thick] (o1) to node[scale=0.8,left]{$\gamma_4x^{\beta}+\eta_4$} (C);
		\draw[->,thick] (D) to node[scale=0.8,right]{$\gamma_5x^{\beta}+\eta_5$} (t1);
		\draw[->,thick] (D) to node[scale=0.8,right]{$\gamma_6x^{\beta}+\eta_6$} (t2);
		\draw[->,thick] (o2) to node[scale=0.8,left]{$\gamma_7x^{\beta}+\eta_7$} (C);
		\end{tikzpicture}
		\subcaption{}
	\end{subfigure}
	\begin{subfigure}{0.30\textwidth}
		\centering
		\begin{tikzpicture}[
		>=latex
		]
			\node[scale=0.4,circle,fill=black,label=left:$o_1$](o1){};
			\node[scale=0.4,circle,fill=black,label=right:$t_1$,right =3of o1](t1){};
			\draw[->,thick] (o1) to [out=90,in=90] node[scale=0.8,above]{$\gamma_1x^{\beta}+\eta_1$} (t1);
			\draw[->,thick] (o1) to  node[scale=0.8,above]{$\gamma_2x^{\beta}+\eta_2$} (t1);
			\node[scale=0.4,circle,fill=black,label=left:$o_2$, below =2of o1](o2){};
			\node[scale=0.4,circle,fill=black,label=right:$t_2$,right =3of o2](t2){};
			\draw[->,thick] (o2) to [out=90,in=90] node[scale=0.8,above]{$(\gamma_7+\gamma_3+\gamma_6)x^{\beta}+\eta_7+\eta_3+\eta_6$} (t2);
			\draw[->,thick] (o2) to  node[scale=0.8,above]{$\gamma_8x^{\beta}+\eta_8$} (t2);
		\end{tikzpicture}
		\subcaption{}
	\end{subfigure}
	\begin{subfigure}{0.30\textwidth}
		\centering
		\begin{tikzpicture}[
		>=latex
		]
		\node[scale=0.4,circle,fill=black,label=left:$o_1$](o1){};
		\node[scale=0.4,circle,fill=black,label=right:$t_1$,right =3of o1](t1){};
		\draw[->,thick] (o1) to [out=90,in=90] node[scale=0.8,above]{$\gamma_1x^{\beta}$} (t1);
		\draw[->,thick] (o1) to  node[scale=0.8,above]{$\gamma_2x^{\beta}$} (t1);
		\node[scale=0.4,circle,fill=black,label=left:$o_2$, below =2of o1](o2){};
		\node[scale=0.4,circle,fill=black,label=right:$t_2$,right =3of o2](t2){};
		\draw[->,thick] (o2) to [out=90,in=90] node[scale=0.8,above]{$(\gamma_7+\gamma_3+\gamma_6)x^{\beta}$} (t2);
		\draw[->,thick] (o2) to  node[scale=0.8,above]{$\gamma_8x^{\beta}$} (t2);
		\end{tikzpicture}
		\subcaption{}
	\end{subfigure}
	\caption{Counterexample to the conjecture of \cite{O2016Mechanisms}}
	\label{fig:BPR-Rate-Counterexample}
\end{figure} 

Let $\epsilon\in [0,1)$ be a constant and $d^{(n)}=\big(d_1^{(n)}=n^\epsilon,d_2^{(n)}=n\big)$
for each $n\in \N,$ where $d_k^{(n)}$ denotes the 
demand of OD pair $(o_k,t_k),$ $k\in \mathcal{K}=\{1,2\}.$  
Obviously, $(d^{(n)})_{n\in \N}$ is decomposable 
because $\lim_{n\to \infty}d_k^{(n)}= \infty$
for $k\in \mathcal{K}$ and 
$\lim_{n\to \infty}\frac{d_1^{(n)}}{d_2^{(n)}}=0.$
Let $g_n^{(1)}=n^{\epsilon\cdot \beta}$
and $g_n^{(2)}=n^{\beta}$ for each $n\in \N.$
Then, for each $k\in \mathcal{K},$ the
demand sequence $\big(d_{|\{k\}}^{(n)}=(d_k^{(n)})\big)_{n\in \N}$ 
is scalable \Wuu{in} the \Wuuuu{singleton subgame} $\Gamma_{|\{k\}}$
formed by the
OD pair $(o_k,t_k)$ w.r.t. the scaling sequence $(g_n^{(k)})_{n\in \N}$.
Thus $\Gamma\asymp_{d^{(n)}}
\Gamma_{|\{2\}}\oplus \Gamma_{|\{1\}}.$

We now analyze the convergence rate of PoA$(d^{(n)}).$

For each $n\in \N,$ let $\tilde{f}^{(n)}$ and $f^{*(n)}$ be 
an NE profile and an SO profile of the game $\Gamma$ 
for $d^{(n)},$ respectively. 
Denote by $u$ and $\ell$ the upper path
and the middle path of OD pair $(o_1,t_1).$
Then $\tau_u(x)=\gamma_1\cdot x^{\beta}+\eta_1$
and $\tau_\ell(x)=\gamma_2\cdot x^{\beta}+\eta_2.$
Moreover, denote by $u'$ and $\ell'$
the upper path (\Wuu{i.e.,}\ $o_2\to C\to D\to t_2$) and the bottom path of OD pair
$(o_2,t_2).$

The NE profile $\tilde{f}^{(n)}$ and
the SO profile $f^{*(n)}$ will not use the 
path $o_1\to C \to D\to t_1$ when $n$ is large enough.
This follows since $d^{(n)}_1= n^{\epsilon}\in o(d_2^{(n)}=n)$, and since
$(C,D)$ is an arc of the path
$u'$ 
whose \Wuu{cost}
is much larger than that of the two arcs $u$ and $\ell$
for large enough $n$.
So $\tilde{f}^{(n)}$
and $f^{*(n)}$ will only use the four parallel paths 
$u,\ell,u',\ell',$ and are thus also the NE profile and the SO profile
of the routing game $\Gamma'$ in Figure~\ref{fig:BPR-Rate-Counterexample}(b), respectively.
So, the PoA$(d^{(n)})$ of the game $\Gamma$ equals
that of the game $\Gamma'$.
We can thus, for $k\in \mathcal{K}$, consider the 
subgame $\Gamma_{|\{k\}}$ as the
OD pair $(o_k,t_k)$ of the game 
$\Gamma',$ and the limit game $\Gamma_{|\{k\}}^{(\infty)}$
of the subgame $\Gamma_{|\{k\}}$
(for the demand sequence $(d_{|\{k\}}^{(n)})_{n\in \N}$)
as the OD pair $(o_k,t_k)$ in 
Figure~\ref{fig:BPR-Rate-Counterexample}(c).

Since $\Gamma_{|\{1\}}$
and $\Gamma_{|\{2\}}$ are disjoint and since we can equivalently consider them as the two OD pairs 
in Figure~\ref{fig:BPR-Rate-Counterexample}(b) for large enough $n$, we obtain that
$
C(\tilde{f}^{(n)})=\sum_{k=1}^{2}C_{\Gamma_{\{k\}}}(\tilde{f}^{(n)}_{|\{k\}})
\quad \text{and}\quad 
C(f^{*(n)})=\sum_{k=1}^{2}C_{\Gamma_{\{k\}}}(f^{*(n)}_{|\{k\}}),
$
where $C_{\Gamma_{|\{k\}}}(\cdot)$
is the cost of subgame $\Gamma_{|\{k\}},$
$\tilde{f}^{(n)}_{|\{1\}}=(\tilde{f}_u^{(n)},\tilde{f}_\ell^{(n)}),$
$f^{*(n)}_{|\{1\}}=(f^{*(n)}_u,f^{*(n)}_\ell),$ $\tilde{f}^{(n)}_{|\{2\}}=(\tilde{f}_{u'}^{(n)},\tilde{f}_{\ell'}^{(n)}),$
and $f^{*(n)}_{|\{2\}}=(f^{*(n)}_{u'},f^{*(n)}_{\ell'}).$
So, when $n$ is large, 
\begin{equation}\label{eq:NE-SO-BPR}
\begin{split}
C(\tilde{f}^{(n)})-C(f^{*(n)})=\sum_{k=1}^{2} \Big(C_{\Gamma_{|\{k\}}}(\tilde{f}_{|\{k\}}^{(n)})
-C_{\Gamma_{|\{k\}}}(f^{*(n)}_{|\{k\}})\Big)
\ge C_{\Gamma_{|\{1\}}}(\tilde{f}_{|\{1\}}^{(n)})
-C_{\Gamma_{|\{1\}}}(f^{*(n)}_{|\{1\}}).
\end{split}
\end{equation}
Herein, we observe for $n$ large enough that $\tilde{f}^{(n)}_{|\{k\}}$
and $f^{(n)}_{|\{k\}}$ are the NE profile and the SO
profile of the subgame $\Gamma_{|\{k\}}$, \Wuu{i.e.,}
of the OD pair $(o_k,t_k)$ in Figure~\ref{fig:BPR-Rate-Counterexample}(b),
respectively, for $k\in \mathcal{K}.$ 

When $\epsilon=0$, then \eqref{eq:NE-SO-BPR} yields
PoA$(d^{(n)})=1+\Omega(T(d^{(n)})^{-\beta-1})$, since
$d_1^{(n)}=n^{\epsilon}\equiv 1,$
and since $C_{\Gamma_{|\{1\}}}(\tilde{f}_{|\{1\}}^{(n)})
-C_{\Gamma_{|\{1\}}}(f^{*(n)}_{|\{1\}})\in \Theta(1)$
is a positive constant when $\epsilon=0$ and 
$n$ is large. The latter follows by observing that 
$\eta_1>\eta_2$, that $\Gamma_{|\{1\}}$
is not well designed, and that 
$C(f^{*(n)})\in \Theta(T(d^{(n)})^{\beta+1}=(n+n^{\epsilon})^{\beta+1}).$ 

Altogether, the case $\epsilon=0$ shows that the conjecture proposed by
\cite{O2016Mechanisms} does not hold in general.

We now discuss the case that $\epsilon\in (0,1).$

For the subgame $\Gamma_{|\{1\}},$ we obtain with Lemma~3 that 
$\lim_{n\to \infty}\frac{\tilde{f}^{(n)}_{|\{1\}}}{n^{\epsilon}}
=\lim_{n\to \infty}\frac{f^{*(n)}_{|\{1\}}}{n^{\epsilon}}
=\tilde{f}_{|\{1\}}^{(\infty)},$ and that $\tilde{f}_{|\{1\}}^{(\infty)}=(\tilde{f}_u^{(\infty)},\tilde{f}_\ell^{(\infty)})$
is an NE profile of the limit game $\Gamma_{|\{1\}}^{(\infty)}$, 
i.e., of the OD pair $(o_1,t_1)$ in Figure~\ref{fig:BPR-Rate-Counterexample}(c).
Thus 
$\tilde{f}^{(n)}_{|\{1\}}=\big((\tilde{f}^{(\infty)}_u+
x_n)\cdot n^{\epsilon}, (\tilde{f}^{(\infty)}_\ell-
x_n)\cdot n^{\epsilon}\big)$
and $
f^{*(n)}_{|\{1\}}=\big((\tilde{f}^{(\infty)}_u+
y_n)\cdot n^{\epsilon}, (\tilde{f}^{(\infty)}_\ell-
y_n)\cdot n^{\epsilon}\big)
$
for sequences $(x_n)_{n\in \N}$
and $(y_n)_{n\in \N}$ with
$\lim_{n\to \infty}x_n=\lim_{n\to \infty}y_n=0.$

Since $\tilde{f}^{(n)}_{|\{1\}}$ is an NE profile 
of the subgame $\Gamma_{|\{1\}},$  we obtain that 
\[
\tau_u(\tilde{f}_u^{(n)})= \gamma_1\cdot \big(\tilde{f}^{(\infty)}_u+
x_n\big)^{\beta}\cdot n^{\epsilon\cdot \beta}
+\eta_1=\gamma_2\cdot \big(\tilde{f}^{(\infty)}_\ell-
x_n\big)^{\beta}\cdot n^{\epsilon\cdot \beta}
+\eta_2=\tau_{\ell}(\tilde{f}_\ell^{(n)}),
\]
which in turn implies that 
$
x_n=\frac{\eta_2-\eta_1}{\beta\cdot \big(\gamma_1\cdot (\tilde{f}_u^{(\infty)})^{\beta-1}+\gamma_2\cdot (\tilde{f}_\ell^{(\infty)})^{\beta-1}\big)}\cdot 
n^{-\epsilon\cdot \beta}+o(n^{-\epsilon\cdot \beta}).
$
Herein, $\gamma\cdot \big(c+
x\big)^{\beta}=\gamma\cdot c^{\beta}+\beta\cdot\gamma\cdot c^{\beta-1}\cdot x+o(x)$ for any constants
$c,\gamma>0,$ $\tau^{(\infty)}_{u,1}(\tilde{f}_u^{(\infty)})=\gamma_1\cdot \big(\tilde{f}^{(\infty)}_u\big)^{\beta}
=\gamma_2\cdot \big(\tilde{f}^{(\infty)}_\ell\big)^{\beta}
=\tau^{(\infty)}_{\ell,1}(\tilde{f}_\ell^{(\infty)}),$
and $\tau^{(\infty)}_{u,1}(x)=\gamma_1\cdot x^{\beta}$ and $\tau^{(\infty)}_{\ell,1}(x
)=\gamma_2\cdot x^{\beta}$ are the limit cost functions of $\Gamma^{(\infty)}_{|\{1\}}$, i.e., of
the OD pair $(o_1,t_1)$ in Figure~\ref{fig:BPR-Rate-Counterexample}(c).
Therefore,
\[
\frac{\tilde{f}_u^{(n)}}{n^{\epsilon}}=\tilde{f}^{(\infty)}_u-\frac{\eta_1-\eta_2}{\beta\cdot \big(\gamma_1\cdot (\tilde{f}_u^{(\infty)})^{\beta-1}+\gamma_2\cdot (\tilde{f}_\ell^{(\infty)})^{\beta-1}\big)}\cdot 
n^{-\epsilon\cdot \beta}+o(n^{-\epsilon\cdot \beta}),\]\[ 
\frac{\tilde{f}_\ell^{(n)}}{n^{\epsilon}}=\tilde{f}^{(\infty)}_\ell+\frac{\eta_1-\eta_2}{\beta\cdot \big(\gamma_1\cdot (\tilde{f}_u^{(\infty)})^{\beta-1}+\gamma_2\cdot (\tilde{f}_\ell^{(\infty)})^{\beta-1}\big)}\cdot 
n^{-\epsilon\cdot \beta}-o(n^{-\epsilon\cdot \beta}).
\]

Since $f^{*(n)}_{|\{1\}}$ is an NE profile 
of the subgame $\Gamma_{|\{1\}}$ w.r.t.\ 
cost functions 
$c_u(x)=(\beta+1)\cdot\gamma_1\cdot x^{\beta}+\eta_1$
and $c_\ell(x)=(\beta+1)\cdot\gamma_2\cdot x^{\beta}+\eta_2,$
we obtain similarly that 
\[
\frac{f_u^{*(n)}}{n^{\epsilon}}=\tilde{f}^{(\infty)}_u -\frac{\eta_1-\eta_2}{(\beta+1)\cdot\beta\cdot  \big(\gamma_1\cdot (\tilde{f}_u^{(\infty)})^{\beta-1}+\gamma_2\cdot (\tilde{f}_\ell^{(\infty)})^{\beta-1}\big)}\cdot 
n^{-\epsilon\cdot \beta}+o(n^{-\epsilon\cdot \beta}),\]\[
\frac{f_\ell^{*(n)}}{n^{\epsilon}}=\tilde{f}^{(\infty)}_\ell+\frac{\eta_1-\eta_2}{(\beta+1)\cdot\beta\cdot  \big(\gamma_1\cdot (\tilde{f}_u^{(\infty)})^{\beta-1}+\gamma_2\cdot (\tilde{f}_\ell^{(\infty)})^{\beta-1}\big)}\cdot 
n^{-\epsilon\cdot \beta}-o(n^{-\epsilon\cdot \beta}).
\]
So, 
\[
\tau_u(\tilde{f}_u^{(n)})=\tau_\ell(\tilde{f}_u^{(n)})=\gamma_2\cdot (\tilde{f}^{(\infty)}_\ell)^{\beta}\cdot n^{\epsilon\cdot \beta}+\frac{ \eta_1\cdot\gamma_2\cdot (\tilde{f}_\ell^{(\infty)})^{\beta-1}+\eta_2\cdot \gamma_1\cdot (\tilde{f}_u^{(\infty)})^{\beta-1}}{\gamma_1\cdot (\tilde{f}_u^{(\infty)})^{\beta-1}+\gamma_2\cdot (\tilde{f}_\ell^{(\infty)})^{\beta-1}}+o(1)
\]
and 
\[
\begin{split}
\tau_u(f^{*(n)}_u)&=\gamma_1\cdot (\tilde{f}^{(\infty)}_u)^{\beta}\cdot n^{\epsilon\cdot \beta}-\frac{\gamma_1\cdot (\tilde{f}_u^{(\infty)})^{\beta-1}\cdot ( \eta_1-\eta_2)}{(\beta+1)\cdot (\gamma_1\cdot (\tilde{f}_u^{(\infty)})^{\beta-1}+\gamma_2\cdot (\tilde{f}_\ell^{(\infty)})^{\beta-1})}+\eta_1+o(1),\\
\tau_\ell(f_\ell^{*(n)})&=\gamma_2\cdot (\tilde{f}^{(\infty)}_\ell)^{\beta}\cdot n^{\epsilon\cdot \beta}+\frac{\gamma_2\cdot(\tilde{f}_\ell^{(\infty)})^{\beta-1}\cdot ( \eta_1-\eta_2)}{(\beta+1)\cdot (\gamma_1\cdot (\tilde{f}_u^{(\infty)})^{\beta-1}+\gamma_2\cdot (\tilde{f}_\ell^{(\infty)})^{\beta-1})}+\eta_2+o(1).
\end{split}
\]
Thus $\frac{\gamma_2\cdot (\tilde{f}_\ell^{(\infty)})^{\beta-1}\cdot \tau_u(f^{*(n)}_u)+\gamma_1\cdot (\tilde{f}_u^{(\infty)})^{\beta-1}\cdot \tau_\ell(f^{*(n)}_\ell)}{\gamma_1\cdot (\tilde{f}_u^{(\infty)})^{\beta-1}+\gamma_2\cdot (\tilde{f}_\ell^{(\infty)})^{\beta-1}}+o(1)=\tau_u(\tilde{f}_u^{(n)})=\tau_\ell(\tilde{f}_\ell^{(n)}),$ and
\[
\begin{split}
&C_{\Gamma_{|\{1\}}}(\tilde{f}^{(n)}_{|\{1\}})-C_{\Gamma_{|\{1\}}}(f^{*(n)}_{|\{1\}})=
f^{*(n)}_u\cdot \big(\tau_u(\tilde{f}_u^{(n)})-\tau_u(f_u^{*(n)})\big)
+f^{*(n)}_\ell\cdot \big(\tau_u(\tilde{f}_u^{(n)})-\tau_\ell(f_\ell^{*(n)})\big)\\
&=
\Bigg[\frac{\beta\cdot (\eta_1-\eta_2)}{(\beta+1)\cdot (\gamma_1\cdot (\tilde{f}_u^{(\infty)})^{\beta-1}+\gamma_2\cdot (\tilde{f}_\ell^{(\infty)})^{\beta-1})}+o(1)\Bigg]\cdot \big(\gamma_2
\cdot (\tilde{f}_\ell^{(\infty)})^{\beta-1}\cdot f^{*(n)}_\ell-\gamma_1\cdot(\tilde{f}_u^{(\infty)})^{\beta-1}\cdot  f^{*(n)}_u\big)
\end{split}
\]
which is in $\Theta\big(T(d_{|\{1\}}^{(n)})^{-\beta+1}\big)=\Theta(n^{-\epsilon\cdot\beta+\epsilon}).$
Herein, we have used that $T(d_{|\{1\}}^{(n)})=d_1^{(n)}=n^{\epsilon}$ \Wuu{and} $\tau^{(\infty)}_{u,1}(\tilde{f}_u^{(\infty)})=\gamma_1\cdot \big(\tilde{f}^{(\infty)}_u\big)^{\beta}
=\gamma_2\cdot \big(\tilde{f}^{(\infty)}_\ell\big)^{\beta}
=\tau^{(\infty)}_{\ell,1}(\tilde{f}_\ell^{(\infty)}).$ 

Combining this with \eqref{eq:NE-SO-BPR} yields
\begin{equation}\label{eq:LowerBound_BPR_PoA}
\text{PoA}(d^{(n)})=
\begin{cases}
1+\Omega(T(d^{(n)})^{-\beta-1})&\text{if } \epsilon=0,\\
1+\Omega(T(d^{(n)})^{-\epsilon\cdot\beta+\epsilon-\beta-1})&\text{if }
\epsilon>0,
\end{cases}
\end{equation}
since $C(f^{*(n)})\in \Theta(T(d^{(n)})^{\beta+1})=\Theta(n^{\beta+1}).$

Moreover, when $\Gamma_{|\{2\}}$ is well designed , which is the case when  
$\eta_8=\eta_3+\eta_6+\eta_7,$ then \eqref{eq:NE-SO-BPR}
turns into an equation, \Wuu{i.e.,} $C(\tilde{f}^{(n)})-C(f^{*(n)})=C_{\Gamma_{|\{1\}}}(\tilde{f}^{(n)}_{|\{1\}})-C_{\Gamma_{|\{1\}}}(f^{*(n)}_{|\{1\}})$,
and (\ref{eq:LowerBound_BPR_PoA}) becomes
\begin{equation}\label{eq:LowerBound_BPR_PoA_tight}
\text{PoA}(d^{(n)})=
\begin{cases}
1+\Theta(T(d^{(n)})^{-\beta-1})&\text{if } \epsilon=0,\\
1+\Theta(T(d^{(n)})^{-\epsilon\cdot\beta+\epsilon-\beta-1})&\text{if } \epsilon>0.
\end{cases}
\end{equation}

\eqref{eq:LowerBound_BPR_PoA_tight} is the key for the analysis of the convergence rate of the the PoA for different degrees $\beta$ of the BPR cost functions. When $\beta\in (0,1)$, then
there is an unbounded sequence $(d^{(n)})_{n\in \N}$ for each $\theta\in (2\cdot\beta,\beta+1]$ such that
PoA$(d^{(n)})=1+\Theta(T(d^{(n)})^{-\theta}).$
When $\beta\ge 1$, then there is an unbounded sequence $(d^{(n)})_{n\in \N}$ for each $\theta\in [\beta+1,2\cdot\beta)$
 such that
PoA$(d^{(n)})=1+\Theta(T(d^{(n)})^{-\theta}),$ since \eqref{eq:LowerBound_BPR_PoA_tight} holds for an arbitrary $\epsilon\in [0,1),$ and we can thus put $\theta=-\epsilon\cdot\beta+\epsilon-\beta-1.$

A special case occurs when $d_2^{(n)}=0$ and $d_1^{(n)}= n^{\epsilon}$. Then
PoA$(d^{(n)})=1+\Theta(T(d^{(n)})^{-2\cdot \beta})$ for any 
$\beta>0$, which is the conjecture by \cite{O2016Mechanisms}. In our example, $\Gamma$ is then equal to \Wuu{the subgame} induced by the OD pair $(o_1,t_1)$ in 
Figure~\ref{fig:BPR-Rate-Counterexample}(b), since then
$T(d^{(n)})=T(d^{(n)}_{|\{1\}})$ and $C(\tilde{f}^{(n)})-C(f^{*(n)})=C_{\Gamma_{|\{1\}}}(\tilde{f}^{(n)}_{|\{1\}})-C_{\Gamma_{|\{1\}}}(f^{*(n)}_{|\{1\}})$.
\hfill$\square$

\end{document}